\documentclass[11pt,a4paper,x11names]{article}
\usepackage{jheppub}

\usepackage{amsfonts}
\usepackage{tikz-cd,tikz}
\usepackage{comment}
\usepackage{physics}
\usepackage{float}
\usepackage{listings}
\usepackage{calc}
\usepackage{subcaption}

\usepackage{subcaption}

\definecolor{deepgreen}{rgb}{0.0, 0.5, 0.0}

\newcommand{\simoint}{
  \mathop{\!\!\!\!\!\vcenter{\hbox{
    \begin{tikzpicture}[baseline=-0.75ex]
      
      \node at (0, 0) {\scalebox{1}{$\displaystyle\int$}};
      
      \draw[semithick] (0, -0.08) arc[start angle=-92, end angle=88, radius=0.1];
    \end{tikzpicture}
  }}\!\!\!}}

\DeclareFixedFont{\ttb}{T1}{txtt}{bx}{n}{12}
\DeclareFixedFont{\ttm}{T1}{txtt}{m}{n}{12}
\DeclareFixedFont{\ttms}{T1}{txtt}{m}{n}{9}
\DeclareFixedFont{\ttmss}{T1}{txtt}{m}{n}{7}

\definecolor{deepblue}{rgb}{0.0, 0.0, 0.55} 
\definecolor{shadecolor}{rgb}{0.96,0.96,0.91}
\definecolor{shadecolor2}{rgb}{1,1,0.99}

\newcommand{\mcomment}[1]{}
\newcommand\mathstyle{\lstset{
language=Mathematica,
basicstyle={\scriptsize\def\fvm@Scale{.5}\fontfamily{fvm}\selectfont},
otherkeywords={self},
keywordstyle=\ttb\scriptsize\color{deepblue},
emph={MyClass,__init__},
emphstyle=\ttb\color{deepred},
backgroundcolor=\color{pink!20!white},
stringstyle=\color{deepgreen},
commentstyle=\color{SkyBlue3!70!PaleGreen4},
frame=tb,
showstringspaces=false
}}

\newwrite\todofile
\immediate\openout\todofile=\jobname.tdo

\newcounter{todocounter}

\newcommand{\printtodos}{
        \section*{To-Do List}
        \immediate\closeout\todofile
        \input{\jobname.tdo}
}
\lstnewenvironment{mathematica}[1][]
{
\mathstyle
\lstset{#1}
}
{}

\newcommand{\la}[1]{\label{#1}}
\newcommand{\eq}[1]{(\ref{#1})}

    \newcommand{\beq}{\begin{equation}}
    \newcommand{\eeq}{\end{equation}}
    \newcommand\beqa{\begin{eqnarray}}
    \newcommand\eeqa{\end{eqnarray}}
\newcommand\bea{\begin{array}}
\newcommand\eea{\end{array}}

\newcommand{\ii}{i}

\newcommand{\hrho}{\hat{\rho}}
\newcommand{\heta}{\hat{\eta}}

\newlength{\widthLOne}
\newlength{\widthLTwo}
\newcommand{\bP}{\mathbf{P}}

\newcommand{\bQ}{\mathbf{Q}}

\newcommand{\lO}{\mathcal{O}}

\newcommand{\bbPP}{\mathbb{P}}
\newcommand{\bbQQ}{\mathbb{Q}}

\newcommand{\algsl}{\mathfrak{sl}}

\newcommand{\lN}{\mathcal{N}}

\def\[{\left[}
\def\]{\right]}
\def\({\left(}
\def\){\right)}
\def\<{\left<}
\def\>{\right>}
\def\d{\partial}

\def\Xint#1{\mathchoice
   {\XXint\displaystyle\textstyle{#1}}
   {\XXint\textstyle\scriptstyle{#1}}
   {\XXint\scriptstyle\scriptscriptstyle{#1}}
   {\XXint\scriptscriptstyle\scriptscriptstyle{#1}}
   \!\int}
\def\XXint#1#2#3{{\setbox0=\hbox{$#1{#2#3}{\int}$}
     \vcenter{\hbox{$#2#3$}}\kern-.5\wd0}}

\def\dashint{\Xint-}

\title{New Approach to Strongly Coupled ${\cal N}=4$ SYM via Integrability}
\author{Simon Ekhammar, Nikolay Gromov and Paul Ryan}
\affiliation{
Department of Mathematics, King's College London
}\emailAdd{simon.ekhammar@kcl.ac.uk}
 \emailAdd{nikolay.gromov@kcl.ac.uk}
 \emailAdd{paul.1.ryan@kcl.ac.uk}

\abstract{
Finding a systematic expansion of the spectrum of free superstrings on AdS$_5 \times$S$^5$, or equivalently strongly coupled $\mathcal{N}=4$ SYM in the planar limit, remains an outstanding challenge. No first principle string theory methods are readily available, instead the sole tool at our disposal is the integrability-based Quantum Spectral Curve (QSC). For example, through the QSC the first five orders in the strong coupling expansion of the conformal dimension of an infinite family of short operators have been obtained. However, when using the QSC at strong coupling one must often rely on numerics, and the existing methods for solving the QSC rapidly lose precision as we approach the strong coupling regime.

In this paper, we introduce a new framework that utilises a novel set of QSC variables with a regular strong coupling expansion. We demonstrate how to use this approach to construct a new numerical algorithm that remains stable even at a 't Hooft coupling as large as $10^6$ (or $g\sim 100$). 

Employing this approach, we derive new analytic results for some states in the $\mathfrak{sl}(2)$ sector and beyond. We present a new analytic prediction for a coefficient in the strong coupling expansion of the conformal dimension for the lowest trajectory at a given twist $L$. For non-lowest trajectories, we uncover a novel feature of mixing with operators outside the $\mathfrak{sl}(2)$ sector, which manifests as a new type of analytic dependence on the twist.

}

\date{September 2023}

\begin{document}

\maketitle

\section{Introduction}

Integrability of planar $\lN=4$ SYM provides a variety of tools to compute numerous observables: the spectrum of anomalous dimensions, correlation functions, amplitudes, Wilson-Loops etc \cite{Minahan:2002ve,Lipatov:2009nt,Beisert:2010jr,Gromov:2009bc,Drukker:2012de,Basso:2015zoa,Komatsu:2017buu,deLeeuw:2017cop}.
In particular with the Quantum Spectral Curve (QSC)~\cite{Gromov:2013pga,Gromov:2014caa} one can explore the spectrum in a wide variety of regimes: powerful analytic methods have been developed for the weak coupling expansion  to high orders  \cite{Marboe:2014gma,Gromov:2015vua}, expansions in near-BPS regimes are under good control \cite{Gromov:2014bva,Gromov:2015dfa}, and high precision numerical packages are readily available \cite{Gromov:2015wca,Gromov:2023hzc}.
Intriguingly, the QSC is also becoming used more and more in the computation of observables beyond the spectrum, e.g. higher point correlation functions \cite{Giombi:2018qox,Giombi:2018hsx,Cavaglia:2018lxi,McGovern:2019sdd,Cavaglia:2021mft,Basso:2022nny,Bercini:2022jxo,Giombi:2022anm}.

At the same time other regimes where one expects interesting physics remain challenging. The strong coupling regime is one of them, corresponding to short strings with large string tension propagating on ${\rm AdS}_5\times $S$^5$.

Currently, no systematic analytic 
techniques exist to solve the QSC at large `t Hooft coupling $g=\frac{\sqrt\lambda}{4 \pi}$. One instead needs to either rely on the extrapolation of numerical data or on extrapolation from the long quasi-classical strings regime, an approach which is likely to fail in general. Previously developed numerical approaches lose efficiency when the coupling is increased and the computational 
costs increase rapidly. Thus the numerical data at hand is still at relatively small values of the coupling and its collection is far too time-consuming if more than a few states need to be considered\footnote{Nevertheless, recently, the first $219$ states were studied systematically in a wide range of coupling in \cite{Gromov:2023hzc}.}. Similarly, from the string theory side, no systematic method exists 
which would produce the string spectrum in this regime (for some partial successes see~\cite{Roiban:2009aa,Roiban:2011fe,Vallilo:2011fj,Gromov:2011de,Frolov:2013lva}).

At the same time, recently new indirect methods became available~\cite{Alday:2022uxp,Alday:2023mvu,Alday:2023flc} based on the conformal bootstrap 
in combination with several non-trivial 
structural observations and input from localization
which allowed to constrain, or in some cases, 
compute the conformal data at strong coupling analytically. These 
results are in agreement with the information extracted from 
integrability for the spectrum but also provide rich data on 
OPE coefficients. 
Unfortunately, this method by  
itself is not constraining enough to provide a systematic 
way of extracting both the spectrum and the 3-point functions 
order by order in the coupling due to the degeneracy of the spectrum at strong coupling. One may hope that combining 
this method with strong coupling integrability techniques
may allow to push these results to higher orders or even eventually lead to the solution of the theory.
This requires a better understanding of the strong coupling regime of the QSC, 
which is the main goal of this paper.

Despite the challenges, by combining various methods some 
strong coupling data has been successfully extracted and 
appears to have a rather simple analytic form. For example 
for the Konishi operator the conformal dimension is known 
to take the form \cite{Roiban:2009aa,Gromov:2009zb,Gromov:2011de,Roiban:2011fe,Vallilo:2011fj,Gromov:2011bz,Frolov:2013lva,Gromov:2014bva}:
\begin{equation}\label{DeltaKonishi}
	\Delta_{\rm Konishi}=2 \lambda^{\frac{1}{4}}-2+2 \left(\frac{1}{\lambda}\right)^{1/4}+\left(\frac{1}{2}-3 \zeta _3\right) \left(\frac{1}{\lambda }\right)^{3/4}+\left(6 \zeta _3+\frac{15 \zeta _5}{2}+\frac{1}{2}\right) \left(\frac{1}{\lambda }\right)^{5/4}+\dots\,.
\end{equation}
With such a simple structure, it seems natural that there 
should be a way of computing it systematically within a concise analytic framework. 

\paragraph{Goal of the paper.} 
In this paper, we propose a novel way to approach the spectral problem, particularly suited for large $g$, by parametrising the QSC in terms what we call \textit{densities} -- functions which are localised in the spectral parameter and which have a regular $1/\sqrt{g}$ expansion. 
Based on this new parameterisation, we construct a new 
numerical algorithm for solving the QSC and with it we are able to reach huge 
values of the `t Hooft coupling $g\sim 100,\;\lambda \sim 10^{6}$
without any instabilities or uncontrollable growth in the 
number of numerical parameters. In fact, the number of parameters 
does not need to be changed when increasing $g$ while keeping 
the numerical error almost the same. 
With this we are able to 
confirm the expansion \eqref{DeltaKonishi} and obtain new 
predictions for higher-order terms:
\beq
\left(-\frac{81 \zeta _3^2}{4}+\frac{\zeta _3}{4}-40 \zeta _5-\frac{315 \zeta _7}{16}-\frac{27}{16}\right)\lambda^{-7/4}\;
\eeq
which can be added to the previously known expansion \eqref{DeltaKonishi}. A generalisation of this result for arbitrary twist $L$ and spin $S$ can be found in equation \eqref{Deltan1LS}. While for the leading trajectory for each twist $L$ in the ${\mathfrak{sl}}(2)$ sector we found polynomial dependence on the quantum numbers, we found that this property no longer holds for higher trajectories. Higher trajectories are distinguished by having so-called mode numbers larger than $1$. The most straightforward definition of these mode numbers are as integers appearing in the logarithmic form of the 1-loop Bethe equations, see \cite{Gromov:2023hzc} for further details. For example for mode number $2$ we found a new type of dependence on the twist $\sqrt{L^4-4L^2+36}$ which we argue is due to mixing at strong coupling with states outside ${\mathfrak{sl}}(2)$ sector.

\paragraph{Idea behind the new method.}

The simplest Q-functions of the QSC are denoted 
$\bP_a(u)$. These are complex functions with power-like asymptotics, $\bP_{a} \sim u^{\texttt{powP}_a}$ and are analytic outside of a branch cut at $(-2g,2g)$. It is often useful to resolve this branch cut, a task accomplished by introducing the Zhukovsky variable $x$ defined as $x+\frac{1}{x}=\frac{u}{g}$. Due to their analytic structure we can always express $\bP_{a}$ as a Laurent series in $x$
\begin{equation}\label{eq:LaurentExpansionWeak}
    \bP_{a} = \sum_{n=-\texttt{powP}_a}^{\infty} \frac{c_{a.n}}{x^{n}}\,,
\end{equation}
which converges until the first branch points located at $|x|<1$. 

The parameterization \eqref{eq:LaurentExpansionWeak} is the standard way to treat $\bP_a$ and is utilized in almost all studies of the QSC.  At weak coupling the coefficients scale as $c_{n>0} \sim g^{n}$ and only a finite number of terms contribute at a fixed order in $g$. Intuitively we can think about this weak coupling expansion as a perturbation around a rational spin chain, which is known to correspond to the case when $\bP_{a}$ is a rational function of $u$.

When we go to strong coupling the series re-organises itself qualitatively as
\begin{equation}\label{eq:LaurentExpansionStrong}
    \bP_{a} \simeq \sum_{m,n} \frac{c_{a,m,n}}{x^{m}(x^2-1)^{n}}
\end{equation}
for $x$ away from the branch points at $\pm 1$ \cite{Hegeds_2016}. To understand \eqref{eq:LaurentExpansionStrong} we recall that $\bP_{a}$ are known to encode the quasi-momenta, $\tilde{p}_a$, of classical string theory solutions as $\bP_{a} \sim \exp\left(-g\int^{x} dy \,\left(1-\frac{1}{y^2}\right)\tilde{p}_{a}(y)\right)$ and $\tilde{p}_{a}(x)$ are functions with poles at $x=\pm 1$ as follows from the classical Lax matrix construction. For example, for the BMN string one finds \cite{Berenstein:2002jq,Kazakov:2004qf,Beisert:2005bm}
\begin{equation}
    \tilde{p}_{1,2} = -\tilde{p}_{3,4} = 2\pi \mathcal{L}\frac{x}{x^2-1}\,,
\end{equation}
with $\mathcal{L}$ measuring the spin around a big circle in S$^{5}$ .

One needs an infinite number of terms in \eqref{eq:LaurentExpansionWeak} to reproduce \eqref{eq:LaurentExpansionStrong}, which is the simple reason that previous numerical methods become increasingly slow at large $g$. Simply switching to \eqref{eq:LaurentExpansionStrong} is not good enough, because if we zoom close to $x=\pm 1$ the singularity should disappear
and thus \eqref{eq:LaurentExpansionStrong} does not cover the important domains near the branch points $u=\pm 2g$ and the series \eqref{eq:LaurentExpansionStrong} has to be resummed.

A better parametrisation which works in all regimes is based on the spectral representation of the form
\begin{equation}
    \bP_{a}(x) \propto \oint \frac{dy}{2\pi \ii} \frac{\rho_{a}(y)}{x-y} + \dots\;,
\end{equation}
where the integration is going over the unit circle. The omitted term is a Laurent polynomial in $x$ which ensures $\bP_a$ have the correct asymptotics. The spectral representation is one of the main tools used in our new approach.

\paragraph{Results of the paper.}
In this paper we give a rigorous definition of the density $\rho$ which is localized near the branch point with the support squeezing towards $x=\pm 1$, thus naturally leading to \eqref{eq:LaurentExpansionStrong}.

\begin{figure}[H]
    \begin{center}
    \scalebox{0.8}{\begin{tikzpicture}
        \node[] (n1) at (-2-5/100,1+0.4){\includegraphics[scale=0.55]{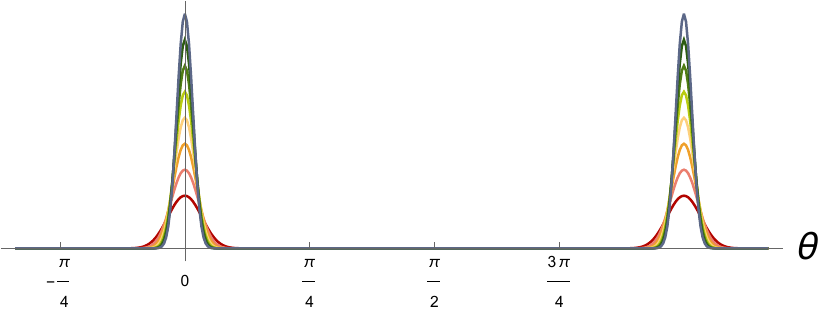}};
        \draw[thick] (-2.67-2,0.5) rectangle (-1.75-1.8,2.75);
        \draw[dashed] (-2.67-2,0.5)--(-6,-6.5-1+1);
         \draw[dashed] (-1.75-1.8,0.5)--(4.8-3.5,-6.5-1+1);
         \draw[dashed] (-2.67-2,2.75)--(-6,-1.6-1+1);
         \draw[dashed] (-1.75-1.8,2.75)--(4.8-3.5,-1.6-1+1);
        \draw[thick,fill=white] (-6,-6.5-1+1) rectangle (4.8-3.5,-1.6-1+1);
         \node[] (n2) at (-2-0.5,-4){
            \includegraphics[scale=0.4]{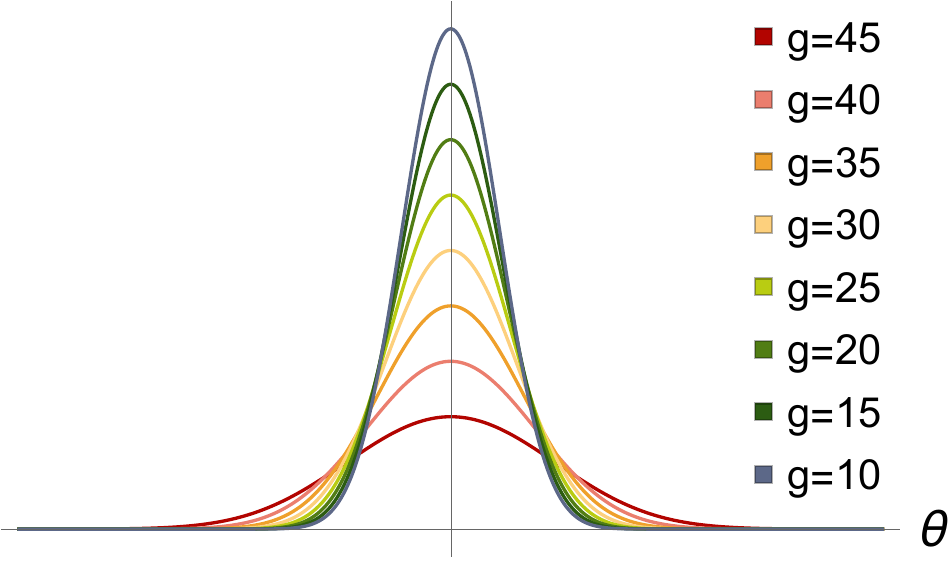}
            }; 
         \draw[->,thick] (1.75,-4)--(3,-4) node[midway,anchor=north] {\scalebox{1.2}{$\theta=\frac{t}{\sqrt{2\pi g}}$}};
         \draw[thick,fill=white] (-6+9.5,-6.5-1+1) rectangle (4.8-3.5+9.5,-1.6-1+1);
         \draw[dashed] (-6+9.5+3,-6.5-1+1+3.5)--(-6+9,-6.5-1+1+5.5);
         \draw[dashed] (-6+9.5+3,-1.6-1+0.5)--(-6+9,-1.6-1+1+3);
         \draw[dashed] (4.8-3.5+9.5-3.5,-6.5-1+1+3.5)--(4.8-3.5+8.5-3,-6.5-1+1+5.5);
         \draw[dashed] (4.8-3.5+9.5-3.5,-1.6-1+0.5)--(4.8-3.5+8.5-3,-1.6-1+1+3);
         \node[] (n3) at (-2-0.5+9.5,-4)
         {\includegraphics[scale=0.35]{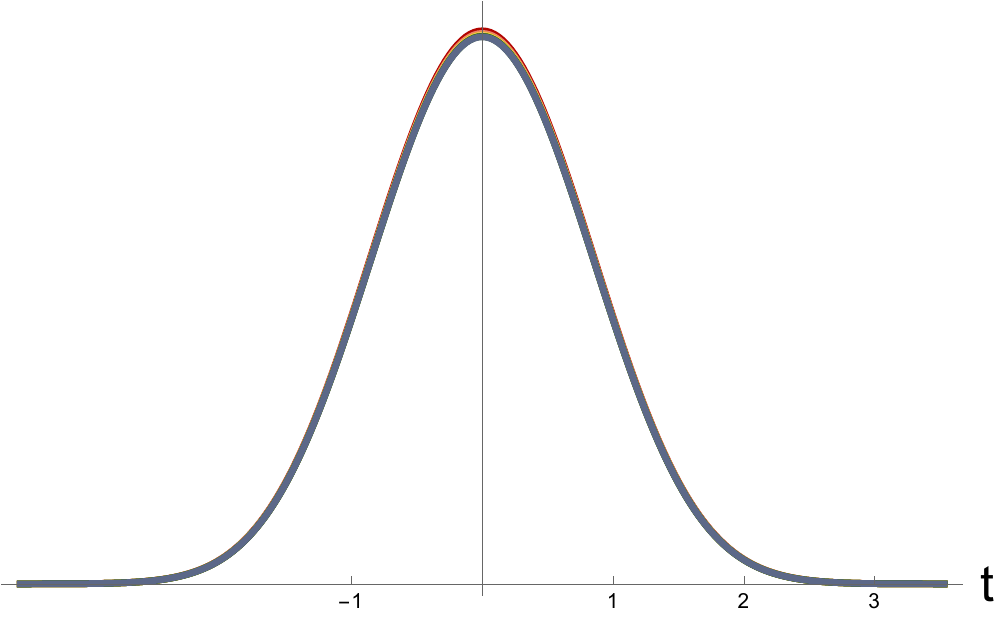}
         };
         \node[] (title 1) at (-6+1.4-0.0,-1.6-1-0.6) {\scalebox{1.4}{$\text{Im}\,\rho_1(\theta)$}};
         \node[] (title 2) at (-6+1.4+9.5,-1.6-1-0.6) {\scalebox{1.4}{$\text{Im}\,\frac{\rho_1(t)}{g}$}};
         \draw[thick,fill=white] (-6+9,-6.5-1+1+5.5) rectangle (4.8-3.5+8.5-3,-1.6-1+1+3);
        \node[] (n4) at (-2-0.5+7.5,-4+4){
            \includegraphics[scale=0.2]{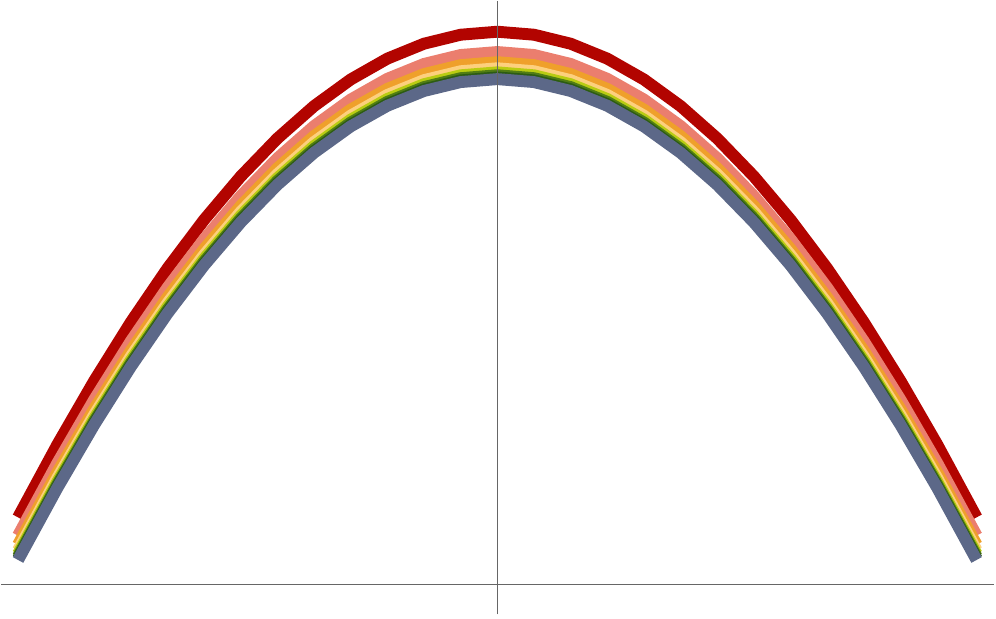}
            };
        \draw[thick] (-6+9.5+3,-6.5-1+1+3.5) rectangle (4.8-3.5+9.5-3.5,-1.6-1+0.5);
    \end{tikzpicture}}
\end{center}
   \caption{$\rho_1$ for $g=10,15,\dots,45$ with short-hand notation $\rho(\theta)\equiv \rho(e^{\ii \theta})$ and $\rho(t)\equiv \rho(e^{\frac{\ii t}{\sqrt{2\pi g}}})$. We see that the lumps get more and more squeezed towards $x=\pm 1$ or $\theta = 0,\;\pi$. The plot for $\rho(t)$ has been rescaled with a factor of $g$ to more clearly illustrate the appearance of a limiting density.}
    \label{fig:RhoPlotIntro}
\end{figure}
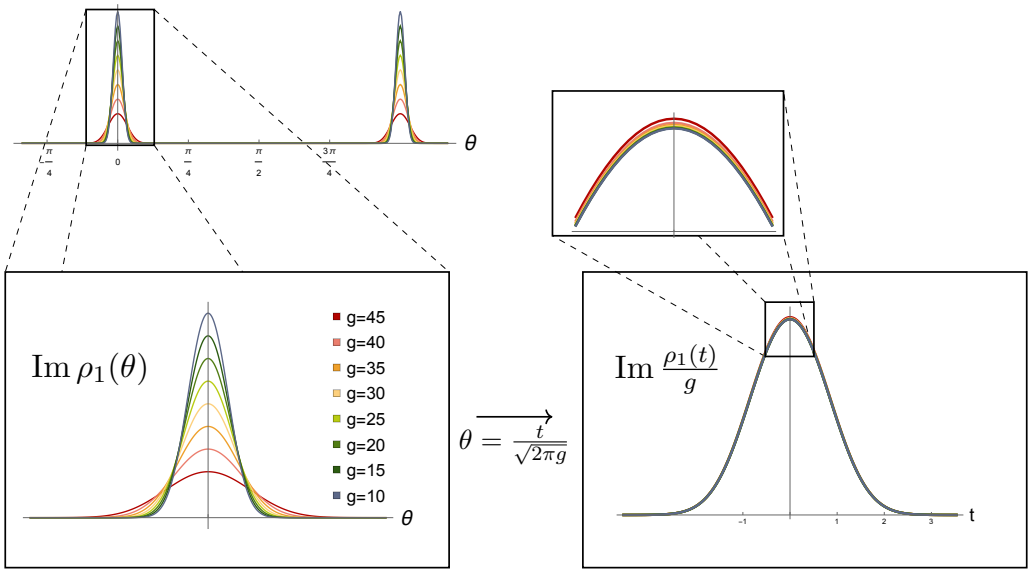

Furthermore, we found that the density has a regular well-defined expansion in $1/\sqrt{g}$.
Switching variables to the angular variable $t$, defined as $x = e^{\frac{\ii t}{\sqrt{2\pi g}}}$, we find a limiting density, see Figure~\ref{fig:RhoPlotIntro} for an illustration. 

Another novelty of our approach is that we introduce a similar density-based parametrization for another set of quantum analogues of quasi-momenta $\bQ_i$, corresponding to AdS$_5$ degrees of freedom. This allows to bypass another bottle-neck of the existing numerical approaches slowing the calculations at strong coupling.

\paragraph{Paper outline.}
In the following sections we introduce these objects in full detail
and explain how to close the QSC equations to obtain a very efficient numerical algorithm, allowing us to reach previously unreachable gigantic values of $g$. 

The remainder of 
the paper is organised as follows:
In Section \ref{sec:Basics} we review the basics of the QSC. In Section \ref{sec:introdensities} we introduce the densities and show that the QSC equations can be closed in terms of them. In Section \ref{sec:NumericalImplementation} we
discuss the new numerical algorithm. 
In Section \ref{sec:Results} we present the results of our 
numerical computations and new analytic predictions. 
In Section \ref{sec:AnalysisDensities} we present results for analytic expansion of the densities.
In Section \ref{sec:discussions} 
we discuss the results and possible future directions.
Appendices contain some technical details of the derivations.

\section{Basics of QSC}\label{sec:Basics}
Here we review the main notations and conventions needed for the next section, where we present the derivation of the new method. 
Some in-depth technical details are relegated to Appendix~\ref{app:QSC}. For an in-depth introduction to the QSC see \cite{Gromov:2014caa} and the reviews \cite{Gromov:2017blm,Kazakov:2018ugh,Levkovich-Maslyuk:2019awk}.

\subsection{Q-functions and Quantum Numbers}

States in planar $\lN=4$ SYM or, equvivalently, free strings on AdS$_5\times$S$^5$ can be labelled by six quantum numbers $[\bar{\Delta},S_1,S_2,J_1,J_2,J_3]$: three $R$-symmetry charges $(J_1,J_2,J_3)$ and three ${\rm AdS}_5$ quantum numbers $(\bar{\Delta},S_1,S_2)$ -- the conformal dimension $\bar{\Delta}$ and two Lorentz spins $S_1$ and $S_2$. For simplicity, we restrict our attention to the $\mathfrak{sl}(2)$ sector, that is states with quantum numbers $[\bar{\Delta},S,0,L,0,0]$ which in the gauge theory have the schematic form 
\begin{equation}
    \lO = {\rm Tr}\left(D^S Z^L\right) + \text{permutations}
\end{equation}
where $D$ is a light-cone covariant derivative and $Z$ is a complex scalar. Nevertheless, the results of this paper can be generalised to general states straightforwardly. 

We introduced the notation $\bar{\Delta}$ for the conformal dimension in $\mathfrak{sl}(2)$ to distinguish it from the dimension of the superconformal primary $\Delta=\bar{\Delta}-2$. For example for the Konishi multiplet with $S=2$, $L=2$  we have at weak and strong coupling
\begin{equation}
    \bar{\Delta} = 4+\lO(g^2)\;,
    \quad
    \bar{\Delta} = 2\,\lambda^\frac{1}{4}+ \lO\left(\lambda^{-\frac{1}{4}} \right)\,.
\end{equation} 

The QSC is a system of $256$ Q-functions: functions of one complex variable $u$ that encodes an infinite number of conserved charges. Among the Q-functions there are $4+4$ Q-functions, $\bP_{a},\;\bQ_i$ with $a,i=1,\dots,4$ that serves as building blocks. 
They have powerlike asymptotics encoding the quantum numbers,
\begin{equation}\label{eqn:PQlargeu}
    \bP_{a} \simeq \mathbb{A}_a\,u^{\texttt{powP}_a}\;,
    \quad
    \bQ_{i} \simeq \mathbb{B}_i\,u^{\texttt{powQ}_i}\;,
\end{equation}
with
\begin{align}\label{eqn:Pasympt}
\mathtt{powP}_a &= \left(-1-\frac{L}{2},-\frac{L}{2},-1+\frac{L}{2},\frac{L}{2}\right)_a\;,\\
\mathtt{powQ}_i&=\left(\frac{\bar{\Delta}}{2}-\frac{S}{2},\frac{\bar{\Delta}}{2}+\frac{S}{2}-1,-\frac{\bar{\Delta}}{2}-\frac{S}{2},-\frac{\bar{\Delta}}{2}+\frac{S}{2}-1 \right)_i\,.
\end{align}

As is natural from their asymptotics, $\bP_{a}$ can be associated to the S$^{5}$ degrees of freedom and $\bQ_i$ to the AdS$_5$ degree of freedom. However, these functions are not independent but related through a 4-order \emph{Baxter equation} whose explicit form we recall in Appendix~\ref{app:QSC}.

The properties so far described are expected to hold for a variety of different types of integrable models with $\mathfrak{psu}(2,2|4)$ symmetry algebra. It is the analytic properties of the Q-functions that distinguishes the QSC from those other models with the same symmetry group. In the remainder of this section we briefly recall these properties.  

\subsection{$\bP\mu$-system}

\begin{figure}[H]
    \centering
    \includegraphics[scale=0.4]{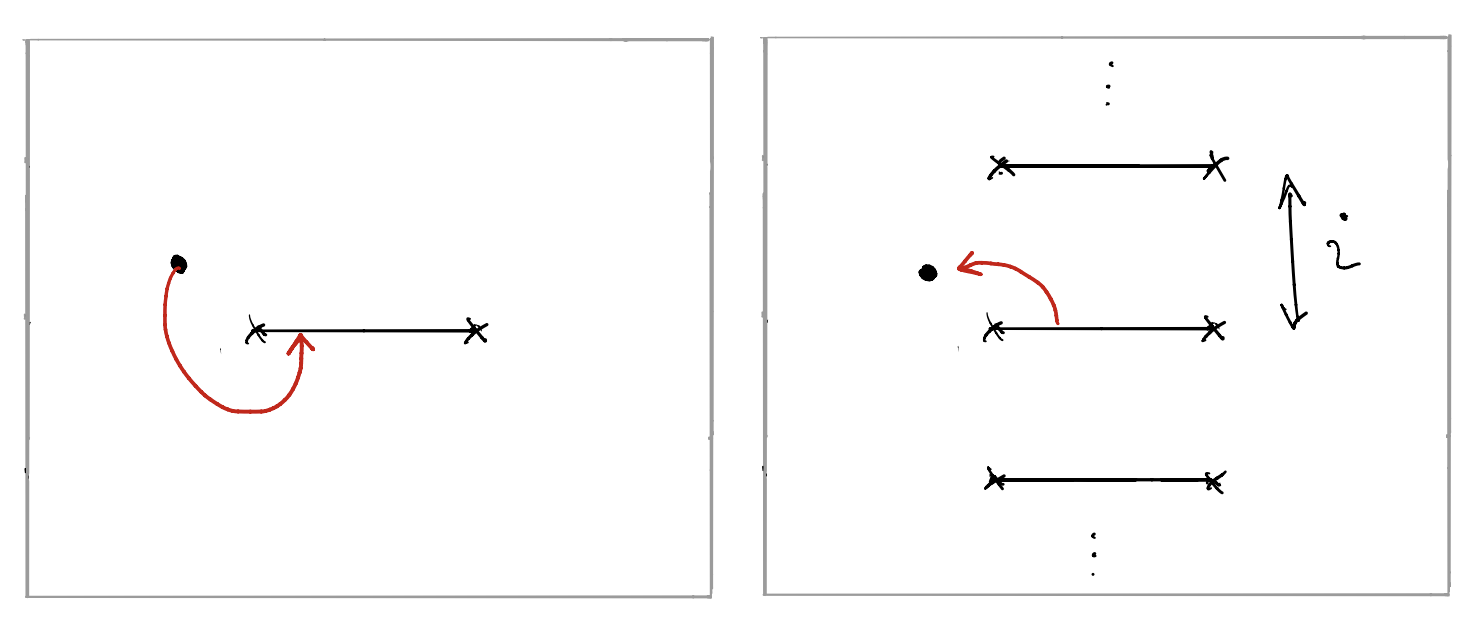}
    \caption{On their defining sheet the $\bP$-functions have a single short cut (left) connecting branch points at $\pm 2g$. Analytically continuing through the cut brings us to a new sheet with an infinite ladder of cuts separated by $i$ (right).}
    \label{fig:Pfns2}
\end{figure}

$\bP_{a}(u)$ only have two square root type branch points at $\pm 2g$ connected by a single short cut. Passing through this cut to the other sheet reveals an infinite number of square-root type branch points located at 
$\pm 2g+i \mathbb{Z}$ connected by short branch cuts 
$[-2g+i\,n,2g+i\,n]$, $n\in \mathbb{Z}$, see Figure 
\ref{fig:Pfns2}. 

Denoting by $\tilde{f}$ the analytic continuation of a 
function around the branch points at $\pm 2g$ we 
have~\cite{Gromov:2013pga,Gromov:2014caa}\footnote{In this discussion we limit ourselves to the left-right-symmetric sector.}
\begin{equation}\label{eqn:Pmueqns}
    \widetilde{\bP}_a = \mu_{a}^{\ b}\,\bP_b\,,
\end{equation}
where $\mu_{a}^{\ b}$ are new functions naturally defined with an infinite ladder of long cuts 
(i.e. cuts $(-\infty+i n,-2g+in]\cup[2g+in,+\infty+i n)$, $n\in \mathbb{Z}$) 
and are periodic functions $\mu_{a}^{\ b}(u+i)=\mu_{a}^{\ b}(u)$.

 Not all $\mu_a^{\ b}$ are independent, but satisfy 
\begin{equation}\label{eq:MuConstraints}
    \mu_{a}^{\ c}\chi_{cb}=-\mu_{b}^{\ c}\chi_{ca}\;,
    \quad
    \mu_{1}{}^{1} = -\mu_{2}{}^{2}\;,
\end{equation}
and the Pfaffian condition
\begin{equation}\label{eq:Pfaffian}
    \mu_{1}{}^{1}\mu_{1}{}^{1} + \mu_{1}{}^{2}\mu_{2}{}^{1} + \mu_{1}{}^{3}\mu_{3}{}^{1}  = 1 
\end{equation}
where $\chi_{ab}$ is a constant antisymmetric tensor defined in \eqref{eqn:defnchi}.

\subsection{$\bQ$-functions}
\begin{figure}[h]
    \centering
    \includegraphics[scale=0.3]{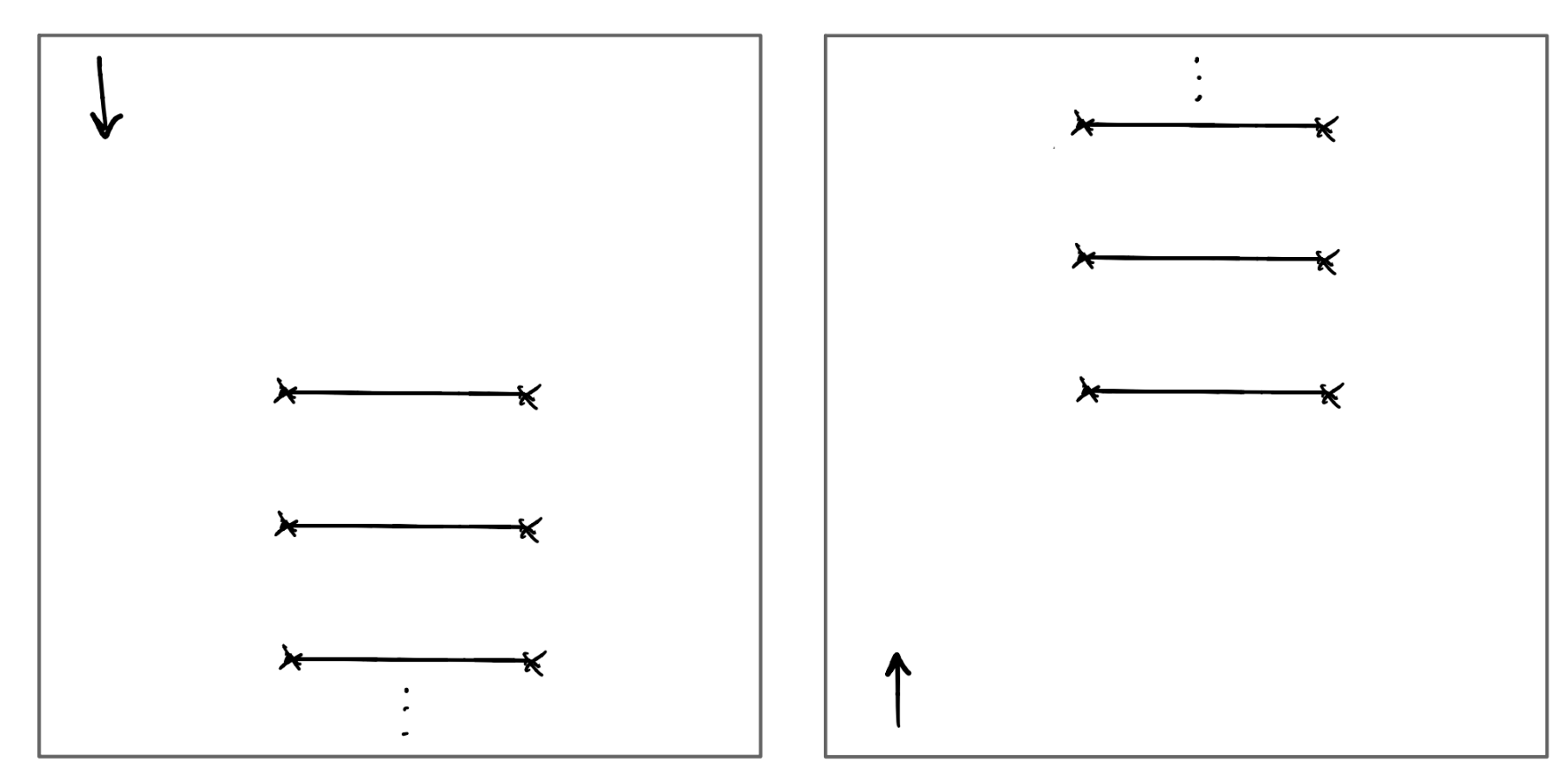}
    \caption{$\bQ^\downarrow$ have an infinite ladder of short cuts in the lower half plane starting on the real axis (left), whereas $\bQ^\uparrow$ have an infinite ladder of cuts in the upper half plane (right).}
    \label{fig:Qupdown}
\end{figure}

The $\bQ$-functions obtained from solving the Baxter equation have an infinite ladder of short cuts. The solutions can be chosen such that they are analytic in either the upper or lower half planes, which we denote as $\bQ_i^\downarrow$ and $\bQ_i^\uparrow$ respectively\footnote{The arrows denote the direction one needs to go to find the ladder of cuts.}, see Figure~\ref{fig:Qupdown}. We refer to these solutions as UHPA (upper half-plane analytic) or LHPA (lower half-plane analytic) correspondingly. Since a fourth-order difference equation can only have $4$ linearly independent solutions, $\bQ_i^\downarrow$ and $\bQ_i^\uparrow$ must be related by an $i$-periodic matrix which we denote $\Omega_i^{\ j}(u)$
\begin{equation}\label{eqn:Omega}
    \bQ_i^\uparrow = \Omega_i^{\ j}\bQ_j^\downarrow\,.
\end{equation}

\paragraph{Gluing conditions.}
The analytic continuation of the UHPA $\bQ$ functions $\bQ_i^\downarrow$ through the cut on the real axis produces functions which are LHPA \cite{Gromov:2014caa}. From the Baxter equation these should be a linear combination of the $\bQ^\uparrow_i$, given by so-called gluing conditions. In particular, the overall normalisation of the $\bQ$-functions can be chosen so that the gluing conditions are given by 
\begin{equation}\label{eqn:gluing}
    \widetilde{\bQ^\downarrow_1} = \bQ_3^\uparrow,\quad \widetilde{\bQ^\downarrow_2} = \bQ_4^\uparrow\,.
\end{equation}
The gluing condition also ensures that one can switch the picture to long cuts, where $\bQ$ and $\bP$ interchange their roles so that $\bQ$ become a function with one cut and $\bP$ and $\tilde\bP$ become UHPA and LHPA functions with long cuts ensuring equivalence between these two descriptions.

\section{Detailed Construction of QSC Densities}\label{sec:introdensities}

We now introduce the main new quantity to which we refer as {\it densities}. The key feature, which we derive in the remainder of this section, is that the whole QSC can be reconstructed from a simple set of densities localised near the branch points. Furthermore, these densities are shown to be useful variables for the future analytic strong coupling analysis as they have a finite limit up to a simple re-scaling when $g\to\infty$. We now proceed to construct these densities.

\subsection{$\bP$ and $\rho$}

Recall that we can parameterise the $\bP_{a}$ by their series expansion in $1/x$ as in \eq{eq:LaurentExpansionWeak}.
This is a convergent series with the radius of convergence determined by the first branch-point located at $|x|< 1$.

This parameterisation is very convenient at weak coupling, as only finite number of terms remains, however, at strong coupling we find qualitatively that $c_{a,n} \sim (1+\frac{1}{\sqrt{g}})^n$ which means the truncation of the sum \eqref{eq:LaurentExpansionWeak} become increasingly less efficient.
At the same time the behaviour of $\bP_a$ in the vicinity of the branch point and away become drastically different with increased $g$, which indicates a need for a different representation, capturing the strong coupling features effectively.

We define the \textit{densities} $\rho_1$ and $\rho_2$  in the following way
\begin{equation}\label{eq:DensitiesFirstDef}
    \rho_1(x) = x^{\frac{L}{2}-1}\left(\bP_1(x) - \bP_3\left(\frac{1}{x}\right)\right)\;,\quad \rho_2(x) = x^{\frac{L}{2}-1}\left(\bP_2(x) - \bP_4\left(\frac{1}{x}\right)\right)\;.
\end{equation}
These densities contain the full information about the functions $\bP_a$ and for $|x|>1$ we can write
\begin{equation}\label{eqn:Pfromrho}
\begin{split}
   & \bP_1(x) = x^{-\frac{L}{2}+1}\displaystyle \oint\frac{ {\rm d}y }{2\pi i}\frac{\rho_1(y)}{x-y}\;,\\
   & \bP_3(x) = x^{\frac{L}{2}-1}\displaystyle \oint \frac{{\rm d}y }{2\pi i}\frac{\rho_1(y)}{\frac{1}{x}-y}\;,\\
   & \bP_2(x) = x^{-\frac{L}{2}}A+x^{-\frac{L}{2}+1}\displaystyle \oint \frac{{\rm d}y}{2\pi i} \frac{\rho_2(y)}{x-y}\;,\\
   & \bP_4(x) = x^{\frac{L}{2}}A+x^{\frac{L}{2}-1}\displaystyle \oint\frac{ {\rm d}y }{2\pi i}\frac{\rho_2(y)}{\frac{1}{x}-y}\;,\\
\end{split}
\end{equation}
with the contour of integration taken counter-clockwise around the unit circle.

The particular linear combinations in \eqref{eq:DensitiesFirstDef} was picked to ensure that $\rho$ can be made to only have support near the branch points. On the unit circle we have $\tilde{\bP}_a(x) = \bP_{a}(1/x)$ and so using the $\bP \mu$-system equations \eqref{eqn:Pmueqns} it follows that
\begin{equation}\label{eqn:rhodensities}
\begin{split}
   & x^{1-\frac{L}{2}}\rho_1(x) =(1-\mu_{3}^{\ 1})\bP_1-\mu_{3}^{\ 3}\bP_3-\mu_{3}^{\ 4}\bP_4\,, \\
   & x^{-\frac{L}{2}+1}\rho_2(x)= (1-\mu_{4}^{\ 2})\bP_2-\mu_{4}{}^{3}\bP_3-\mu_{4}^{\ 4}\bP_4 \,.
\end{split}
\end{equation}
The key idea, which we explain in detail in the next paragraph, is that we can perform a linear (gauge) transformation of $\bP_a$ and $\mu_{a}{}^{b}$ so that at strong coupling
\beq\label{eq:muConstant}
\mu_{a}^{\ b}\simeq \left(
\begin{array}{cccc}
    0 & 0 & 1 & 0 \\
    0 & 0 & 0 & 1 \\
    1 & 0 & 0 & 0 \\
     0 & 1 & 0 & 0 \\
\end{array}
\right) + \mathcal{O}(e^{-4\pi g})\,.
\eeq
for $u$ of order $\mathcal{O}(g^{0})$. This means that $\rho_a$ is exponentially suppressed away from $x=\pm 1$.

\paragraph{Fixing the gauge.}
In this paragraph we explain how to achieve  \eqref{eq:muConstant}. When $g$ is large and $u\sim 1$ the branch points are very far away and $\mu_{a}^{\ b}(u)$ can be expanded into a Fourier 
series along the imaginary axis of the form 
$\sum_{n} c_n\, e^{2\pi n u} \,G_0^{|n|}$. This series should converge until the first branch-points at $u= \pm 2g$, then  $G_0=e^{-4\pi g}$ and is thus an exponentially small factor, implying that between the branch points, and sufficiently far from them, $\mu$ is a constant matrix with exponential precision.

Next, in order to bring $\mu$ to the form \eq{eq:muConstant}, we can act on P-functions with linear transformations $\bP_{a} \mapsto H_{a}{}^{b}\;\bP_{b}\;, \mu_{a}{}^{b} \mapsto H_{a}{}^{c}\;\mu_{c}{}^{d}(H^{-1})_{d}{}^{b}$. $H$ must be a constant lower-triangular  matrix to preserve the leading asymptotics of $\bP_a$ and it must satisfy $H_{a}{}^{c}\chi^{ab} H_{b}{}^{d} = \chi^{cd}$ to preserve the constant tensor $\chi^{ab}$. Using  \eqref{eq:MuConstraints} and \eqref{eq:Pfaffian} it is easy to see that $H$ can, indeed, bring $\mu$ to the form \eqref{eq:muConstant}.

\begin{figure*}[h]
    \centering
    \begin{subfigure}[t]{0.5\textwidth}
        \centering
        \includegraphics[scale=0.35]{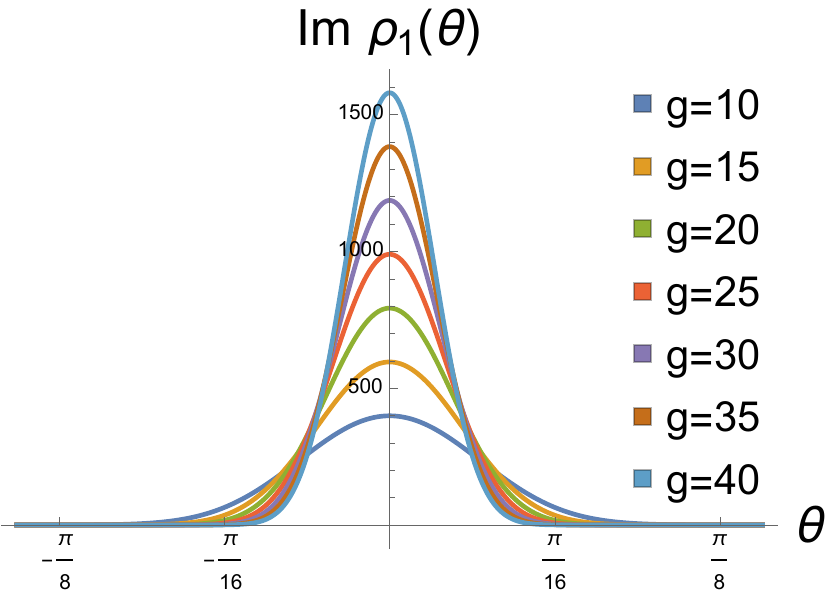}        
    \end{subfigure}%
    ~ 
    \begin{subfigure}[t]{0.5\textwidth}
        \centering
        \includegraphics[scale=0.35]{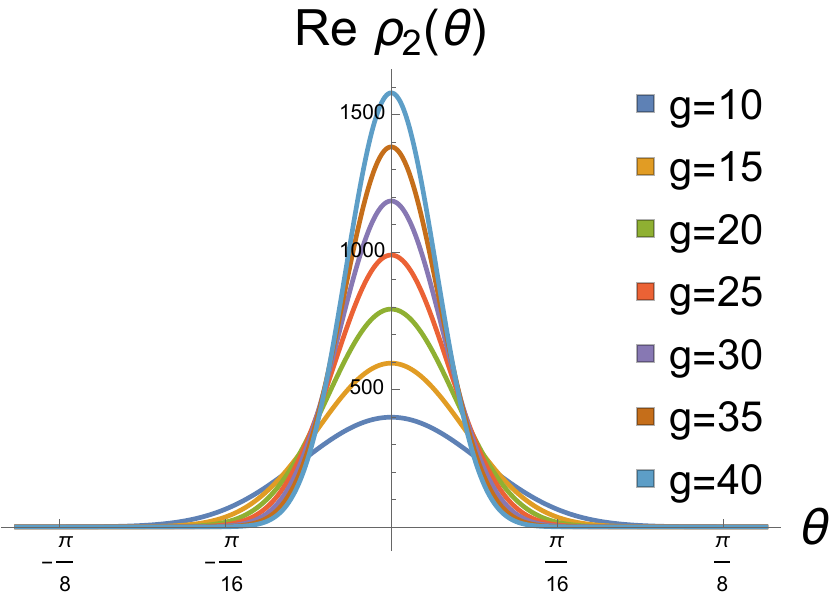}
    \end{subfigure}
    \begin{subfigure}[t]{0.5\textwidth}
        \centering
        \includegraphics[scale=0.36]{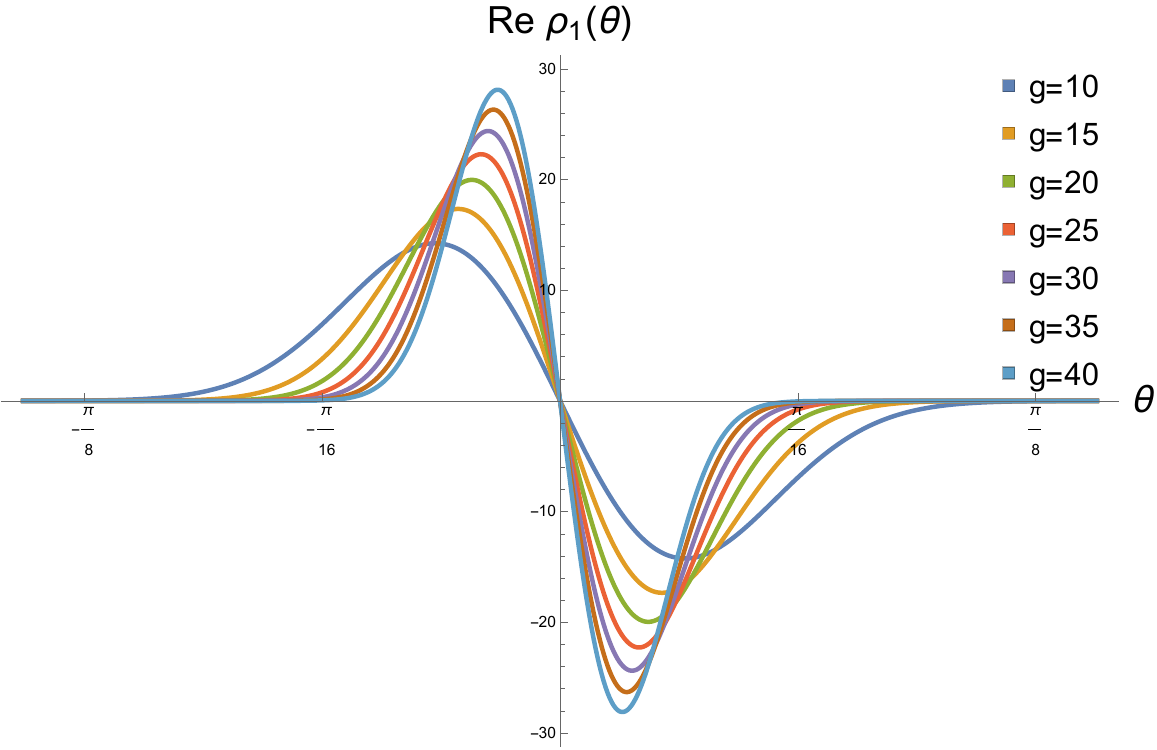}
    \end{subfigure}%
    ~
    \begin{subfigure}[t]{0.5\textwidth}
        \centering
        \includegraphics[scale=0.28]{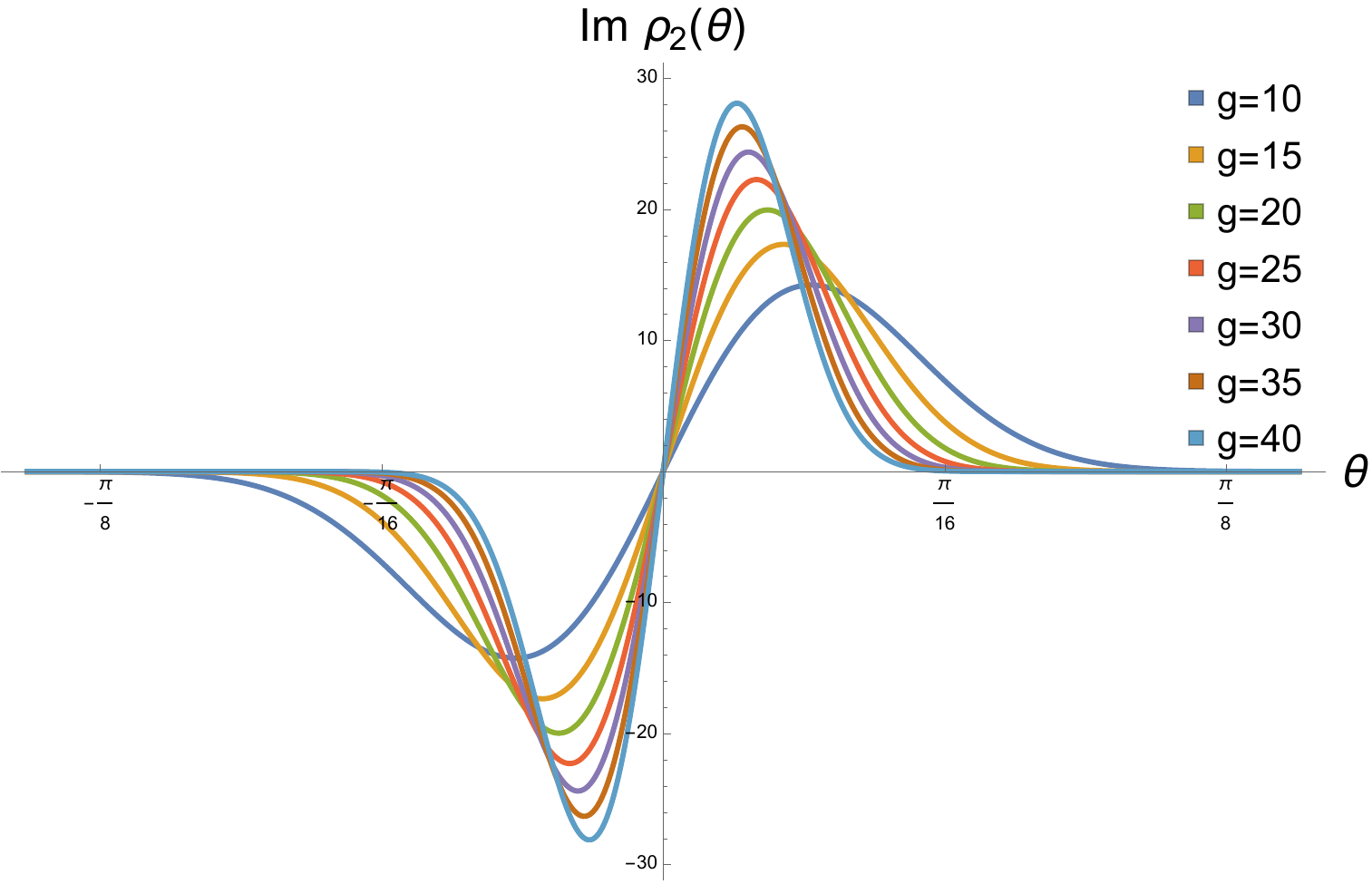}
    \end{subfigure}
    \caption{The densities $\rho_a$ for Konishi for $g=10,15,\dots,40$ using the numerical algorithm of Section~\ref{sec:NumericalImplementation}, for $x=e^{i\theta}$. }
    \label{fig:rhoplotsmain}
\end{figure*}

\paragraph{Plots of the densities from numerics.} As a concrete example we plot $\rho$ for the Konishi multiplets, $L=2, S=2$, in Figure~\ref{fig:rhoplotsmain}. The density data used in the plot has been acquired using the algorithm to be detailed in Section~\ref{sec:NumericalImplementation}.

\subsection{$\bQ$ and $\eta$}
Similar to the densities $\rho$ parameterising $\bP$, we now introduce densities $\eta$ parameterising $\bQ$. Our goal is to use linear combinations of $\bQ^{\downarrow}$ and $\bQ^{\uparrow}$ to find a density with an exponential fall-off away from $x=\pm 1$, mimicking the properties of $\rho$. As a first step, consider the following combination: 
\begin{equation}\label{eqn:Qdownmup}
    \bQ^\downarrow_j(u)-\bQ_j^\uparrow(u)\,.
\end{equation}
For ${\rm Re}\, u>2g$ this combination is exponentially suppressed $\lO\left(e^{-2\pi u} \right)$. Indeed, since both $\bQ_j^\downarrow$ and $\bQ_j^\uparrow$ solve the Baxter equation and have the same asymptotics at $u\to +\infty$ the $i$-periodic matrix $\Omega(u)$ relating them, see \eqref{eqn:Omega}, must be of the form
\begin{equation}
    \Omega_{i}^{\ j}(u) = \delta_i^{\ j} + \lO(e^{-2\pi u})\;,\quad u\rightarrow \infty\;,\quad u>2g\;.
\end{equation}

Now let's investigate \eqref{eqn:Qdownmup} for ${\rm Re}\, u<-2g$. For this one can analytically continue the large $u$ asymptotic of $\bQ$-functions from $u\rightarrow +\infty$ to $u\rightarrow -\infty$. Due to the infinite ladder of cuts in the $\bQ$-functions this requires some care: since $\bQ^\downarrow$ is UHPA the large $u$ asymptotic should be analytically continued along a large semi-circle in the upper half plane counterclockwise. Similarly, the asymptotics of $\bQ^\uparrow$ should be analytically continued with a clockwise semi-circle in the lower-half plane. Since at large $u$ $
  \bQ_j(u)\, \sim\,   u^{\mathtt{powQ}_j}$ we get
\begin{equation}
    \bQ_j^\downarrow(-\infty) \sim e^{i\pi \mathtt{powQ}_j}\bQ^\downarrow(+\infty),\quad \bQ_j^\uparrow(-\infty) \sim e^{-i\pi \mathtt{powQ}_j}\bQ^\uparrow(+\infty)\,,
\end{equation}
implying that with exponential precision 
\begin{equation}
    \bQ_j^\downarrow(u) \simeq e^{2i\pi \mathtt{powQ}_j}\bQ_j^\uparrow(u)\;,\quad u\to-\infty\;,
\end{equation}
which then implies that the combination \eq{eqn:Qdownmup}
is not exponentially decaying for $u<-2g$.
In order to solve this problem one can introduce a $u$-dependent factor, multiplying $\bQ$'s. More precisely we consider 
\begin{equation}\label{eqn:smallq}
\begin{split}
    & q_1(x) = x^{-\frac{\Delta}{2}+\frac{S}{2}-2}\;\bQ_1(x)\;,\\
    & q_2(x) = x^{-\frac{\Delta}{2}-\frac{S}{2}}\;\bQ_2(x)\;,\\
    & q_3(x) = x^{\frac{\Delta}{2}-\frac{S}{2}+2}\;\bQ_3(x)\;,\\
    & q_4(x) = x^{\frac{\Delta}{2}+\frac{S}{2}}\;\bQ_4(x)\;,
\end{split}\,.
\end{equation}
The $q$-functions are defined to have integer asymptotics
\begin{align}\label{eqn:Pasymptq}
q_i\sim u^{\left(-1,0,1-S,S-2\right)_i}\,,
\end{align}
which allows us to introduce the key quantities of this subsection, $\eta_j(x)$ as follows
\begin{equation}
    \eta_j(x) = q_j^\downarrow(x) - q_j^\uparrow(x)\,,
\end{equation}
which is indeed exponentially suppressed for $|{\rm Re}\, x|>1$.

\paragraph{Moving inside the unit circle.}
We now investigate what happens when we analytically continue $\eta_j(x)$ inside the unit circle. In terms of the $q_j(x)$ functions, the gluing conditions \eqref{eqn:gluing} take the form, in a suitably chosen gauge
\begin{equation}\la{eq:gluing}
\begin{split}
   & q_1^\downarrow\left(x\right) = q_3^\uparrow\left(\frac{1}{x}\right),\quad q_2^\downarrow(x) = q_4^\uparrow\left(\frac{1}{x} \right)\,, \quad |x|\leq 1,\ {\rm Im}\;x>0\,,\\
   & q_1^\uparrow\left(x\right) = q_3^\downarrow\left(\frac{1}{x}\right),\quad q_2^\uparrow(x) =  q_4^\downarrow\left(\frac{1}{x} \right)\,,\quad |x|\leq 1,\ {\rm Im}\;x<0\,,\\
\end{split}
\end{equation}
and as a result we find for $x$ inside the unit circle on the real axis that 
\begin{equation}
    \eta_1(x)= -\eta_3(1/x),\quad  \eta_2(x)= -\eta_4(1/x),\; |x|\leq 1\,.
\end{equation}
In other words, the densities $\eta_j(x)$ are also exponentially suppressed on the real axis for $|x|\to 0$. 

\paragraph{Constructing the Riemann-Hilbert problem.}

\begin{figure}[H]
    \centering
    \includegraphics[scale=0.4]{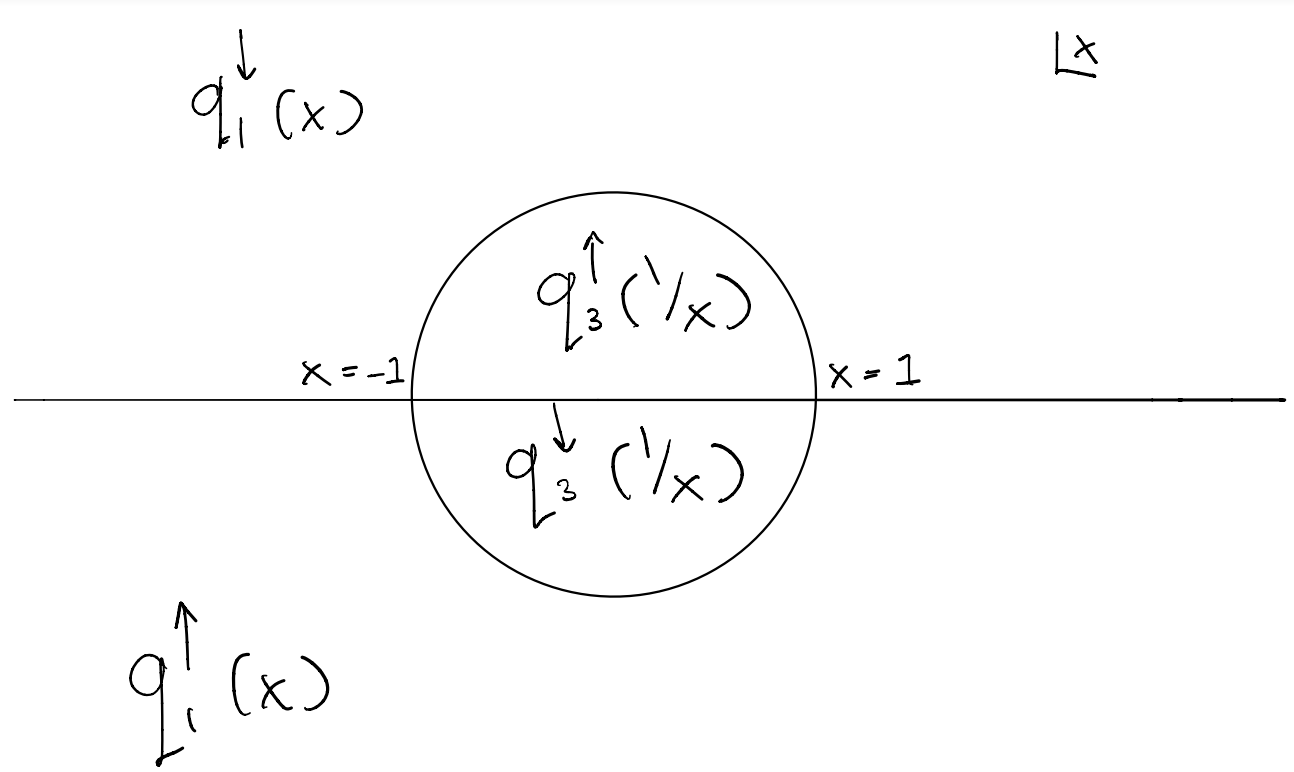}
    \caption{The sectionally analytic function $q_{13}(x)$ defined in \eqref{q13eq}. Note that there is no discontinuity across the unit circle thanks to the gluing condition~\eq{eq:gluing}. }
    \label{fig:q13interstellar}
\end{figure}

We now show that all $\bQ$-functions can be reconstructed from $\eta_j$ by solving a simple set of Riemann-Hilbert problems. To demonstrate the idea we focus on $q_1$ and $q_3$. Consider the following sectionally-analytic function, in the $x$-plane, 
\beq\la{q13eq}
q_{13}(x)\equiv 
\left\{
\bea{ll}
q_1^\downarrow(x)\;\;&,\;\;|x|>1\;\;{\rm and}\;\;{\rm Im}\;x>0\\
q_1^\uparrow(x)\;\;&,\;\;|x|>1\;\;{\rm and}\;\;{\rm Im}\;x<0\\
q_3^\uparrow(1/x)\;\;&,\;\;|x|<1\;\;{\rm and}\;\;{\rm Im}\;x>0\\
q_3^\downarrow(1/x)\;\;&,\;\;|x|<1\;\;{\rm and}\;\;{\rm Im}\;x<0
\eea
\right.\;,
\eeq
see Figure \ref{fig:q13interstellar}.

This function has a discontinuity on the whole real axis, but is regular across the unit circle due to the gluing condition~\eq{eq:gluing}. Furthermore, it is decaying at infinity and its discontinuity on the real axis is exponentially suppressed away from $x=\pm 1$. 

This provides a very simple Riemann-Hilbert problem whose solution is given by
\begin{equation}\label{eqn:q13solution}
    q_{13}(x) = \displaystyle \int^{\infty}_{-\infty} \frac{{\rm d}y}{2\pi i} \frac{\eta_{13}(y)}{y-x}
\end{equation}
where we have defined the discontinuity $\eta_{13}(x)$ by
\begin{equation}
    \eta_{13}(x) = \begin{cases} 
      \eta_1(x) & |x|>1 \\
     -\eta_3(x) & |x|<1 
     \end{cases}
\end{equation}
which is regular on the real $x$-axis thanks to the gluing conditions.

\paragraph{Checking asymptotics.} 
Both $q_1(x)=q_{13}(x)$ and $q_3(x)=q_{13}(1/x)$ are decaying at infinity, meaning that no polynomial in $x$ can be added to \eq{q13eq}. At the same time, $q_3(x)\sim x^{1-S}$ implying that some moments of the density $\eta_{13}(x)$ should vanish:
\begin{equation}\label{eqn:vanishingmoments}
    \displaystyle\int^{\infty}_{-\infty} \frac{\eta_{13}(y)}{y^{n}}=0,\quad n=1,\dots,S-1\,.
\end{equation}
This should be imposed additionally in our numerical procedure. For the particular case of $S=2$ with parity symmetry \eq{eqn:vanishingmoments} is satisfied automatically. 

\paragraph{Reconstructing $\bQ_1$ and $\bQ_3$.}

From here it is trivial to construct the original $\bQ$-functions 
\begin{equation}\label{eqn:Q1Q3}
\begin{split}
   & \bQ^\downarrow_1(x) = \frac{x^{\frac{\Delta}{2}-\frac{S}{2}+2}}{2\pi i}\displaystyle \int^{\infty}_{-\infty} {\rm d}y \frac{\eta_{13}(y)}{y-x},\; |x|>1,\; {\rm Im}\; x>0\,, \\
   & \bQ^\downarrow_3(x) = \frac{x^{-\frac{\Delta}{2}+\frac{S}{2}-2}}{2\pi i}\displaystyle \int^{\infty}_{-\infty} {\rm d}y \frac{\eta_{13}(y)}{y-\frac{1}{x}},\; |x|>1,\; {\rm Im}\; x>0\,, \\
\end{split}
\end{equation}
and we swap $\downarrow$ for $\uparrow$ if ${\rm Im}\; x<0$.

\paragraph{Constructing $\bQ_2$ and $\bQ_4$.}

We can now repeat exactly the same type of argument to construct $\bQ_2$ and $\bQ_4$. We define 
\beq
q_{24}(x)\equiv 
\left\{
\bea{ll}
q_2^\downarrow(x)\;,\;&\;\;|x|>1\;\;{\rm and}\;\;{\rm Im}\;x>0\\
q_2^\uparrow(x)\;,\;&\;\;|x|>1\;\;{\rm and}\;\;{\rm Im}\;x<0\\
q_4^\uparrow(1/x)\;,\;&\;\;|x|<1\;\;{\rm and}\;\;{\rm Im}\;x>0\\
q_4^\downarrow(1/x)\;,\;&\;\;|x|<1\;\;{\rm and}\;\;{\rm Im}\;x<0
\eea
\right.
\eeq
with the discontinuity $\eta_{24}(x)$ on the real axis given by 
\begin{equation}
    \eta_{24}(x) = \begin{cases} 
      \eta_2(x) & |x|>1 \\
     -\eta_4(x) & |x|<1 
     \end{cases}\,.
\end{equation}
Then we have
\begin{equation}
    q_{24}(x) = R_{24}(x)+\frac{1}{2\pi i}\displaystyle \int^{\infty}_{-\infty} {\rm d}y \frac{\eta_{24}(y)}{y-x}
\end{equation}
where $R_{24}(x)$ is a function without discontinuity, which can only have a singularities at zero or at infinity i.e. is 
a Laurent polynomial. The difference with the case of $q_{13}$ is that $q_2(x)$ is constant at infinity and $q_4(x)$ goes like $x^{S-2}$ so $R_{24}(x)$ must be of the form
\begin{equation}
    R_{24}(x) = \displaystyle \sum_{n=0}^{S-2} \frac{r_n}{x^n}
\end{equation}
for some constants $r_0,\dots r_{S-2}$. These are extra parameters in addition to the density which are needed to recover the Q-functions.

Finally, $\bQ_2$ and $\bQ_4$ are then given by 
\begin{equation}\label{eqn:Q2Q4}
\begin{split}
   & \bQ^\downarrow_2(x) = x^{\frac{\Delta}{2}+\frac{S}{2}}q_{24}(x),\; |x|>1,\; {\rm Im}\; x>0\,, \\
   & \bQ^\downarrow_4(x) = x^{-\frac{\Delta}{2}-\frac{S}{2}}q_{24}\left(\frac{1}{x}\right),\; |x|>1,\; {\rm Im}\; x>0 \,,
\end{split}
\end{equation}
and we swap $\downarrow$ with $\uparrow$ if ${\rm Im}\; x<0$.

The equations \eqref{eqn:Q2Q4} and \eqref{eqn:Q1Q3} recover the 4 $\bQ$-functions from the local densities and a finite set of constants $r_0,\dots,r_{S-2}$. 

\begin{figure*}[h]
    \centering
    \begin{subfigure}[t]{0.5\textwidth}
        \centering
        \includegraphics[scale=0.35]{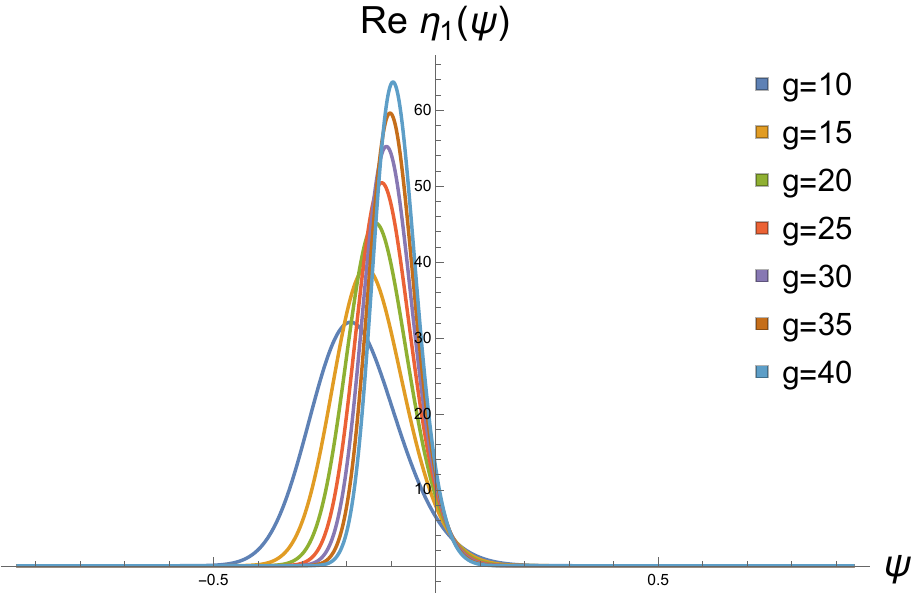}
    \end{subfigure}%
    ~ 
    \begin{subfigure}[t]{0.5\textwidth}
        \centering
        \includegraphics[scale=0.35]{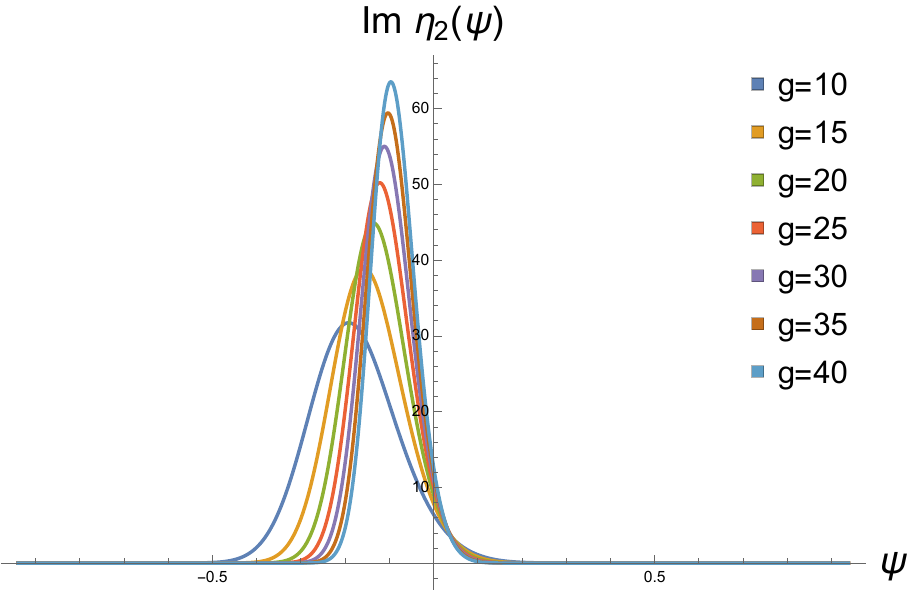}        
    \end{subfigure}
    \caption{The densities $\eta_i$ for Konishi for $g=10,15,\dots,40$ using the numerical algorithm of Section~\ref{sec:NumericalImplementation}. The densities scale as $g^{\frac{1}{2}}$ and, as we discuss in Section~\ref{sec:AnalysisDensities}, to order $\mathcal{O}(g^{\frac{1}{2}})$ the densities are equal, but deviates from each other at subleading orders.}
    \label{fig:etaplotsmain}
\end{figure*}

\paragraph{Plots of the densities from numerics.} We display $\eta$ for the Konishi multiplets in Figure~\ref{fig:etaplotsmain}. The data used in the plot is from the algorithm to be explained in the next section.

\section{Details of the Numerical Algorithm} \label{sec:NumericalImplementation}

In the previous section we described how to express $\bP$ and $\bQ$ in terms of the densities $\rho$ and $\eta$. We now discuss how to set up a numerical algorithm that utilizes this parameterisation. The crucial advantage of this algorithm, as compared to the previous one developed in \cite{Gromov:2015wca}, is due to the fact that it requires much fewer parameters at large coupling, making it also much faster in this regime. Furthermore, since we parameterise both $\bP$ and $\bQ$ on equal footing we do not need to solve the $\bQ$ in terms of $\bP$ using finite difference equations, which was another factor in the old method becoming increasingly slow at large $g$'s.

Before presenting the technical tools  needed we first concisely summarize the algorithm in four steps.

\paragraph{Step 1.} Parameterise $\rho$ and $\eta$ using ``measure" factors $m(x),n(x)$ and a set of polynomials $\mathcal{P}_{n},\mathcal{Q}_{n}$. It is convenient to use the parameterisations $x=e^{i\theta}$, $\theta\in[-\pi,\pi]$ and $x=e^{\psi}$, $\psi\in(-\infty,\infty)$ for the unit circle and real line respectively. That is we use
\begin{equation}\label{eqn:rhomeasure}
    \rho^{(i)}_a(e^{\ii \theta})=m(e^{\theta}) \sum_{n=0}^{N} c^{(i)}_{a,n}\,\mathcal{P}_{n}(\theta)\;,
    \quad
    \eta^{(i)}_i(e^{\psi}) = \ii^{a+1}n(e^{\psi})\,\sum_{n=0}^{N}d^{(i)}_{a,n}\mathcal{Q}_n(\psi)\;,
\end{equation}
where $(i)$ labels the $i$-th iteration, $c^{(i)}_{a,n}$ and $d^{(i)}_{a,n}$ are the parameters at this iteration to be fixed and $N$ is a cut-off (which for historical reasons we denote \verb"ChPW" in the code, which stands for \verb"Ch"op-\verb"P"o\verb"W"er).  

\begin{figure}[h]
    \centering
    \includegraphics[scale=0.4]{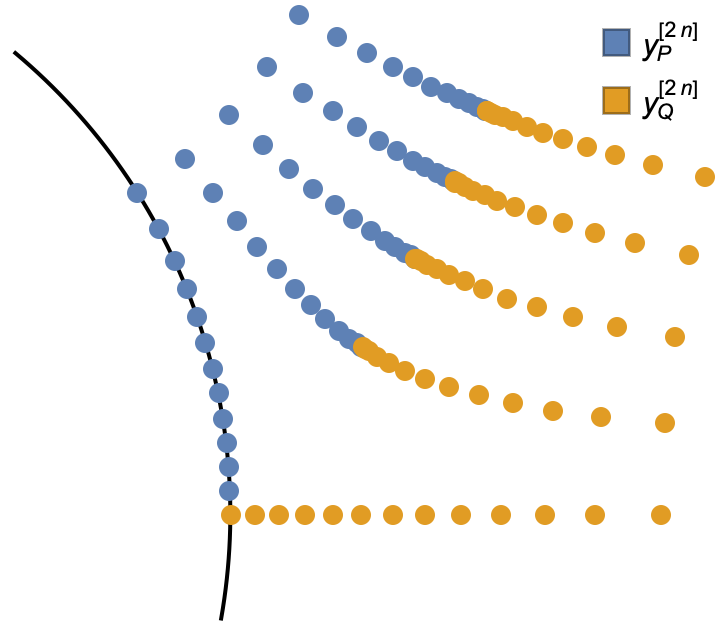}
    \caption{Example of probe points used in the numerical algorithm with $n=0,\dots,4$ for $g=10$. The picture shows $125$ points corresponding to $\mathtt{ChPW}=24$, see Subsection~\ref{subsec:SteByStep}. Those are designed to probe the information near the branch point in the optimal way.}
    \label{fig:ProbePoints}
\end{figure}

\paragraph{Step 2.} Construct $\bP^{(i)}$ and $\bQ^{(i)}$ from the densities \eqref{eqn:rhomeasure} using \eqref{eqn:Pfinal} and \eqref{eqn:Qfinal} at a set of sampling points $y_{P,Q}^{[2n]}\equiv x(u_{P,Q}+\ii n)$ with $n=0,\dots,4$. 
Due to the symmetries present we can restrict to $y_Q$ on the real line with $y_Q > 1$ and $y_P = e^{\ii \theta_P}\,,\,\theta_P\in [0,\pi/2]$. Note that both $\bP^{(i)}$ and $\bQ^{(i)}$ are to be evaluated at both $y_{P}^{[n]}$ and $y_{Q}^{[n]}$, see Figure~\ref{fig:ProbePoints}. 

\paragraph{Step 3.} Define $\bP^{\text{B},(i)}_a(y_{P}+i0)$ in terms of $\bP^{(i)}_a(y_{P}^{[2n]})\,,n=1,\dots,4$ and
$\bQ^{(i)}_a(y_{P}^{[2n]})\,,n=0,\dots 4$ using the Baxter equation \eq{eq:BaxterEqBQ}. Find $\bQ^{\text{B},(i)}(y_Q+\ii 0)$ in the same manner. Build $\rho^{\text{B},(i)}$ and $\eta^{\text{B},(i)}$ from $\bP^{\text{B},(i)}$ and $\bQ^{\text{B},(i)}$, see \eqref{eq:RhoFromP} and \eqref{eq:EtaFromQ} for explicit expressions.

\paragraph{Step 4.} If the values of the coefficients $c^{(i)}_{a,n}$ and $d^{(i)}_{a,n}$ as well as $\Delta$ at a given iteration are sufficiently close to their actual values then $\rho^{\text{B},(i)}$ and $\eta^{\text{B},(i)}$ should match well with the corresponding iteration of the densities $\rho^{(i)}$ and $\eta^{(i)}$. We can phrase this as a minimisation problem for the vector of mismatches of the densities at the probe points
\begin{equation}
\rho_a^{\text{B},(i)}(y_P)-\rho^{(i)}(y_P)\;\;,\;\;
\eta^{\text{B},(i)}(y_Q)-\eta^{(i)}(y_Q)\;.
\end{equation}
Note that $\rho_a^{\text{B},(i)}$ and $\eta^{\text{B},(i)}$ are complicated functions of all parameters $c,\;d$ and $\Delta$.
We use Newton's method to find $c,\;d$ and $\Delta$ which set the mismatch vector to zero within the numerical tolerance limit. In practice we should also make sure that at each step the gauge conditions, given below in \eqref{eqn:gaugeconds}, are satisfied, as otherwise zero modes can appear and dramatically decrease the convergence rate. 

\indent While the algorithm outlined above should in principle work for many different choices of $m(x),n(x),y_P\dots$ it is crucial to pick these objects with care in order for the algorithm to be efficient. In the rest of this section we discuss in detail appropriate choices when $g$ is large and then finally in Subsection~\ref{subsec:SteByStep} give a step by step implementation of the algorithm in \verb"Mathematica". 

\paragraph{Fixing the gauge.}
In order to have a well-defined numerical algorithm we need to ensure all gauge freedom of the QSC has been completely fixed. In our algorithm we impose the following conditions for $L$ even, in the $x$-plane: 
\begin{equation}\label{eqn:gaugeconds}
    -\ii\rho_1(1)=\pm\rho_2(1),\quad \rho_2(i)=0,\quad {\rm Re}(-\ii\rho_1(e^i) \mp \rho_2(e^i))=0,\quad 
    {\rm Re}(\eta_1(0) \mp \ii\eta_2(0))=0\,.
\end{equation}
For odd $L$ there is an additional gauge parameter which we fix with additional ${\rm Im}(\rho_1(e^i) \mp \rho_2(e^i))=0$ condition. The $\pm$ signs are not correlated and depends on the particular state in question. In practice we fixed them by looking at the inital densities obtained from \cite{Gromov:2023hzc}.

\subsection{Parameterising and Reconstructing $\rho$ and $\eta$}
\begin{figure}[h]
    \centering
        \includegraphics[scale=0.35]{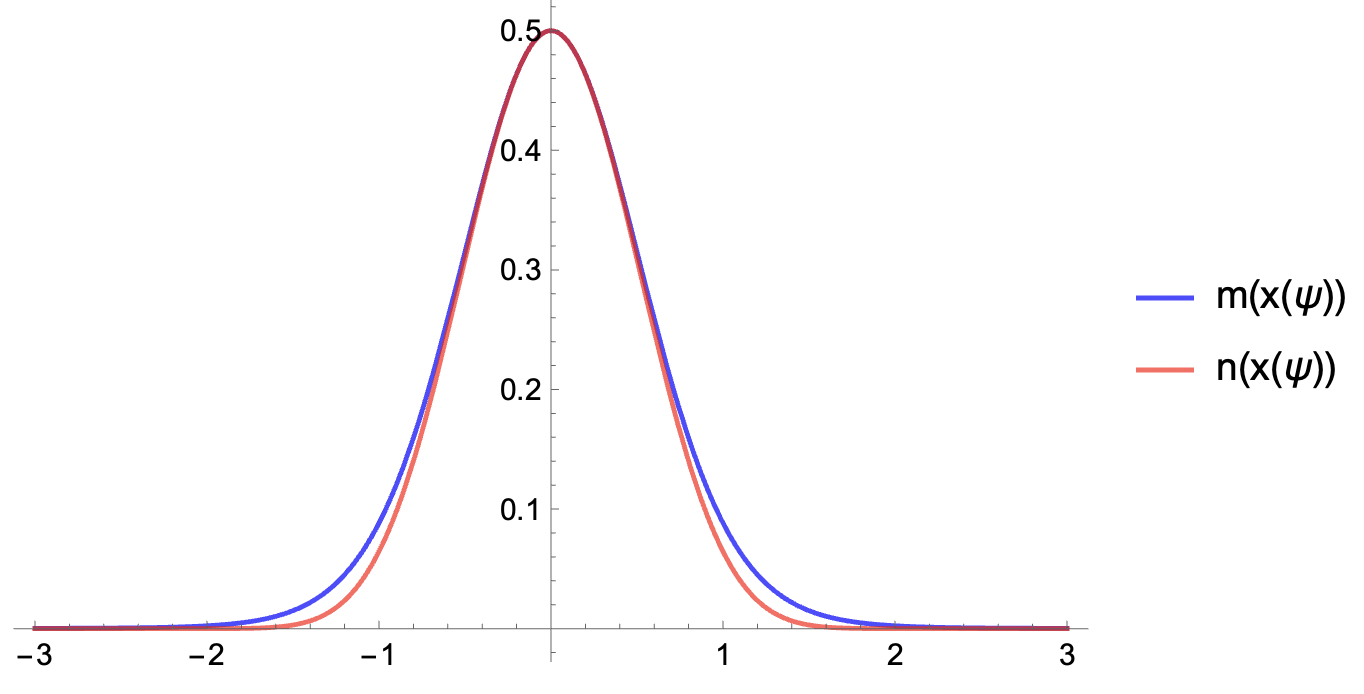}
\caption{Plot of the measures $m(x)$ for $x=e^{i\psi}$ and $n(x)$ for $x=e^\psi$, $\psi\in[-3,3]$ at $g=0.3$. The measures coincide more and more as $g$ increases, but this difference becomes important for high-precision numerics.}
\label{fig:measureplot}
\end{figure}

In this subsection we explain how to accomplish \textbf{Step 1} and \textbf{Step 3} when the coupling is large. Recall that then we can ensure that $\rho$ and $\eta$ are exponentially suppressed away from the branch points at $u=\pm 2g$, or equivalently $x=\pm 1$, on the unit circle and real line respectively. We can make the exponential suppression away from $x=+1$ manifest by writing
\begin{equation}\label{eqn:rhomeasure}
    \rho_a(x)=m(x) \,R_a(x)\;,\;\;\; \eta_a(x) = i^{a-1}n(x)\,E_a(x)\;,\;\;\;a=1,2\;,
\end{equation}
where $R_a(x)$ and $E_a(x)$ are order $1$ on the support of the densities, whereas the exponential decay away from the branch points is determined by $m(x)$ and $n(x)$, given by
\beqa\label{eqn:measures}
    m(x) &\equiv& \frac{1}{2}e^{+2\pi (u-2g)}= \frac{1}{2}{\rm exp}\left(+2\pi g\left(x+\frac{1}{x}-2\right)\right),\\
    n(x) &\equiv& \frac{1}{2}e^{-2\pi (u-2g)}=\frac{1}{2}{\rm exp}\left(-2\pi g\left(x+\frac{1}{x}-2\right)\right)\,, 
\eeqa
see Figure \ref{fig:measureplot}. 

The reality properties \eqref{eqn:Preality} imply that real and imaginary parts has definite parity
\begin{equation}
    R_{a}(\theta) = R_{a,1}(\theta)+i R_{a,2}(\theta),\quad R_{a,c}(-\theta) = (-1)^{a+c-1}R_{a,c}(\theta),\quad c=1,2\;,
\end{equation}
where $R_{a,c}(\theta)$ are real polynomials. Clearly this means that ${\rm Re}\;\rho_1(\theta)$ is  odd and ${\rm Im}\;\rho_1(\theta)$  even, and oppositely for $\rho_2(\theta)$, as reflected in Figure~\ref{fig:rhoplotsmain}.

Note that when $g\to\infty$ the behaviour of the profile function $m$ becomes Gaussian and localised at $\theta\sim 1/\sqrt g$
\beq
m(x)\simeq \frac{1}{2} e^{-2 \pi  g \theta ^2}\;\;,\;\;x=e^{i\theta}\;,
\eeq
and same is true for $n(x)$ for $x=e^{\psi}$
\beq
n(x)\simeq \frac{1}{2} e^{-2 \pi  g \psi ^2}\;\;,\;\;x=e^{\psi}\;.
\eeq
For the purpose of the numerical procedure we introduce an efficient parametrisation of the non-trivial functions $R_a$ and $E_a$.  We found it is convenient to use the basis of orthogonal polynomials ${\cal P}_n(\theta)$ and ${\cal Q}_n(\psi)$, which we describe in the next section. This choice guarantees fast decay of the expansion coefficients, which is important for generating stable starting points from smaller $N$ solution. We take
\begin{equation}
    R_{a,1}(\theta) = \displaystyle \sum_{n=1}^N c_{a,2n-1}{\cal P}_{2n-1}(\theta)\;,\quad R_{a,2}(\theta) = \displaystyle \sum_{n=0}^N c_{a,2n}{\cal P}_{2n}(\theta)\;,
\end{equation}
\begin{equation}
E_a(\psi)=i^{a-1}\displaystyle \sum_{n=0}^N d_{a,n}{\cal Q}_n(\psi)\;.
\end{equation}
As we will be using orthogonal polynomials, ${\cal P}_{n}(\theta)$ and ${\cal Q}_n(\psi)$ has definite parity ${\cal P}_n(-\theta)=(-1)^n {\cal P}_n(\theta)$, ${\cal Q}_n(-\psi)=(-1)^n {\cal Q}_n(\psi)$ and the coefficients $c_{a,n}$ and $d_{a,n}$ are real.

\paragraph{Reconstructing $\rho$ and $\eta$.}
To complete \textbf{Step 3} we need to first build $\rho$ from $\bP_a(x)$ with $x$ on the unit circle in the first quadrant and $\eta$ from $\bQ_i(x)\,, x>1$. 

To find $\rho_{a}$ we can simply use its definition supplemented with reality conditions to find
\begin{equation}\label{eq:RhoFromP}
\begin{split}
    \rho_{1}(x) = x^{\frac{L}{2}-1}\left(\bP_1(x) + \overline{\bP_{3}(x)}\right)\;,
    \quad
    \rho_2(x) = x^{\frac{L}{2}-1}\left(\bP_2(x) - \overline{\bP_{4}(x)}\right)\;,
\end{split}
\end{equation}
for $|x| =1$. Note that it is enough to consider $|x|=e^{\ii \theta}, \theta \in [0,\frac{\pi}{2}]$ to find $\rho$ on the whole unit circle since the remaining parts are fixed by parity and complex conjugation.

To find $\eta$ we use Sokhotsky's formula to find
\begin{align}\label{eq:EtaFromQ}
   &\eta_1(x) = 2\,x^{-\frac{\Delta}{2}+\frac{S}{2}-2} \,\text{Re}\,\bQ^{\downarrow}_{1}(x+\ii 0)\;,
   &
    &\eta_1(1/x) = -2 \,x^{\frac{\Delta}{2}-\frac{S}{2}+2}\,\text{Re}\,\bQ^{\downarrow}_{3}(x+\ii 0)\;, \\
    &\eta_2(x) = 2\,i\,x^{-\frac{\Delta}{2}-\frac{S}{2}} \,\text{Im}\,\bQ^{\downarrow}_{2}(x+\ii 0)\;,
   &
    &\eta_2(1/x) = -2\,i \, x^{\frac{\Delta}{2}+\frac{S}{2}}\,\text{Im}\,\bQ^{\downarrow}_{4}(x+\ii 0)\;, \nonumber
\end{align}
for $x>1$.

\subsection{Orthogonal Polynomials and Gaussian Quadrature}
\label{sec:ortpoly}
In order to carry out the numerical algorithm we need to be able to efficiently perform integrals of the form 
\begin{equation}
    \displaystyle \int^{\pi/2}_{-\pi/2}{\rm d}\phi\, m(\phi) f(\phi)\;,\quad \displaystyle \int^{\infty}_{0}{\rm d}\psi\, n(\psi) f(\phi)\;,
\end{equation}
where $f$ is some smooth function. 

These type of integrals appear when we construct $R$ and $E$ in \textbf{Step 1} of the algorithm
and when we integrate the densities to construct $\bP^{(i)}$ and $\bQ^{(i)}$ in \textbf{Step 2} at a set of probe points. We use the Gaussian quadrature method to perform these integrals, 
which requires the knowledge of orthogonal polynomials for the measure $m(\phi)$ and $n(\psi)$.

On the other hand, to implement \textbf{Step 3} we need to probe functions of the form $m(\phi)f(\phi)$ and $n(\psi)f(\psi)$ at an optimal set of points.
As we discuss in Section~\ref{sec:OptimalInterpolationMain} 
and go into detail in Appendix~\ref{app:optpoly}, 
the optimal set of points is given by the roots of the orthogonal polynomials for the measure $m^2(\phi)$ and $n^2(\psi)$.

Below we review the basic theory of orthogonal polynomials and Gaussian quadrature, and then discuss the specific choice of orthogonal polynomials for the measure $m(\phi)$ and $n(\psi)$.

\paragraph{Gauss quadrature integration and orthogonal polynomials.}

Let us recall the Gauss quadrature method. Given an integral of the form
\begin{equation}\label{eqn:sampleint}
    \displaystyle \int_a^b {\rm d}x\; w(x) f(x)
\end{equation}
where $w(x)$ is some measure factor and $f(x)$ is a smooth function. 
The Gauss quadrature method allows us to approximate this integral by a sum of the form
\begin{equation}\label{eqn:gaussquad}
    \displaystyle \int_a^b {\rm d}x\; w(x) f(x) \simeq \displaystyle \sum_{i=1}^n w_i f(x_i)\;,
\end{equation}
where the nodal points $x_i$ and weights $w_i$ are chosen in such a way 
that the approximation is exact for all polynomials of degree $2n-1$.
Since for smooth functions polynomial approximation is usually very efficient, 
the Gauss quadrature method is a very fast and precise way to evaluate these integrals numerically.

First one can show that the points $x_i$ are the roots of 
the orthogonal polynomials $p_n(x)$ for the measure $w(x)$, by considering $f(x)=p_n(x)x^k$
with $k=0,1,\dots,n-1$. 
Since the $p_n(x)$ for the weight $w(x)$ is orthogonal, we have 
\begin{equation}\label{eqn:orthogpoly}
    \displaystyle \int_a^b {\rm d}x\; w(x) p_n(x)x^k = 0\;\;,\;\;k<n\,,
\end{equation}
which is consistent with \eq{eqn:gaussquad} assuming that $x_i$ are $n$ zeroes of $p_n(x)$. 
The weights $w_i$ can be determined by the fact that \eq{eqn:gaussquad} 
is exact for all polynomials of degree $2n-1$. In our code we impose
\begin{equation}
  m_k \equiv \displaystyle \int_a^b {\rm d}x\; w(x) x^k = \displaystyle \sum_{i=1}^n w_i x_i^k,\quad k=0,1,\dots,n-1
\end{equation}
which provides a linear system for the $w_i$. In the next section we describe how to generate the orthogonal 
polynomials $p_n$.

\paragraph{Building orthogonal polynomials.}

We now review a very efficient method for generating the orthogonal 
polynomials $p_n(x)$ for a given measure $w(x)$ on an interval $[-a,a]$.
In what follows, we assume the measure $w(x)$ is an even function implying that
the polynomials 
have definite parity, that is $p_n(-x)=(-1)^n p_n(x)$.

In addition to the orthogonality property  \eq{eqn:orthogpoly} we also impose the normalisation
\begin{equation}
    \displaystyle \int_{-a}^{a} {\rm d}x\, w(x)p_n(x)p_m(x) = \delta_{nm}\,.
\end{equation}
We also use the monic polynomials $\Pi_n(x)=x^n+\dots$ related to $p_n(x)$ by
\begin{equation}
  \Pi_n(x)= \mathfrak{n}_np_n(x)\,.
\end{equation}
Due to the parity we of course know the first two such polynomials
\begin{equation}
 \Pi_0(x) = 1,\quad \Pi_1(x)=x\,.
\end{equation}
In order to generate the $n$th orthogonal polynomial we use the following procedure.
Firstly, we define the moments $m_n$ of the measure $w(x)$
\begin{equation}\label{moms}
    m_n \equiv \displaystyle \int {\rm d}x\, w(x)x^n
\end{equation}
from which we construct the $(n+1)\times (n+1)$ determinant
\begin{equation}\la{Dndef}
    D_n = \displaystyle\det_{1\leq i,j\leq n+1} m_{i+j-2}\,. 
\end{equation}
We notice that the coefficient $\mathfrak{n}_n$ is given by
\begin{equation}
    \mathfrak{n}_n\equiv\sqrt{\frac{D_n}{D_{n-1}}} \,.
\end{equation}

Secondly, the orthogonal polynomials $\Pi_n(x)$ satisfy the 
three-term recurrence relation, which for the monic polynomials with even measure is given by
\begin{equation}\label{Pin}
    \Pi_n(x)= x \Pi_{n-1}(x)-\beta_{n-1}\Pi_{n-2}(x),\quad \beta_n(x) = \frac{D_{n-2}D_n}{D_{n-1}^2}\,.
\end{equation}
As all terms in the recursion relation are fixed in terms of the moments $m_n$ 
we can generate the orthogonal polynomials $\Pi_n(x)$, and also the normalised polynomials $p_n(x)$ for any $n$.

In our code we use $4$ sets of orthogonal polynomials. Two are for the measures $m(\phi),\;\phi\in[-\pi/2,\pi/2]$ and $n(\psi),\;\psi\in[0,\infty]$ - in order to set up the Gaussian quadrature 
for the integrals \eqref{eqn:sampleint} -- and another two sets to generate the optimal probe points with measures $m^2(\phi)$ and $n^2(\psi)$ as we briefly explain in the next section.

The recursion relation \eq{Pin} contains determinants $D_n$ of the moments $m_n$ so we need to also have an efficient way of evaluating these.

\paragraph{Efficient algorithm for computing the determinants.}
Given an $n \times n$ dense matrix without any additional structure, the available algorithms for computing determinants have a complexity of ${\cal O}(n^3)$, and this cannot be significantly improved. Since we also need approximately $n$ such determinants, the overall complexity of the polynomial orthogonalisation becomes ${\cal O}(n^4)$.
For a large number of parameters, this becomes a bottleneck of our algorithm.

Fortunately, the matrix \eqref{Dndef} has a Toeplitz structure (up to a trivial redefinition). For such matrices, one can use the Levinson-Durbin recursion, which has a complexity of ${\cal O}(n^2)$. Furthermore, one can arrange the recursion such that all $D_i, \; i \leq n$ are computed in one go, resulting in an overall complexity improvement by order $n^2$.

Let us explain the idea of the adopted Levinson-Durbin recursion method briefly. Denote by $M^{(n)}_{ij}=m_{i+j-2}$ $1\leq i,j,\leq n+1$ the matrix in the r.h.s. of \eqref{Dndef}. Then introduce a vector $f^{(n)}$ of size $n+1$ such that
\beq
M^{(n)}f^{(n)}=e_{n+1}
\eeq
where $e_{n+1}=(0,0,\dots,0,1)$. First notice that the last component of the vector $f^{(n)}$ gives the ratio of the determinants
\beq\label{fnn}
f^{(n)}_{n+1} = \frac{D_{n-1}}{D_{n}}\;.
\eeq
Furthermore, the vector $f^{(n)}$ satisfies a simple recursion relation. Notice that we can concatenate $f^{(n-1)}$ and $f^{(n)}$ with extra zero components to
obtain a vector $\tilde f^{(n+1)}$ such that $M^{(n+1)}\tilde f^{(n+1)} = c_{n+1} e_{n+2}$, for some constant $c_{n+1}$, more precisely
\beq
\tilde f^{(n+1)} = (f^{(n-1)},0,0)-(0,f^{(n)})\;.
\eeq
Note that for the above to hold we have to use that $m_{2k+1}=0$ for an even measure due to the definition \eq{moms}.
Since $f^{(n+1)} = \tilde f^{(n+1)}/c_{n+1}$ we only need to determine the constant $c_{n+1}$, which can be obtained simply as $c_{n+1}=\sum_{i=1}^{n+2}M_{n+1,i}^{(n+1)}\tilde f_i^{(n+1)}$. Thus obtaining a recursion relation for the ratio of the determinants \eq{fnn}, which can be supplemented with the initial condition $D_0=1$.

\subsection{Dealing with Singular Integrals}
In the previous section we discussed how to perform integrals of the form 
\eq{eqn:sampleint} using the Gaussian quadrature method. 
However, we also need to be able to compute integrals of the form
\begin{equation}
    \displaystyle \int {\rm d}x\; w(x)\frac{f(x)}{x-y}
\end{equation}
which become singular when $y$ approaches the contour of integration.
Even if $y$ is not right on the contour, the integrand is not sufficiently 
smooth to be able to use the Gaussian quadrature method efficiently.

In order to overcome this difficulty we use an efficient subtraction method.
We  write the integral as
\begin{equation}\label{subtraction}
    \displaystyle \int {\rm d}x\; w(x)\frac{f(x)}{x-y} = \displaystyle \int {\rm d}x\; w(x)\frac{f(x)-f(y)}{x-y} + f(y)\displaystyle \int {\rm d}x\; w(x)\frac{1}{x-y}\;.
\end{equation}
The advantage of this rewriting is that the first integral 
is now smooth and can be evaluated using the Gaussian quadrature method.
The second integral is a principal value integral, however it does not depend on $f(x)$
and can be precomputed. 

In our code we implement a slightly different version of the subtraction trick \eq{subtraction}, which we found to give better precision and speed. Namely we write
\begin{equation}\label{subtractionAdvanced}
    \displaystyle \int {\rm d}x\; \frac{w(x)f(x)}{x-y} = \displaystyle \int {\rm d}x\; w(x)\[\frac{f(x)}{x-y}-\frac{1}{y i}\frac{f(y)}{\frac{1}{i}\log x-\frac{1}{i}\log y}\] + \frac{f(y)}{y i}\displaystyle \int {\rm d}x\; \frac{w(x)}{\frac{1}{i}\log x-\frac{1}{i}\log y}
\end{equation}
where the integration goes over the right half of the unit circle. The integrand in the square brackets is smooth as the singularity at $x=y$ is cancelled and in the last term the dependence on the parameters sitting inside $f(y)$ factorised.

\subsection{Building Optimal Interpolation Points.}\label{sec:OptimalInterpolationMain}

In addition to being able to efficiently evaluate integrals \eqref{eqn:sampleint} numerically, 
we also need to be able to efficiently probe functions of the form $w(x)f(x)$.
By that we mean that the information contained in the values of the function of this form should be maximised in order to be able to reconstruct the function $f(x)$ as accurately as possible.

More precisely, the interpolation polynomial $r(x)$ going thorough the points 
$\{x_i, f(x_i)\}$, multiplied by the weight function $w(x)$ 
should be as close as possible to the combination $w(x)f(x)$.
Technically, we are trying to minimize the $L_\infty$ norm of the difference
$\max_{x\in[-a,a]} |w(x)f(x)-w(x)r(x)|$. For the flat measure on the interval $[-1,1]$ the optimal interpolation points are given by the Chebyshev points, which are the roots of the Chebyshev polynomials of the first kind.

As we argue in the Appendix~\ref{app:optpoly} for the general measure $w(x)$ 
the close to optimal interpolation points are given by the roots of the orthogonal polynomials for the measure $w(x)^2$. We could not find a reference for this observation in the literature -- we have included a derivation in Appendix \ref{app:optpoly}. 
We should notice, that unlike the Chebyshev points, which give the nice behaviour of the error function even near the end points of the interval, our argument only applies in the bulk of the interval. This is something which can be further improved, but at the same time in may increase the computational costs of finding perfectly optimal points. The points we use are the roots of the orthogonal polynomials, for which we have highly optimised algorithms and which have only slight increase of the error near the boundaries.

Having discussed the general theory we now move 
on to the specific details of our implementation.

\subsection{Step by Step Implementation}\label{subsec:SteByStep}
Above we described the general outline of the numerical procedure. Now we go through
the \verb"Mathematica" code which implements the procedure and discuss the specific details of the implementation.

\paragraph{The key parameters.}
As described earlier in the text, we are focusing on the $\mathfrak{sl}(2)$ sector with additional parity symmetry. In this section we furthermore set the spin to its lowest non-trivial value $S=2$ (even though we have also implemented $S=4$ case). There are still infinitely many states in this sector.
The majority of the information about a particular state enters through 
the initial data for the densities $\rho$ and $\eta$, which can be extracted at smaller $g\sim 4$, in the regime where many states have been studied already in \cite{Gromov:2023hzc}.
The remaining parameter encoding the quantum numbers of the states is the length $L$.
The way the code is structured allows us to change $L$ at the last stage of Newton iterations, having all the previous steps done without fixing $L$.
One has to also specify the value of the coupling $g$, which we denote \verb"g0" in the code.
We also need to set the cutoff in the number of parameters \verb"ChPW" 
such as the maximal degree of the polynomial in $\rho$ and $\eta$, and the number of the probe points.
Finally, the last parameter is the working precision \verb"WP", used at all steps of the calculation.

\paragraph{Gaussian integration and probe points.}
First we define the orthogonal polynomials and define the Gaussian quadrature function which we use to evaluate the integrals of the form \eq{eqn:sampleint}
as well as a set of probe points \verb"XQ" at which we evaluate the Baxter equation.
\begin{mathematica}
(* Define a function to find Gaussian weights *)
FindGaussianWeights := Block[{},
  (* Finding optimal probe points XQ  *)
  XQ = SortBy[SetPrecision[XQoptimalNew[g0, ChPW], 4*WP] // Re, Re];
      
  (* Build the Gaussian integration points and the corresponding weights *)
  {psii, wpsii} = BuildGaussIntegrationQ[g0, ChPW];
      
  (* Define a Gaussian numerical integration function NIntK *)
  NIntK[a_] := (
    Sum[a wpsii[[i]] 
       /. Association[{px -> psii[[i]], x -> Exp[psii[[i]]]}]
    , {i, Length[psii]}]
  )
];
\end{mathematica}
where \verb"XQoptimalNew"  is a function which generates the optimal probe points, 
depends on the value of the coupling \verb"g0" 
and the cutoff in the number of parameters \verb"ChPW"
\begin{mathematica}
(* Define a function XQoptimalNew that generates optimal probe points *)
XQoptimalNew[g0_?NumericQ,Chpw_?NumericQ]:=Block[{tab,me2,IM2},    
    (* Define me2 as a square of the measure *)
    me2=1/Exp[4 \[Pi] g0 (x+1/x-2)]/2/.x->Exp[psi]//Simplify;
    
    (* Precomputing the momnets of the measure *)
    IM2[n_]:=IM2[n]=If[EvenQ[n],
        NIntegrate[me2 psi^n,{psi,-\[Infinity],\[Infinity]}
        ,Method->"DoubleExponential",WorkingPrecision->4WP,
        AccuracyGoal->2WP,MaxRecursion->100],0];
    
    (* Use parallelization to generate the moments *)
    tab=ParallelTable[IM2[n],{n,0,2Chpw+4}];
    Do[IM2[n]=tab[[n+1]],{n,0,2Chpw+4}];
    
    (* Generate polynomials from precomputed moments *)
    GeneratePolynomialsFromMoments[IM2,\[GothicCapitalP]2,Chpw];
    
    (* Getting probe points by finding the zeroes of the polynomials *)
    Sort[Exp[psi]/.Solve[\[GothicCapitalP]2[Chpw+1]==0,psi]//Re]
]
\end{mathematica}
\verb"BuildGaussIntegrationQ" is a function which generates the Gaussian quadrature points and weights,
defined in a similar way as follows
\begin{mathematica}
(* Define a function BuildGaussIntegrationQ that takes two numeric inputs g0 and Chpw *)
BuildGaussIntegrationQ[g0_?NumericQ,Chpw_?NumericQ]:=Block[{tab,me,
    IM,psii,wpsii},
    (* Define the measure *)
    me=1/Exp[ 2 \[Pi] g0 (x+1/x-2)]/2/.x->Exp[psi]//Simplify;
    
    (* Define a function to compute the moments of the measure *)
    IM[n_]:=IM[n]=If[EvenQ[n],
        NIntegrate[me psi^n,{psi,-\[Infinity],\[Infinity]},
        Method->"DoubleExponential",WorkingPrecision->4WP,
        AccuracyGoal->2WP,MaxRecursion->100],0];
    
    (* Use parallelization to precomute the moments *)
    tab=ParallelTable[IM[n],{n,0,2Chpw+4}];
    Do[IM[n]=tab[[n+1]],{n,0,2Chpw+4}];
    
    (* Generate polynomials from moments *)
    GeneratePolynomialsFromMoments[IM,\[GothicCapitalP],Chpw];
    
    (* Find nodal points as zeroes *)
    psii=Quiet[psi/.Solve[\[GothicCapitalP][Chpw]==0,psi]]
        /. 0->1/10^(2WP)//Sort;
    
    (* Solve a linear system to find the weights *)
    wpsii=LinearSolve[Table[(psii^m),{m,0,Chpw-1}],Table[IM[m]
        ,{m,0,Chpw-1}]];
    
    (* Return nodal points and weights *)
    {psii,wpsii}
]
\end{mathematica}
Finally both functions depend on \verb"GeneratePolynomialsFromMoments" 
which generates the orthogonal polynomials from the moments of the measure

\begin{mathematica}
GeneratePolynomialsFromMoments[IM_(*moments*),Po_(*name of the poly*),
                               Chpw_(*max degree*)]:= Block[{d,N0,B0,PI},
    Clear[d];
    
    (* Define d[Nc] as the determinant of a table of moments *)
    d[Nc_]:=d[Nc]=Table[IM[n+m-2],{n,Nc+1},{m,Nc+1}]//Det;
    d[-1]=1;
    
    (* Define the leading coefficien in the normalized polynomial *)
    N0[Nc_]:=Sqrt[d[Nc]/d[Nc-1]];
    
    (* compute quantity appearing in the recursion relation *)
    B0[Nc_]:=(d[Nc-2]d[Nc])/d[Nc-1]^2;
    
    (* Setting initial conditions for the recursion *)
    Clear[PI,Po];
    PI[0]=1;
    PI[1]=x;
    
    (* Define PI[n] using recursion relation *)
    PI[n_]:=PI[n]=Expand[x PI[n-1]-B0[n-1]PI[n-2]];
    
    (* Define normalized polynomial Po[nn], which is a global variable *)
    Do[Po[nn]=Collect[PI[nn]/N0[nn],x]/.x->[\Psi],{nn,0,Chpw+1}]
]\end{mathematica}
As explained in Section~\ref{sec:ortpoly} the part of the computing the determinant of the moments of the measure can be further improved, which is implemented with just a few lines of code
\begin{mathematica}
pr=Precision[IM[0]]; (*deducing current precision*)
Clear[f, d];
f[-1]={};
f[0]={1/SetPrecision[f[0],4pr]};
fv[n_]:=(f[n-2]~Join~{0,0}-{0}~Join~f[n-1]);
(*the round-off error accumulates quite fast in this recursion*)
f[n_]:=f[n]=SetPrecision[fv[n]/Table[SetPrecision[f[n-1+i],4pr],{i,n+1}].fv[n],4pr];
d[-1]=1;
d[n_]:=d[n]=(1/f[n][[-1]])d[n-1];    
\end{mathematica}

We repeat the above procedure for the measure $m(\theta)$ \eqref{eqn:measures} on the interval $[-\pi/2,\pi/2]$. We skip the details of the code, as it is essentially identical to the one above. The complete code is available in the ancillary files of the arxiv submission of this paper.

Finally, we have to compute the values of $\bP$ and $\bQ$ on a set of 
points related to the probe points by a shift by $i n$ in $u$ plane, which we
denote as \verb"ys" in the code.
\begin{mathematica}
(* Selecting probe one the main sheet in u plane and above the cut *)
{xQ, xP} = {
    Select[XQ, Re[#] > 1 &], 
    Select[XP, Im[#] >= 0 &]};
(* The probe points in u plane on the real axis *)
us = Flatten[{
    {g0 (xP + 1/xP) // Re, g0 (xQ + 1/xQ) // Re} // Flatten // Union}];
(* Shifted points at which we need to evaluate P and Q *)
ys = SetPrecision[
    Flatten[Table[X[us + I n] /. g -> g0, {n, 0, 4}]], 5WP];
\end{mathematica}

\paragraph{Setting parametrisation of the densities.}
Next we define the parametrisation of the densities $\rho$ and $\eta$ in terms of the orthogonal polynomials.
To remind we have the parameter \verb"ChPW" which sets the maximal degree of the polynomial in the parametrisation.
As we have $4$ polynomials $R_a$ and $E_a$ to parametrize we get in total
$4({\verb"ChPW"}+1)$ parameters. In addition we have the parameter $\Delta$ which we also include in the list of parameters.
Below function lists the $4{\verb"ChPW"}+5$ parameters names
\begin{mathematica}
(* Define a function Prm that returns a list of parameters *)
Prm := Block[{},
    Flatten[{
        \[CapitalDelta],
        Table[d1[i], {i, 0, ChPW}],
        Table[d2[i], {i, 0, ChPW}],
        Table[c1[i], {i, 0, ChPW}],
        Table[c2[i], {i, 0, ChPW}]
    }]
];
\end{mathematica}
We also define the projector matrices which project the whole set of parameters to $(\verb"ChPW"+1)$
relevant for a particular density. For example for $\rho_1$ we define
the following $(\verb"ChPW"+1)\times(4\verb"ChPW"+5)$ matrix
\begin{mathematica}
ToRO1=Table[Coefficient[Table[(rp[n]),{n,0,ChPW}],p],{p,prm}]\[Transpose];
\end{mathematica}
Denoting \verb"ROvec" and \verb"ETvec" the set of the first $(\verb"ChPW"+1)$ 
 orthogonal polynomials 
for the measure $m(\phi)$ and $n(\psi)$ respectively,
we can write the parametrisation of the densities as a matrix multiplication, e.g. $R_1(x)$ becomes
\begin{mathematica}
ROvec[x].ToRO1.params
\end{mathematica}

\paragraph{Integration kernels defining $\bP$ and $\bQ$ at the probe points.}
Next we define the integration kernels which we use to 
evaluate the functions $\bP$ and $\bQ$ at the probe points from the set or parameters.
As the integrals depends linearly on the densities the goal is to
precomute the integrals of the form of a large matrix which converts the list of parameters
\verb"params" to the values of the functions $\bP$ and $\bQ$ at the probe points \verb"ys".

To explain the procedure we focus on the $\bP$ function, the $\bQ$ function is defined in a similar way.
When computing the integrals \eq{eqn:Pfinal} we can immediately apply the Gaussian quadrature method to the second term, which does not have a singularity at $x=y$.
\begin{mathematica}
RRp[y_]=NIntX[(-ROvecLog[px]/(2\[Pi] I)) 1/(x+y) I x];
\end{mathematica}
where we keep the argument $y$ symbolic and the function \verb"NIntX" performs the Gaussian quadrature.
At the same time for the first term we follow the subtraction 
method described above in \eq{subtractionAdvanced}. For the smooth part we again apply the function \verb"NIntX" 
\begin{mathematica}
SSp[y_,yp_]=NIntX[((-ROvecLog[px]/(2\[Pi] I)) x/(x-y))I]
        -(-ROvecLog[yp])NIntX[(1/(2\[Pi] I)) 1/(1/I Log[x]-yp)];    
\end{mathematica}
where we keep two types of arguments $y$ and $yp=\frac{1}{i}\log y$ symbolic.
Finally the integral in the last term of \eq{subtractionAdvanced} can be precomputed using build-it function
\begin{mathematica}
NIntegrate[wn[psi]/(2\[Pi] I) 1/(psi-psi0),{psi,-\[Pi]/2,Re[psi0],\[Pi]/2}
    ,WorkingPrecision->WP(1+1/2),AccuracyGoal->WP,MaxRecursion->200]]   
\end{mathematica}
Then we combine these parts together, and evaluate them at the points \verb"ys" to get the matrix
\verb"P1mat" $\dots$ \verb"P4mat". 
Those matrices do not depend on the parameters $c_{a,n}$ and $d_{a,n}$, or $L$ and $\Delta$ 
and can be precomputed before the Newton iterations. 
For example, in order to get the vector of values of $\bP_1$ at the points \verb"ys" we can write
\begin{mathematica}
P1vec= ys^(-L/2+1) P1mat . ToRO1 . params;  
\end{mathematica}
And similarly for the other functions $\bP_2,\bP_3,\bP_4$. The procedure for $\bQ$ is almost identical
and we do not repeat it here.

\paragraph{Forming the equations for the parameters.}
Next we use the Baxter equation \eq{eq:BaxterEqBQ}
to express the values of $\bP$ and $\bQ$
at the probe points on the real axis of $u$ in terms of the $\bP$'s and $\bQ$'
at the shifted up points, for which we have already precomputed the values
and then deduce the values of the densities $\rho$ and $\eta$ at the probe points
as explained at the beginning of this section, see \eqref{eq:RhoFromP} and \eqref{eq:EtaFromQ}. In this way we get exactly 
the same number of equations as the number of parameters $c$'s and $d$'s. 
However, we should also impose the gauge conditions \eq{eqn:gaugeconds}.
As for Newton's method, the number of equations should coincide 
with the number of parameters, we exclude some values of the densities to match the number of parameters.
More precisely, the equation appearing from matching the re-computed $\eta_1$
with the initial one is obtained as follows
\begin{mathematica}
(* range of corresponding to the relevant probe points *)
range12=(ChPW+2)/2;;ChPW+1;

(* finding values of eta1 on the probe points from parameters *)
eta1vec=Table[ ETvec[Exp[ps]]. ToET1 . params wm[ps],{ps,Log[ys[[range12]]]}]

(* finding values of eta1 using Baxter equation *)
eta1vecFromBax = 2*Re[Q1vecFromBax[0][[range12]]*ys[[range12]]^(-Delta/2 - 1)]

(* forming equation*)
eq1 = eta1vecFromBax - eta1vec // Re;
\end{mathematica}
The same procedure is repeated for $\bQ_2,\;\bQ_3$ and $\bQ_4$
and also for $\bP_1$ and $\bP_2$. This would give us 
\verb"eq1" $\dots$ \verb"eq6" (in total $4*(\verb"ChPW"+1)$ scalar real equations).
Finally we also need to impose the gauge conditions \eqref{eqn:gaugeconds}.
For that we form the following $4$ equations (here we assume $L$ is even, otherwise one should add one more condition as discussed in the beginning of this section, see below 
\eqref{eqn:gaugeconds}). 
\begin{mathematica}
    gauge1 = Re[rho1[1] - rho2[1]];
    gauge2 = Re[eta1[0] - eta2[0]]; 
    gauge3 = Re[rho1[Exp[I]] - rho2[Exp[I]]]; 
    gauge4 = Im[rho2[I]; 
\end{mathematica}
Finally we combine all the equations into a single vector, making sure that 
the number of equations matches the number of parameters i.e. $4\verb"ChPW"+5$.
We call the function \verb"F[params]" which returns the vector of equations for a given numerical values of the parameters.

\paragraph{Newton's method.}
Finally we are ready to apply Newton's method to solve the system of equations.
We use the following code, which uses parallelization, to compute the gradient of the function \verb"F[params]"
\begin{mathematica}    
(* Define a function J that takes a list of parameters as input *)
J[params_] := J[params] = SetPrecision[
    (* Initialize progress counter *)
    prog = 0;
    
    (* Monitor the progress of a parallel table computation *)
    Monitor[
        tblP = ParallelTable[
            (* Increment progress counter *)
            prog++;
            
            (* Compute the function F at perturbed parameters *)
            (1/eps) * F[params + eps * Table[Boole[i == j], {i, Length[params]}]],
            
            {j, 0, Length[params]}, DistributedContexts -> None
        ],
        
        (* Display a progress indicator *)
        ProgressIndicator[prog, {0, Length[params]}]
    ];
    
    (* Update the function F at the original parameters *)
    F[params] = eps * tblP[[1]];
    
    (* Compute the gradient *)
    Table[tblP[[i]] - tblP[[1]], {i, 2, Length[tblP]}],
    
    (* Set the precision of the result *)
    WP
];
\end{mathematica}
After that we update the parameters using the following code
\begin{mathematica}
paramsUpdated = SetPrecision[Re[params 
              - LinearSolve[Transpose[J[params]], F[params]]], 2 WP]
\end{mathematica}
We iterate the above procedure until the norm of the vector \verb"F[params]" is smaller than a given tolerance.
Finally, the value of the energy is given by the value of the parameter $\Delta$ at the end of the iterations.

We have included a Mathematica notebook implementing this algorithm with the Arxiv submission of this paper, including a data file providing starting points for the parameters for $g=100$, $S=2$ and $L=2,\dots,10$ with mode numbers $n=1,2$.

In the next section we give some examples of the data we generated with the above method.

\section{Data Analysis and New Analytic Results}\label{sec:Results}
In this section we present the results of the numerical implementation of the algorithm described in the previous section and new analytic predictions we managed to make based on these results.

We start by presenting the spectrum of the anomalous dimensions of the operators $\tr D^S Z^{L}$ for $S=2$ 
and $S=4$ and various $L$'s and mode numbers, then we discuss the strong coupling analysis of the data.
With the newly-available high precision strong coupling data as well as analytical insights originating from our method (described in the next section) we can confirm the string theory prediction
for the leading order of the anomalous dimensions 
for mode number $n=1$, we also present new results for $n=2$ case, which disagree, in general, with the quasi-classical string theory extrapolation due to non-polynomiality in charges of the expansion coefficients of $(\Delta+2)^2$, a common assumption made in the literature previously. We argue that this new type of the dependence on the charges is due to the mixing with the operators outside $\mathfrak{sl}(2)$ sector.

\subsection{Numerical Results Overview}
\begin{figure}[h]
    \centering
    \includegraphics[scale=0.5]{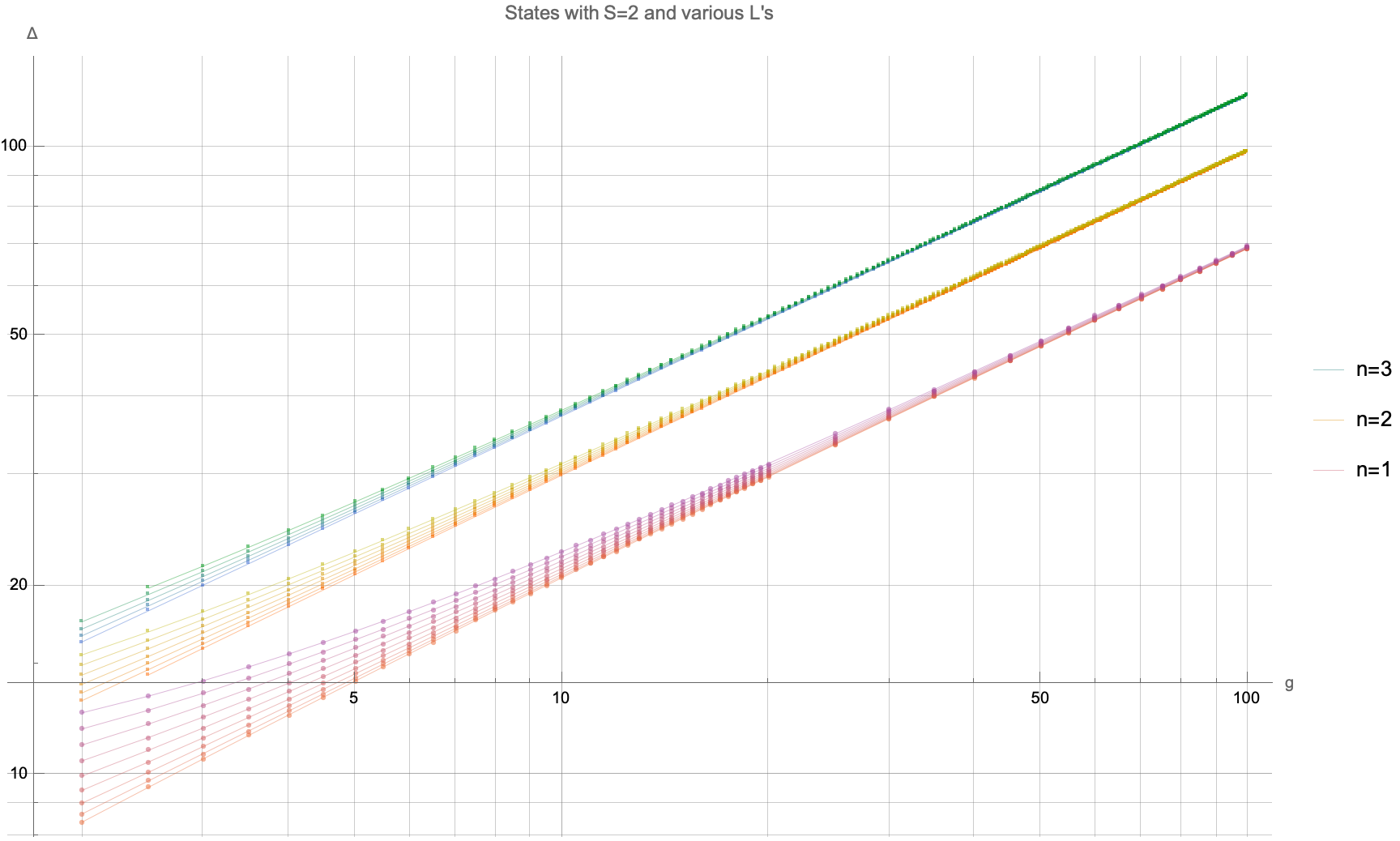}
    \caption{Spectrum of dimensions of $21$ states in the $\mathfrak{sl}(2)$ sector with $S=2$ and $L=4$ and $n=1,2,3$ for $L=2n,\dots,10$ as a function of the 't Hooft coupling $g=\frac{\sqrt\lambda}{4\pi}$.
    The range of the coupling in terms of $\lambda$ is $\lambda\in(650,1500000)$. The data is generated with the precision $\sim 10^{-25}$ in the range $g=2,\dots,20$, $\sim 10^{-40}$ for $n=1,L=4$ in the range $g=2,\dots,20$, and lower precision for other points of about $10^{-10}$, but can be pushed further without visible problems or a significant drop in speed and efficiency.
    }
    \label{fig:Spectrumn123}
\end{figure}
Let us overview some of the data we have generated by the algorithm described in Section~\ref{sec:NumericalImplementation}. 
For simplicity let us first focus on the case $S=2$, 
i.e $\mathcal{N}=4$ operators of type $\tr D^2 Z^{L}$. 
For fixed $L$ the number of states is equal to the number of unique ways to distribute the two derivatives, 
giving $\lfloor \frac{L}{2}\rfloor$. The different states are distinguished by the so-called mode number 
$n=1,\dots,\lfloor \frac{L}{2}\rfloor$.

The slight drawback of the current implementation of our method is that it is sharpened for large values of $g$'s and thus finding the starting points 
from the perturbation theory at weak coupling is at the moment not possible. We hope to eventually have good analytic control over the strong coupling regime which would 
open an exciting possibility to initiate the numerics from the string theory side instead of the gauge theory side.
It is also possible to improve the method to be able to start from the weak coupling side, but this we leave for future work, where
one may hope to create an industrial level code for the numerical analysis of the spectrum for operators of all types.

At the moment, to start our numerical algorithm we utilized the database in \cite{Gromov:2023hzc}. 
This database includes all states up to $\Delta_0 = 6$, which in practice means that 
we can find a starting point for $n=1,2,3$ for a restricted number of $L$. 
Whereas for $n=1$ and $n=2$ there are multiple states with various $L$'s,
 for $n=3$ there is only one state with $L=6$. 

Luckily, given only one state for a fixed $n$ we can using our algorithm easily construct 
all other states with different $L$. 
This works as follows: the parameter $L$ only appears 
explicitly in $\bP$ and $\bQ$ as an overall multiplication by $x^{\frac{L}{2}-1}$. 
Furthermore, all expressions are sensible even when $L$ is not an integer and the algorithm converges steadily when $L$ is changed by a small non-integer amount. 
Thus, we can fix $g$ and move in $L$, which turns out to be very efficient. 
Using this method we found all states for $n=1,2,3,\, L=2n,\dots,10$ which we subsequently extended to the range $g=2,\dots,100$ 
as shown in Figure~\ref{fig:Spectrumn123}. We kept the absolute error at the level $\sim 10^{-25}$ in the range $g=2,\dots,20$.
For the state with $n=1$ and $L=4$ we pushed the precision for $g=2,\dots,20$ to $\sim 10^{-40}$ (which takes around $3$ hours per point).
We give the details of our accuracy/speed test below.
\begin{table}[h]
    \centering
$$    
    \begin{array}{r||r|r||r|r}
 \verb"ChPW" & \text{time, } g=10 & \text{error, } g=10 & \text{time, } g=100 & \text{error, } g=100 \\ \hline 
 50 & \text{1m 55s} & 8.3\times10^{-10} & \text{1m 57s} & 1.0\times10^{-9}\,\, \\
 90 & \text{3m 42s} & 4.5\times10^{-15} & \text{4m \;\,7s} & 6.8\times10^{-14}\\
 130 & \text{6m 27s} &  8.5\times10^{-17} & \text{7m 26s} & 4.8\times10^{-16} \\
 170 & \text{10m 49s} &  4.5\times10^{-21} & \text{12m 57s} & 4.8\times 10^{-20}\\
 210 & \text{17m 38s} &  2.2\times10^{-23}  & \text{21m 47s} & 1.8\times 10^{-21}\\
 250 & \text{26m 58s} &  9.3\times10^{-26} & \text{32m 31s}  & 3.0\times 10^{-23} \\
 290 & \text{51m \;\,5s} &  1.8\times10^{-28} & \text{48m 28s} & 3.1\times 10^{-25}\\
 330 & \text{1h \;\,8m 57s} &  3.4\times10^{-30} & \text{1h \;\,9m 42s} & 1.2\times10^{-26}\\
 370 & \text{1h 35m \;\,2s} &  3.0\times10^{-32} & \text{1h 36m 12s} & 1.0\times 10^{-28} \\
 410 & \text{2h 18m 18s} &  5.5\times10^{-34} & \text{2h \;\,4m 38s} & 7.0\times 10^{-30}\\
 450 & \text{2h 56m 51s} &  8.6\times10^{-36} & \text{2h 40m 48s} & 6.4\times 10^{-31} \\
    \end{array}
$$    
    \caption{Benchmark results for our numerical algorithm with relative error, which was run on $30$ cores and working precision $280$. We considered two values of the coupling $g=10$ 
    and $g=100$ for the state with $n=1,\;L=4$. We performed $2$ iterations for each of the points. We see that increasing the value of the coupling by a factor of $10$ does not lead to a dramatic decline in the performance.}
    \label{tab:my_label}
\end{table}

The data we have generated for $S=2$ is shown in Figure~\ref{fig:Spectrumn123}.
As can be seen the spectrum of the states with different $L$'s merge at strong coupling into a multiplet. 
This can naturally be understood from string theory. 
The AdS/CFT dictionary relates $\sqrt{\lambda}= \frac{R^2}{\alpha'}$ so that at strong coupling, $\lambda\rightarrow\infty$, 
the spectrum approaches that of superstrings in flat spacetime. 
To leading order $\Delta^2 = 4 \,\lambda^{\frac{1}{2}} \,\delta +\mathcal{O}(\lambda^{0})$ with $\delta$ an integer 
labelling the flat-space string mass level~\cite{Gubser:2002tv}. Reassuringly the slope from our numerics perfectly matches the expectation from string theory as was already noticed in previous numerical studies~\cite{Frolov:2010wt,Frolov:2012zv, Gromov:2015wca,Hegeds_2016,Julius:2023hre} but now this is particularly apparent due to the huge range of the `t Hooft coupling available to us.  
In the next section we will be able to compare also sub-leading coefficients to the data available and give analytic predictions for the new orders.

\subsection{Analytic Predictions for the Spectrum}\label{subsec:LargeGFitting}

Here we present the strong coupling analysis of the data we have generated, 
which we managed to convert into concrete analytic predictions for
the strong coupling expansion coefficients of the anomalous dimensions in some cases.

It is useful to introduce the notation of \cite{Gromov:2023hzc} for the expansion coefficients of $\Delta$
\begin{equation}\label{eq:DeltaStrongCoupling}
    \Delta+2 = (\sqrt{\lambda}\,\delta)^{\frac{1}{2}}\,\left(2+\sum _{n=1}^{\infty}\frac{d_{n}}{(\delta \sqrt{\lambda})^{n}}\right)\,.
\end{equation}
In our case
\begin{equation}
    \delta = \frac{S n}{2}\,,
\end{equation}
where $n\in \mathbb{Z}_{>0}$ is the mode number. 

It is also useful to make certain assumption on the behaviour of the coefficients $d_n$ on the spin $S$.
When $S$ and $L$ become of order $\sqrt\lambda$ the energy should scale as $\sqrt\lambda$ too and, assuming there is no
order of limits issue, should be consistent with the classical string prediction for the folded string, which we review in Appendix~\ref{app:foldedstring}.
By observing the classical and quasi-classical results one can also make some analyticity assumption on the dependence of the coefficients 
$d_n$ on the spin $S$, which can be summarised by the following ansatz first proposed in~\cite{Basso:2011rs}
\begin{equation}\label{eq:SquareDeltaStrongMain}
    \bar\Delta^2=L^2 + S(\sqrt{\lambda}\,A_1 + A_2 + \dots) + S^2\(B_1+\frac{B_2}{\sqrt{\lambda}} + \dots \) + S^3\(\frac{C_1}{\sqrt{\lambda}} + \frac{C_2}{\lambda}+\dots\)+\mathcal{O}(S^4)\;.
\end{equation}
Even stronger assumption was used in \cite{Beccaria:2012xm} by further restricting all $A_a,\;B_a,\;C_a,\dots$ to the polynomial in other changers, which is the R-charge $L$ in our case.
One should be, though, careful with the equation \eq{eq:SquareDeltaStrongMain} as we will see for the mode number $n>1$ it does fully hold true.

While this formula suggests the existence of some analytic continuation in $S$ for states with the same mode number $n$ and twist $L$,
at the quantum level it is only known how to make this continuation in the QSC for the case $n=1$. That is another reason on why one should take the 
formula \eqref{eq:SquareDeltaStrongMain} with care at least for the states with $n>1$. Furthermore, it was pointed out in \cite{Gromov:2011bz} that the structure 
\eq{eq:SquareDeltaStrongMain} is inconsitent with the 1-loop quasi-classical correction for $n>1$.
Below we examine different mode numbers separately and show that for $n=2$ for fixed $S$ the dependence on $L$ is non-polynomial and discuss a possible reason for that.

\subsection{Large $g$ Expansion for $n=1$}
The lowest lying $\Delta$ states for each given $L$ and $S$ are the states with mode number $n=1$. When studying their strong coupling expansion instead of $\Delta$ it is more convenient to consider the combination
 $(\Delta+2)^2=\bar \Delta^2$, which has the form \eqref{eq:SquareDeltaStrongMain}.
 The coefficients $A_n,\;B_n,\;\dots$ 
can be additionally assumed to be polynomials in $L^2$ with the maximal degree limited by the 
consistency with the classical scaling $\Delta\sim L\sim S\sim \sqrt\lambda$. 
The coefficients $A_n$ are known for any $n$ from Basso's slope function~\cite{Basso:2011rs}. 
For example
\beqa
\begin{array}{ccl}
 A_1 & = & 2 \\
 A_2 & = & -1 \\
 A_3 & = & -\frac{1}{4}+L^2 \\
 A_4 & = & -\frac{1}{4}+L^2 \\
 A_5 & = & -\frac{25}{64}+\frac{13}{8}L^2-\frac{1}{4}L^{4} \\
 A_6 & = & -\frac{13}{16}+\frac{7}{2}L^2-L^4 \\
 A_7 & = & -\frac{1073}{512}+\frac{1187}{128} L^2-\frac{115}{32}L^4+\frac{1}{8}L^6
\end{array}\,.
\eeqa
All the coefficients $B_n$ are also known in principle from~\cite{Gromov:2014bva} for $L=2,3,4$, 
but in a less-than-explicit way from the curvature function. The result is given in
the form of a dressing phase type of integral and its analytic expansion in large $g$ is rather complicated. In practice one could evaluate the integral with high precision numerically and then decode 
the coefficients analytically assuming they are given by combination of odd zeta values with rational coefficients. This procedure gives
\beqa
\begin{array}{ccl}
 B_1 & = & \frac{3}{2} \\
 B_2 & = & \frac{3}{8}-3 \,\zeta _3 \\
 B_3 & = & \left(\frac{5}{16}-\frac{9 \zeta _3}{2}\right)-\frac{L^2}{2}  \\
 B_4 & = & -\frac{3}{16} \left(62 \,\zeta _3+40\, \zeta _5+1\right)+\frac{3}{16} \left(16\, \zeta _3+20 \,\zeta _5-9\right) L^2  \\
 B_5 & = & -\frac{1}{64} \left(2362\, \zeta _3+1580 \,\zeta _5+203\right)+\frac{1}{8} \left(116 \,\zeta _3+100\, \zeta _5-39\right) L^2+\frac{L^4}{2}
\end{array} \,,
\eeqa
where the last coefficient $B_5$ we obtained for the first time in this paper; it is used in what follows.

Next, starting from $C_n$, systematic knowledge is limited. One can deduce leading and sometimes subleading 
coefficients in $L$ by extrapolating from classical and one-loop semiclassical results, which we review in Appendix \ref{app:foldedstring}.
Let us summarize what we know from the classical and quasi-classical folded string:
\begin{eqnarray}
&&C_1 = -\frac{3}{8} \nonumber \;\;,\;\;
C_2 = \frac{3}{16}\left(20\,\zeta_3+20\,\zeta_5-3 \right)\;\;,\;\;
C_{3} = C_{3,0} + \frac{13}{16}\,L^2\;,  \\
&&D_1 = \frac{31}{64}  \;\;,\;\;
D_2 = \frac{1}{512}\left(-4720\,\zeta_3-4160\,\zeta_5-2520\,\zeta_7+337 \right)\;,\\ \nonumber 
&&E_1 = -\frac{411}{512}\;.
\end{eqnarray}
We see that the coefficient $C_3$ is not known beyond its leading $L^2$ part.
We denoted the constant part by $C_{3,0}$
and this is the only unknown coefficient entering into the $\lambda^{-3/2}$ order of $\bar\Delta^2$. 
Note that to compute it from the string side, one would need to either calculate the $5$-th non-trivial coefficient 
of the short string state or the $2$-loop contribution for the folded string and expand it then for small $L$ and $S$. 
Neither of these calculations are known how to perform from string side.
Nevertheless, we managed to find the $C_{3,0}$ coefficient by comparing the ansatz \eq{eq:SquareDeltaStrongMain} to our numerical results. 
For that we computed the $L=4$ and $S=2$ state with about $40$ digits accuracy in the range $g=2,\dots,20$. As a result we found\footnote{Most of the analytic results in this section are based on our high precision numerical data, so one should leave some room for a doubt that our analytic predictions may not be $100\%$ correct. However, in all cases we tried to convince ourselves in the validity of our prediction by further increasing precision or by making some independent test. So whenever we present an analytic result we are very confident in its validity.}
\beq\la{guessC3}
C_3=-9 \zeta _3^2+21 \zeta _3-\frac{15 \zeta _5}{4}+\frac{131}{128}+\frac{13}{16}L^2\;.
\eeq
We managed to extract this coefficient with the absolute numerical error of $\sim 10^{-20}$ by fitting with our highest precision data.
The simplicity of the result gives further support of the validity of our guess~\eq{guessC3}. After obtaining \eq{guessC3} we further pushed
the precision by several digits to confirm the result, so there is very little doubt in its validity.

Finally, from \eq{eq:SquareDeltaStrongMain} we can extract the coefficients $d_n$.
The coefficients up to $d_3$ are available in the literature~\cite{Gromov:2011bz,Basso:2011rs,Gromov:2011de,Roiban:2011fe,Vallilo:2011fj,
Beccaria:2012xm,
Gromov:2014bva}.
The new coefficient we obtained based on \eq{guessC3} is the one in front of $\lambda^{-7/4}$. Our result for

the general $S$ and $L$ reads
\beqa\label{Deltan1LS}
\begin{array}{l}
 \Delta=\sqrt{2} \sqrt[4]{S^2 \lambda }- 2 +
 \frac{\sqrt{2} \left(\frac{L^2}{4}+\frac{3 S^2}{8}-\frac{S}{4}\right)}{\sqrt[4]{S^2 \lambda }} +
 \frac{2 \sqrt{2} \left(-\frac{L^4}{64}+\frac{S L^2}{32}-\frac{21 S^4}{256}+S^2\left(\frac{5 L^2}{64}-\frac{3}{64}\right) +S^3 \left(\frac{3}{32}-\frac{3 \zeta _3}{8}\right)\right)}{\left(S^2 \lambda \right)^{3/4}} \\
 \frac{\frac{L^6}{512}-\frac{3 S L^4}{512}+\frac{187 S^6}{4096}+ S^2\left(\frac{5 L^2}{512}-\frac{7 L^4}{1024}\right)+S^3 \left(L^2 \left(\frac{3 \zeta _3}{64}+\frac{7}{128}\right)-\frac{11}{512}\right)+S^4 \left(-\frac{73 L^2}{2048}-\frac{21 \zeta _3}{64}+\frac{41}{1024}\right)+S^5 \left(\frac{39 \zeta _3}{128}+\frac{15 \zeta _5}{64}-\frac{129}{2048}\right)}{2^{-5/2}\left(S^2 \lambda \right)^{5/4}} \\
 \frac{8 \sqrt{2}}{\left(S^2 \lambda \right)^{7/4}} \left(-\frac{5 L^8}{16384}+S\frac{5 L^6}{4096}+S^2\left(\frac{9 L^6}{8192}-\frac{21 L^4}{8192}\right)+S^3 \left(\left(-\frac{9 \zeta _3}{1024}-\frac{29}{4096}\right) L^4+\frac{19 L^2}{4096}\right)\right.\\
 +S^4 \left(-\frac{139 L^4}{32768}+\left(\frac{27 \zeta _3}{512}+\frac{419}{8192}\right) L^2-\frac{253}{16384}\right)\\
 +S^5 \left(\left(\frac{63 \zeta _3}{1024}+\frac{45 \zeta _5}{512}-\frac{1015}{16384}\right) L^2-\frac{423 \zeta _3}{1024}-\frac{15 \zeta _5}{64}+\frac{11}{2048}\right)\\
  +S^6 \left(\frac{969 L^2}{32768}-\frac{81 \zeta _3^2}{256}+\frac{99 \zeta _3}{128}-\frac{45 \zeta _5}{512}+\frac{409}{32768}\right)\\
  \left.+S^7 \left(-\frac{1477 \zeta _3}{4096}-\frac{305 \zeta _5}{1024}-\frac{315 \zeta _7}{2048}+\frac{687}{16384}\right)-S^8\frac{9261 }{262144}\right)+{\cal O}(\lambda^{-9/4})\;.\\
\end{array}
\eeqa
which we also checked against $L=2,S=4$ to a precision of $10^{-9}$ in the $\lambda^{-{\frac{7}{4}}}$ term.
In particular for $L=S=2$ i.e. the Konishi operator we get
\beqa\nonumber
\Delta_{\rm Konishi}&=&2 \sqrt[4]{\lambda }-2+2 \sqrt[4]{\frac{1}{\lambda }}+\left(\frac{1}{2}-3 \zeta _3\right) \left(\frac{1}{\lambda }\right)^{3/4}+\left(6 \zeta _3+\frac{15 \zeta _5}{2}+\frac{1}{2}\right) \left(\frac{1}{\lambda }\right)^{5/4}\\
&+&\left(-\frac{81 \zeta _3^2}{4}+\frac{\zeta _3}{4}-40 \zeta _5-\frac{315 \zeta _7}{16}-\frac{27}{16}\right) \left(\frac{1}{\lambda }\right)^{7/4}+\dots\;,
\eeqa
where the last line is our new result.

Finally, an important quantity is the value of $S$ for $\Delta=0$ for the inverse function $S(\Delta)$ to $\Delta(S)$.
This quantity is simply related to the intercept which should control the Regge limit of scattering amplitudes.
Our results thus update the current expansion for the intercept~\cite{Gromov:2014bva,Brower:2014wha}\footnote{
We noticed that the last arXiv version (from 2015) of \cite{Gromov:2014bva} gives the correct result for $L=2$, whereas a typo from the published version seems to propagate into the \cite{Brower:2014wha} coefficient $1/\lambda^{6/2}$.
}
\beqa
\begin{array}{rl}
 \left.S\right|_{\Delta=0}=&-\frac{L^2}{2 \sqrt{\lambda }} -\frac{L^2}{4 \lambda } + \frac{L^2 \left(8 L^2-24\right)}{128 \lambda ^{3/2}} +
 \frac{L^2 \left(L^2 (44+96 \zeta_3)-48\right)}{256 \lambda ^2} +
 \frac{L^2 \left(-4 L^4+(228+576 \zeta_3) L^2-126\right)}{512 \lambda ^{5/2}} \\ &
 \frac{L^2 \left(-12 (3+12 \zeta_3+20 \zeta_5) L^4+(1274+3072 \zeta_3+960 \zeta_5) L^2-432\right)}{1024 \lambda ^3} \\ &
 \frac{L^2 \left(L^6+\left(-227-1504 \zeta_3-2304 \zeta_3^2-4160 \zeta_5\right) L^4+(7953+17896 \zeta_3+9200 \zeta_5) L^2-1899\right)}{2048 \lambda ^{7/2}} \,.
\end{array}
\eeqa
We checked this result by fitting the data for $L=3$ of \cite{Klabbers:2023zdz}, matching several digits for the last coefficients, 
which further supports the validity of our analytic result.

\subsection{Large $g$ Expansion for $n=2$}
Even though introduction of the bigger mode number may seem to be a trivial enterprise 
we will shortly see this is not the case.
For $n>1$ the expression for the large coupling expansion coefficient,
with polynomial dependence on the charges $L$ and $S$ is known to contain inconsistencies as was noticed
in \cite{Gromov:2011bz} such as presence of the negative powers of $L$ in the coefficients of the equation \eq{eq:SquareDeltaStrongMain} and a need to introduce coefficients with negative indexes e.g. $C_{-2}=12/L^4$ which gives a strange $\sim\lambda$ term at $S^3$ order. This means that an ansatz \eqref{eq:SquareDeltaStrongMain} would eventually fail when applied to the $n=2$ case.

Nevertheless, let us assume \eqref{eq:SquareDeltaStrongMain} as before and
try to deduce maximum of information we can about the unknown coefficients.
For our purposes in this section we will only need to find $A_n$, $B_1,B_2$ 
and $C_1$ the reason for this will become clear shortly. 
The coefficients $A_n$ 
are fixed from a simple generalisation of Basso's slope function which  
amounts to shifting $A_m \mapsto n^{2-m}\,A_{m}$ with $n$ the mode number. 
Assuming the polynomial dependence on $L$ as well, we can find $B_{1}$ and $C_{1}$
from the classical string limit 
by matching \eqref{eq:SquareDeltaStrongMain} with the classical expression \eq{eq:DeltaClassic}. $B_2$ was found in \cite{Gromov:2011bz} from a one-loop correction.

\begin{equation}\label{eq:ABCModes}
    B_1 = \frac{3}{2}\,,
    \quad
     B_{2} = 
                -12 \, \zeta_3 - \frac{13}{16} \,,\quad
    C_1 = -\frac{3}{16}\,.
\end{equation}
At the end this procedure gives\footnote{as a spoiler: we found that this formula is incorrect beyond the first $3$ terms!}
\beqa\label{eq:DeltaAns}
&&\Delta^{\rm Ansatz}_{L,S=2,n=2}= 2 \sqrt{2} \sqrt[4]{\lambda }-2+\frac{\left(L^2+4\right) }{4 \sqrt{2}\lambda^{1/4}}+\frac{\left(-4L^4+96 L^2-64(96 \zeta _3+11)\right)}{512 \sqrt{2}\lambda^{3/4}}\\ \nonumber
&&+\frac{\left(L^6-20 L^4+80 L^2+1536 \left(L^2+24\right) \zeta _3+122880 \zeta _5+2048 B_{3,0}+6176\right) }{2048 \sqrt{2}\lambda^{5/4}}\;.
\eeqa
Fitting our numerics we find perfect agreement for the first $3$ terms. However, when we go to the next order something surprising happens. Let us focus on $\lambda^{-3/4}$ coefficient and remove its denominator, then the ansatz \eq{eq:DeltaAns} predicts the following
\begin{equation}\label{eq:FirstGuess}
    512\, \sqrt{2}\Delta^{\text{Anstaz}}_{L,S=2,n=2}\bigg|_{\lambda^{-\frac{3}{4}}} \simeq  -4\,L^4+96 L^2-8089.43761301255 \,.
\end{equation}
To check \eqref{eq:FirstGuess} we collected data for $L=4,\dots,10$ with an estimated error around $10^{-14}$ for the coefficient of $\mathcal{O}(\lambda^{-\frac{3}{4}})$. Fitting an even fourth order polynomial in $L$ to our data we found 
\begin{equation}
    - 4.0386 L^4+ 102.2702 L^2-8370.5094  
\end{equation}
which is close but still definitely in  \emph{disagreement} with \eqref{eq:FirstGuess} at our precision. This indicates that
the prediction \eqref{eq:FirstGuess} is not completely correct. 

Next, we supplemented the fit with additional powers of $\frac{1}{L^2}$, obtaining
\begin{equation}
    - 4.000001 L^4+ 96.0003 L^2 
    -8089.4- \frac{3067.8}{L^2} - \frac{6317.6}{L^4}+\frac{16006.9}{L^6} +\frac{93168.2}{L^8}
\end{equation}
which surprisingly improves the agreement for the first three terms! One can furthermore verify that the relative coefficients between the inverse powers of $\frac{1}{L^2}$, after subtracting the expected coefficients from positive powers of $L$ to increase precision, appears to be be simple rational numbers. 
At the same time there is no indication that the series in the inverse powers of $L^2$ truncates and thus we need to restore the whole function of $L^2$ from some other principle, as we only have finite number of numerical points in $L$ to play with. 

In order to get more insight into the kind of functions/singularities in $L^2$ may appear, we explored the analytic expansion of the densities at strong coupling, as presented in the Section~\ref{eq:ScalingWithL} below. The main output of this analysis is a natural appearance of the nontrivial combination $\sqrt{L^4-4L^2+36}$.
We thus changed the basis for the fit by adding this square root for the basis we use for the linear fit. Fitting against the set $\{L^4,L^2,1,\sqrt{L^4-4L^2+36}\}$ we found
\begin{equation}\label{eq:DeltaN2Lambda}
   512\,\sqrt{2} \Delta\bigg|_{\lambda^{-\frac{3}{4}}} = -4\,L^4+288\,L^{2}-192\,\sqrt{L^4-4L^2+36} -64\left(17+96\zeta_3\right)
\end{equation}
and now with all errors of order $10^{-14}$ in perfect agreement with the precision of our fit.

Furthermore, inspired by this success and after further increasing precision we also managed to find the next order coefficient $d_4$ which reads
\begin{equation}
\begin{split}
    \Delta\bigg|_{\lambda^{-\frac{5}{4}}} &= \frac{3 \left(L^6-64 L^4+180 L^2-432\right)}{128 \sqrt{2} \sqrt{L^4-4 L^2+36}}\\
    &+\frac{L^6-68 L^4+16 \left(96 \zeta _3+191\right) L^2+192 \left(96 \zeta _3+640 \zeta _5+3\right)}{2048 \sqrt{2}}\;.
\end{split}
\end{equation}
We notice that the above result agrees at all positive powers of $L^2$ with the prediction \eq{eq:DeltaAns} as well as the coefficient of $\zeta_5$ in the constant term, but of course also contain infinitely many negative powers of $L^2$.

For completeness let us present the full result for $\Delta$, which get more compact in the square form
\begin{equation}\la{DeltaIni}
    \begin{split}
    (\Delta_{S=2,n=2}+2)^2&=
8 \sqrt{\lambda }+\left(L^2+4\right)+\frac{\frac{5 L^2}{2}-\frac{3}{2} \sqrt{L^4-4 L^2+36}-48 \zeta_3-8}{\sqrt{\lambda }}\\
&+\frac{6 L^2-\frac{3 \left(2 L^4-5 L^2+18\right)}{\sqrt{L^4-4 L^2+36}}+240 \zeta_5+24 \zeta_3-1}{\lambda}+{\cal O}(\lambda^{-3/2})\,.
\end{split}
\end{equation}

Let us emphasise that
the key for sucessully finding the above expansion is the knowlege 
of the square root structure $\sqrt{L^4-4L^2+36}$ coming from the 
analytic analysis of the densities at strong coupling as we present in Section~\ref{sec:AnalysisDensities}, combined with the information 
from the classical limit and one-loop correction and the Basso's slope function.

\paragraph{Analytic continuation of $n=2$.} Our answer \eqref{eq:DeltaN2Lambda} 
immediately makes another prediction: there should be another state with the 
same $\Delta$ up to a sign in front of $\sqrt{L^4-4L^2+36}$. 
Denoting the dimension of this 
state as $\Delta^*$ we have
\begin{equation}\la{Deltastar}
    \begin{split}
    (\Delta^*_{S=2,n=2}+2)^2&=
8 \sqrt{\lambda }+\left(L^2+4\right)+\frac{\frac{5 L^2}{2}+\frac{3}{2} \sqrt{L^4-4 L^2+36}-48 \zeta_3-8}{\sqrt{\lambda }}\\
&+\frac{6 L^2+\frac{3 \left(2 L^4-5 L^2+18\right)}{\sqrt{L^4-4 L^2+36}}+240 \zeta_5+24 \zeta_3-1}{\lambda}+{\cal O}(\lambda^{-3/2})\,.
\end{split}
\end{equation}

However, there is some puzzle in this natural statement. We see that this state should have the same level as the original state and the same quantum numbers. 
In  the $\algsl(2)$ sector the string mass-level is given by $S n/2$ there is no other state within this ``closed'' sector, which could then put some doubt on our proposal of the existence of such state.
However, the closeness of the $\algsl(2)$ sector to all loops order does not imply that the $\algsl(2)$ operators cannot mix with other sectors in strong coupling expansion!

Indeed, by observing the table of states from \cite{Gromov:2023hzc} we identify the state with \verb"St.No."$109$ which has the same quantum numbers and the same first subleading order in $\Delta$ at strong coupling, but
at weak coupling behaves as $\Delta=6+g^2\; 4.524563121$. There is also yet another state, \verb"St.No."$107$, with the same quantized quantum numbers but with different subleading coefficient in $\Delta$. It seems that this state does not play a role in the discussion to follow. It would be interesting to better understand why and potentially make a connection with KK-towers.

\begin{figure}[h]
    \centering
    \includegraphics[scale=0.7]{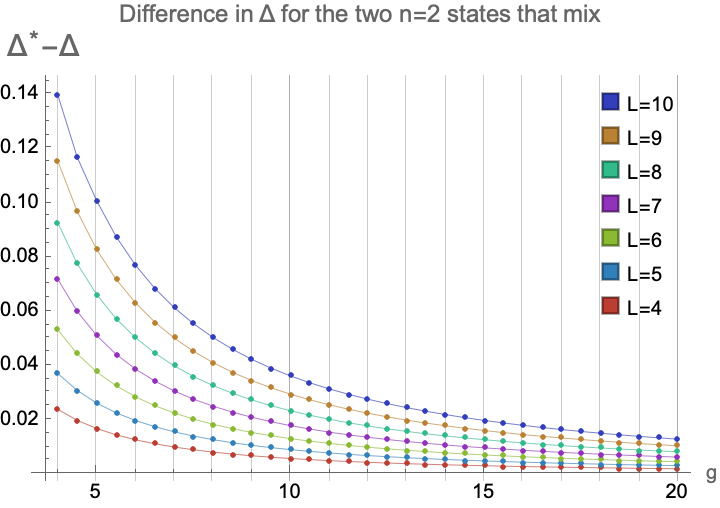}
    \caption{Comparison of the states $\Delta_{n=2,S=2}$ and $\Delta^*_{n=2,S=2}$ for various values of $L$. The difference decreases as $g^{-3/2}$ at strong coupling indicating degeneracy, as the states are indistinguishable by their quantum numbers. The mixing of these states results in a branch-cut in the $L$ plane breaking down the common analyticity assumption.}
    \label{fig:MixingPicture}
\end{figure}

In particular, since all quantum numbers are equivalent to the initial state, the $\bP$-functions scale with the same powers. We analysed \verb"St.No."$109$ in our numerical algorithm and found agreement with \eqref{Deltastar} with a precision of $10^{-9}$ for the last term. We display the difference between $\Delta$ and $\Delta^{*}$ in Figure~\ref{fig:MixingPicture}.

Schematically one can think of this state as the addition of an extra Laplacian to the Konishi-like operators, which does not affect the quantum numbers but changes the bare dimension. Such states also appear on the analytic continuation in spin $S$.

The possibility of the mixing opens up an option of restoring the potentiality of the ansatz \eq{eq:SquareDeltaStrongMain}, if instead of the dimensions themselves we assume the polynomiality of a mixing matrix.
Indeed, consider the simple polynomial matrix ${\cal M}_{S=2,n=2}$
\beqa\nonumber
{\cal M}_{S=2,n=2}&=&
\(8 \sqrt{\lambda }+4+
L^2+\frac{\frac{5 L^2}{2}-48 \zeta_3-8}{\sqrt{\lambda }}+\frac{6 L^2+240 \zeta_3+24 \zeta_3-1}{\lambda }\) {\mathbb I}_{2\times 2}\\
&+&
\left(
\begin{array}{cc}
 -\frac{\frac{L^2}{2}-9}{\sqrt{\lambda }}-\frac{2 L^2-9}{\lambda } & \frac{\sqrt{2} L^2}{\sqrt{\lambda }}+\frac{4 \sqrt{2} L^2}{\lambda } \\
 \frac{\sqrt{2} L^2}{\sqrt{\lambda }}+\frac{4 \sqrt{2} L^2}{\lambda } & +\frac{\frac{L^2}{2}-9}{\sqrt{\lambda }}+\frac{2 L^2-9}{\lambda } \\
\end{array}
\right)\;.
\eeqa
One can check that its eigenvalues reproduce 
the squared dimensions $(\Delta+2)^2$ and
$(\Delta^*+2)^2$. Note that the off-diagonal elements are of order $\sqrt{\lambda}$ in the classical scaling, meaning that they should be in principle be computable by some leading order quasi-classical method. 
That would be interesting to investigate this direction. 

Lastly, it is natural to expect that for states with larger mode numbers, the mixing matrix should become larger. This suggests an intriguing possibility: a finite-dimensional spin-chain-like picture could emerge at strong coupling, describing these mixing matrices as integrable Hamiltonians.

\subsection{$n=2$, $S=4$ States}
Another case we considered is $n=2$ and $S=4$ state. Naively, one would expect that 
the situation is very similar to the $S=2$ case, where the quasi-classical prediction correctly reproduces the first $3$ orders and all terms with positive powers of $L$ when expanded in $1/L^2$ up to the order $1/\lambda^{5/4}$. However, as was already found in \cite{Gromov:2023hzc} already the $\lambda^{-1/4}$ deviates from the prediction. By fitting our numerical data we found
\beq
\Delta_{S=4,n=2}=4 \sqrt[4]{\lambda }-2+\frac{\frac{L^2}{8}+4}{\sqrt[4]{\lambda }}-\frac{\frac{L^4}{2}+16 L^2+144 \sqrt{L^4-4 L^2+36}+6144 \zeta_3+608}{256 \lambda ^{3/4}}+\dots
\eeq
where the 3rd term is expected to be $(\frac{L^2}{8}+\frac{5}{2})\lambda^{-1/4}$.
Similarly, the large $L^2$ expansion of the $\lambda^{-3/4}$ term gives
\beq
-\frac{L^4}{512}-\frac{5 L^2}{8}-24 \zeta_3-\frac{5}{4}+{\cal O}(1/L^2)
\eeq
whereas the quasi-classical ansatz would give $-\frac{L^4}{512}+\frac{11 L^2}{64}-24 \zeta_3-\frac{127}{32}$ i.e. only the leading in $L$ terms and the $\zeta_3$ terms agree. This suggests that not only the polynomial structure in $L$ is lost but also dependence on $S$ is more complicated than in \eq{eq:SquareDeltaStrongMain} and requires further future detailed investigation.

\section{Analysis of the Densities at Strong Coupling}\label{sec:AnalysisDensities}
In this section we present our initial analysis of the densities $\rho$ and $\eta$ 
at strong coupling.\footnote{Some results in this section were obtained in collaboration with Nicolò Primi in the early stage of this project.}
In particular we will deduce crucial clues about the structure of the strong coupling expansion of the anomalous dimensions, which we have already used in the previous section.

\subsection{Expansion of Densities}
\begin{figure*}[ht]
    \centering
    \begin{subfigure}[t]{0.5\textwidth}
        \centering
        \includegraphics[scale=0.25]{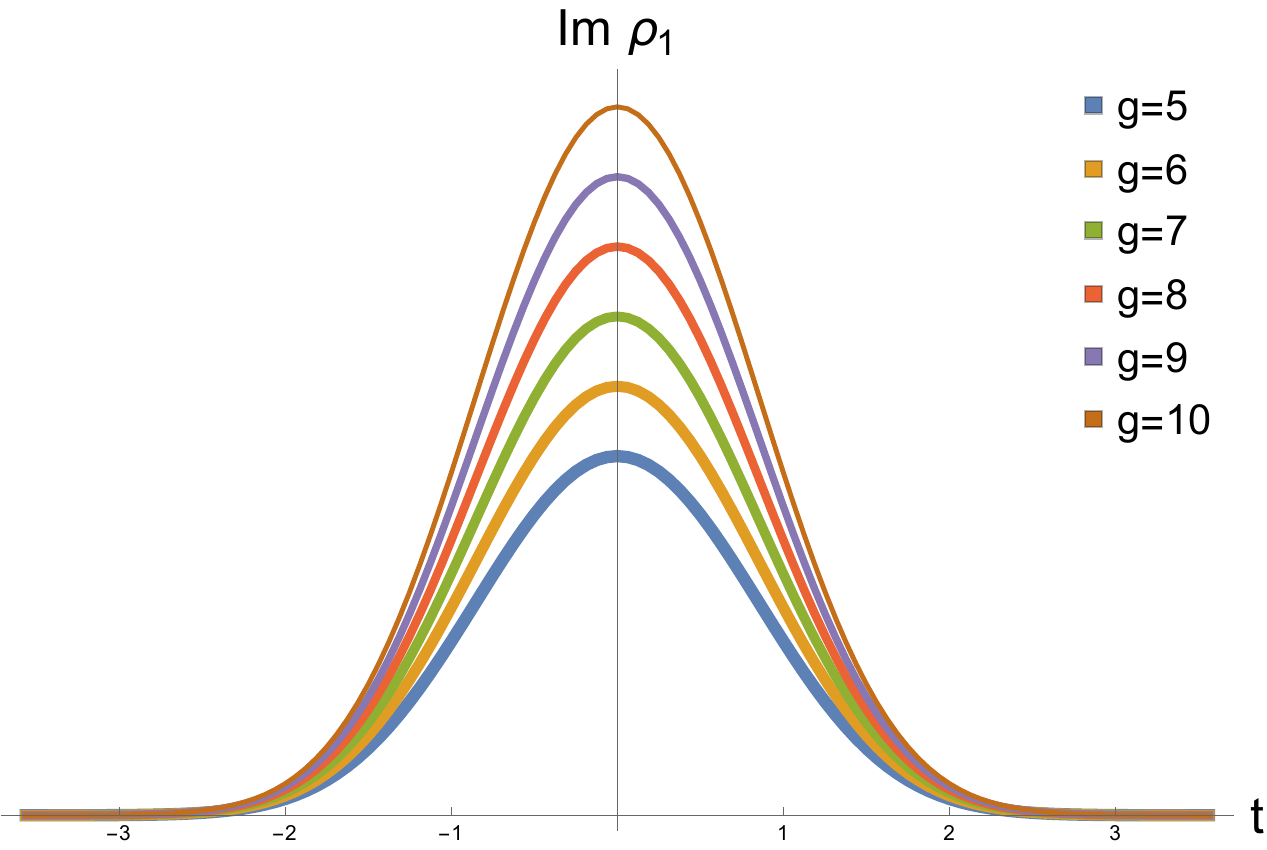}
        
    \end{subfigure}%
    ~ 
    \begin{subfigure}[t]{0.5\textwidth}
        \centering
        \includegraphics[scale=0.25]{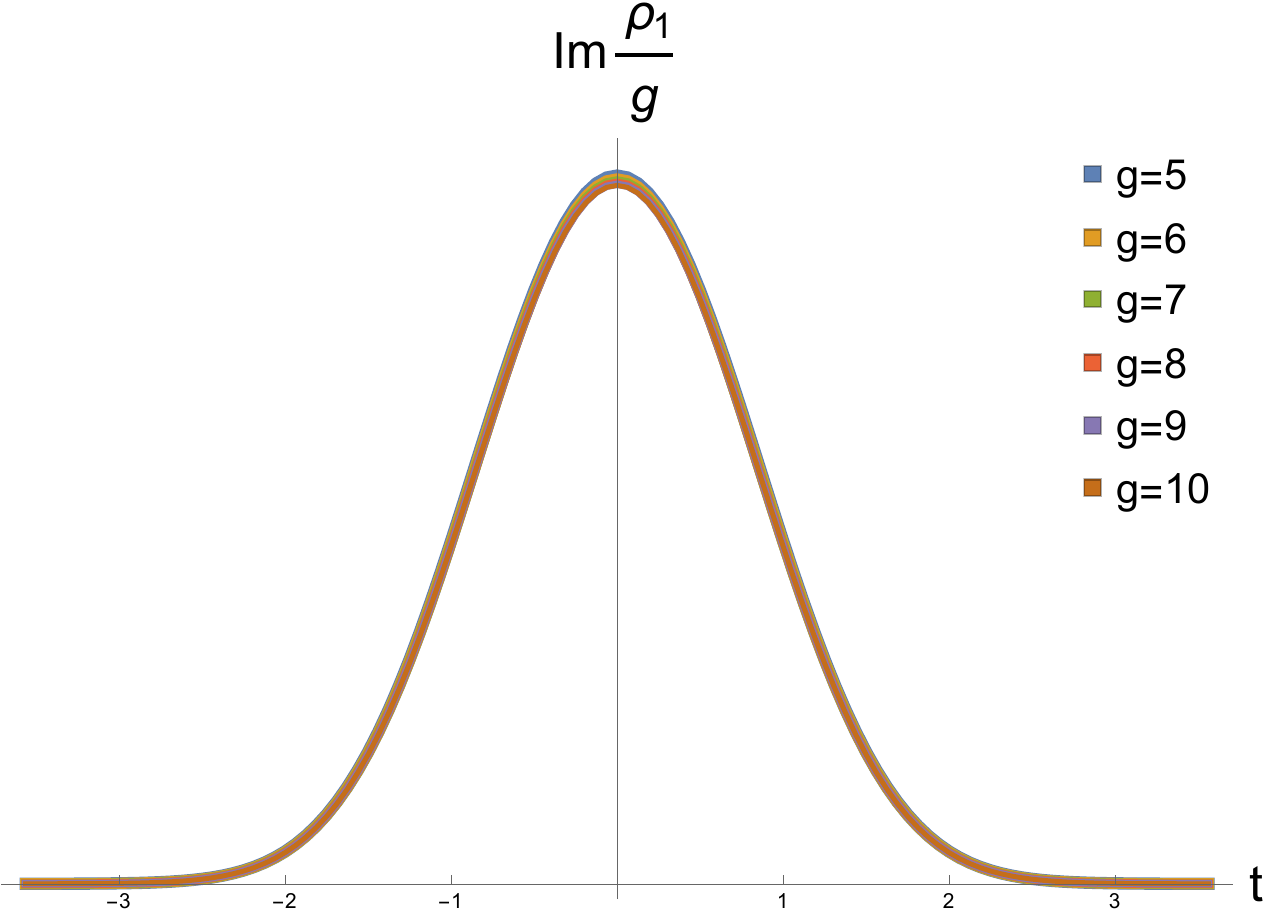}
        
    \end{subfigure}
    \caption{The imaginary part of the density $\rho_1$ for $n=1,L=2$ scales as $g$.}
    \label{fig:ScalingRho1}
\end{figure*}

In order to define the expansion we need to change variables to ones that are more suitable at strong coupling.
As discussed in detail in Section~\ref{sec:introdensities} the densities $\rho$ and $\eta$ are 
sharply peaked at $x=\pm 1$ at strong coupling. The width of the peaks is shrinking as $g$ increases.
To better probe the non-trivial part of the density we 
switch variables to
\begin{equation}
    \hat{x} = e^{\frac{\ii t}{\sqrt{2\pi g }}}\;,
    \quad
    \check{x} = e^{\frac{s}{\sqrt{2\pi g}}}\;,
\end{equation}
where $\hat{x}$ and $\Check{x}$ parameterise the unit circle and the real line respectively where the support of $\rho$ and $\eta$ are located.
As we can see on the right panel of Figure~\ref{fig:ScalingRho1} the densities now look similar for different $g$'s, furthermore if 
we rescale the densities by a suitable power of $g$ we see that the densities almost exactly coincide.

In the coming two sections we illustrate some features in our data for mode number $n=1,2,3$. 
While all plots and observations are based on data up to $n=3$ it is natural to expect that 
the patterns observed extended to higher $n$. For simplicity we mainly focus on the simplest $S=2$ case in this section,
reserving the general $S$ case for future work.

\subsection{Scaling of $\rho$ for Various Mode Numbers}
\begin{figure*}[h]
    \centering
    \begin{subfigure}[t]{0.3\textwidth}
        \centering
        \includegraphics[scale=0.34]{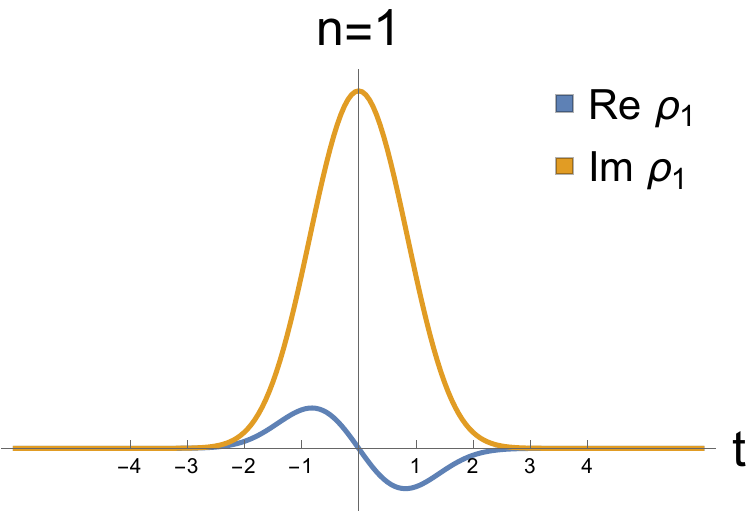}
        
    \end{subfigure}
    ~ 
    \begin{subfigure}[t]{0.3\textwidth}
        \centering
        \includegraphics[scale=0.34]{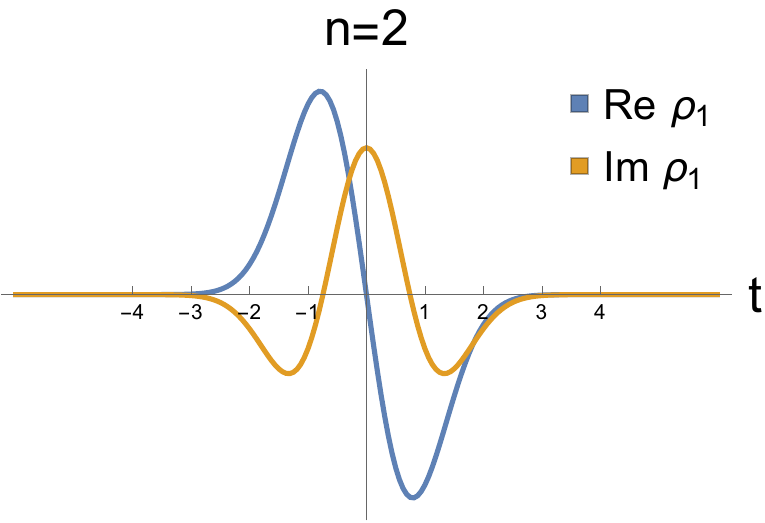}
        
    \end{subfigure}
    ~ 
    \begin{subfigure}[t]{0.3\textwidth}
        \centering
        \includegraphics[scale=0.34]{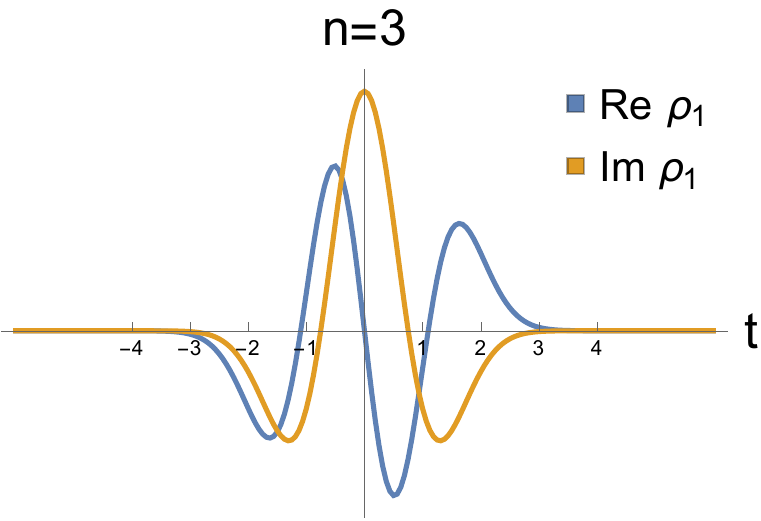}
        
    \end{subfigure}
    \caption{The shape of the density depends on the mode number $n$. The picture only displays $\rho_1$ for the minimal possible choices of $L$ i.e. $2,\;4$ and $6$ respectively, however changing $L$ does not lead to any new features.}
    \label{fig:rhoDifferentN}
\end{figure*}

We depict an example of the real and imaginary parts of the 
densities for various $n$'s at large $g$  
in Figure~\ref{fig:rhoDifferentN}. 
One of the crucial observations is that the mode number determines the 
overall scaling of the density with $g$.
In addition $n$ also changes the structure of the densities as we depict in the same figure.

By comparing the densities for different values of $g$ we found the scaling $\rho(t) \sim g^{\frac{1}{2}+\frac{n}{2}}$ where $n=1,2,3$ is the mode number. The real and imaginary parts of $\rho$ scale differently. From our numerical data we found that the densities furthermore have a natural expansion in $g^{-\frac{1}{2}}$, explicitly
\begin{equation}\label{eq:RhoExpansion}
    \rho_{a}(t) = \sum_{m =0,\frac{1}{2},1,\dots}^{\infty} \frac{g^{\frac{n+1}{2}}}{(2\pi g)^m}\,\rho^{(\frac{n+1}{2}-m)}_{a}\;\;,\;\;n=1,2,3 \,.
\end{equation}
We explicitly verified the above expansion for $n=1,2,3$ but it is natural to expect this to hold for bigger $n$'s as well.
In this expansion $\rho^{(m)}_2,\rho_1^{(m-\frac{1}{2})},m\in \mathbb{Z}$ are real while $\rho^{(m)}_1,\rho_2^{(m-\frac{1}{2})},m\in \mathbb{Z}$ are imaginary so that the real and imaginary parts of $\rho$ each have only integer or half-integer powers of $g$ in their expansion.

\begin{figure*}[h]
    \centering
    \begin{subfigure}[t]{0.3\textwidth}
        \centering
        \includegraphics[scale=0.3]{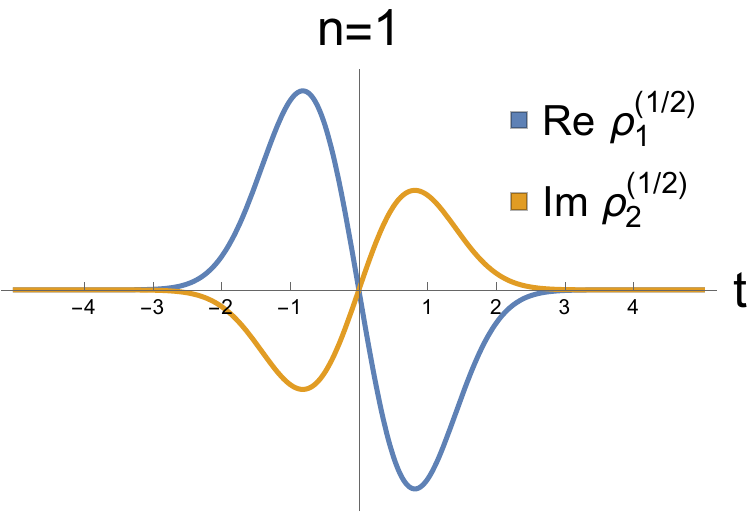}
        
    \end{subfigure}
    ~ 
    \begin{subfigure}[t]{0.3\textwidth}
        \centering
        \includegraphics[scale=0.3]{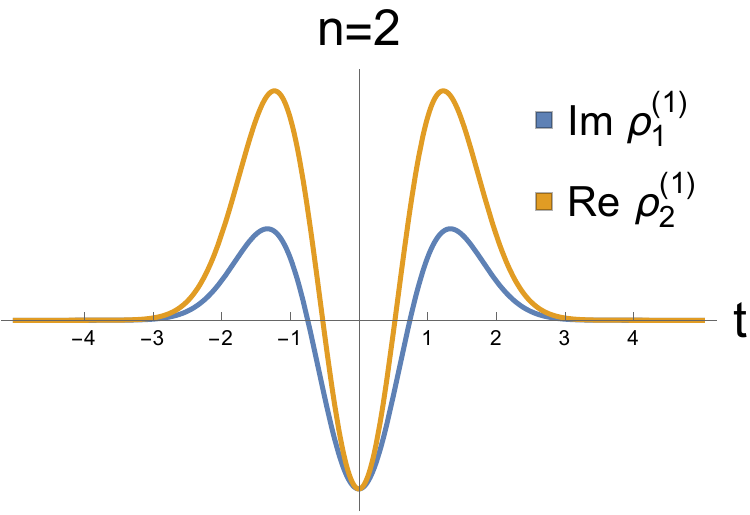}
        
    \end{subfigure}
    \begin{subfigure}[t]{0.3\textwidth}
        \centering
        \includegraphics[scale=0.3]{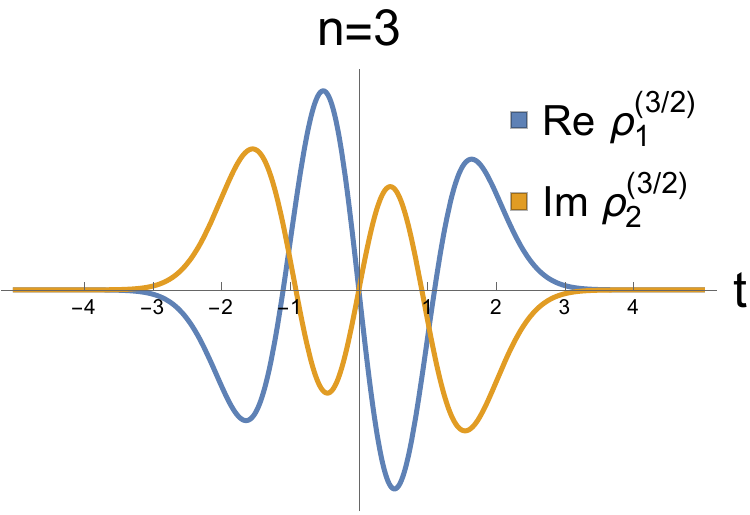}
        
    \end{subfigure}
    \caption{Whereas at the leading order the densities $\rho_{1}(t)$ and $\rho_{2}(t)$ are equal up to a (purely imaginary) constant, the relation between the sub-leading terms in general involves the leading order densities too. For $n=1$, however, they are still proportional to each other. We discuss these relations in Section \ref{sec:ScalingInL}. }
    \label{fig:SubRho}
\end{figure*}

In Figure~\ref{fig:rhoDifferentN} we only plot $\rho_{1}$ because we found that at large $g$ both densities are indistinguishable  $\rho_{1}^{(\frac{n+1}{2})} \propto \rho_{2}^{(\frac{n+1}{2})}$.
In the next section we give an analytic proof of this relation. However, at the next order in $g$ the densities starts to differ for $n\geq 2$ which we display in Figure~\ref{fig:SubRho}.

We hope that the expansion \eqref{eq:RhoExpansion} could be a natural starting point for future analytic investigations into the strong coupling regime of the QSC. While we will not pursue such an analytic analysis in this paper we will take some first steps in Section~\ref{sec:ScalingInL} to deduce how some of $\rho^{(a)}$ depend on the parameter $L$.

\subsection{Scaling of $\eta$ for Various Mode Numbers}\label{sec:ScalingEta}
\begin{figure*}[h]
    \centering
    \begin{subfigure}[t]{0.3\textwidth}
        \centering
        \includegraphics[scale=0.3]{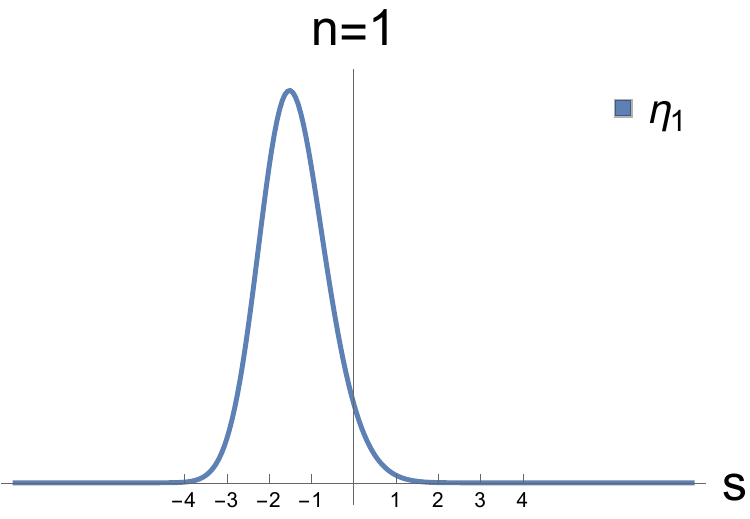}
        
    \end{subfigure}
    ~ 
    \begin{subfigure}[t]{0.3\textwidth}
        \centering
        \includegraphics[scale=0.3]{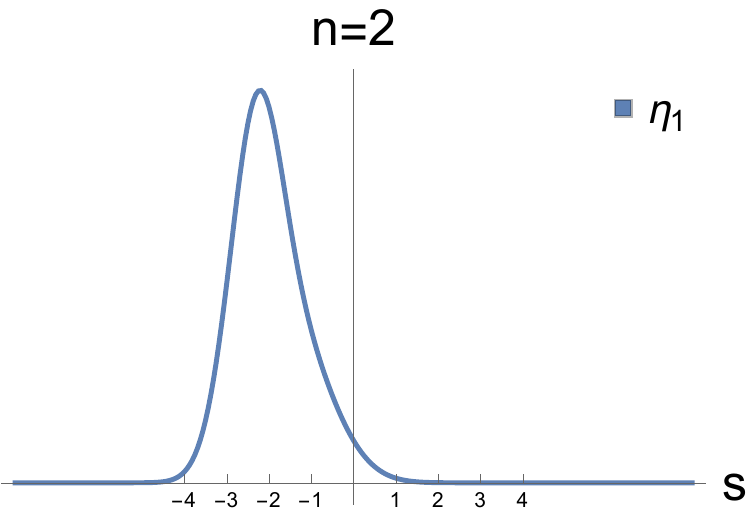}
        
    \end{subfigure}
    \begin{subfigure}[t]{0.3\textwidth}
        \centering
        \includegraphics[scale=0.3]{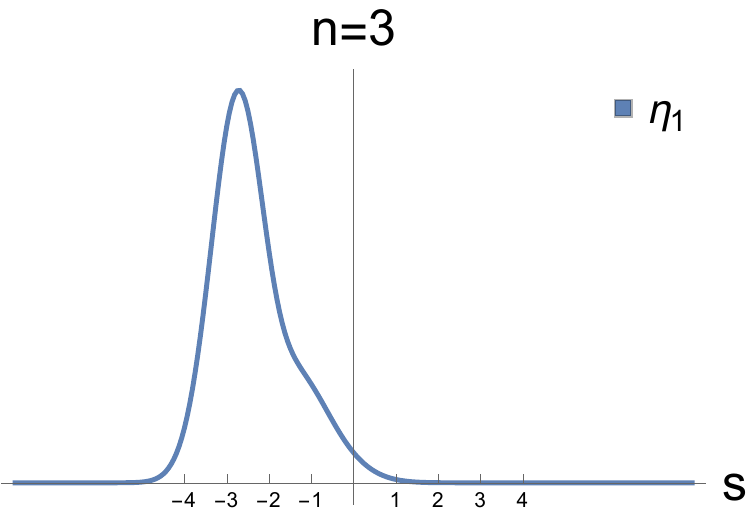}
        
    \end{subfigure}
    \caption{We plot the densities $\eta_{1}(s)$ for the mode numbers $n=1,2,3$.}
    \label{fig:etaDifferentN}
\end{figure*}

\begin{figure*}[h]
    \centering
    \begin{subfigure}[t]{0.3\textwidth}
        \centering
        \includegraphics[scale=0.25]{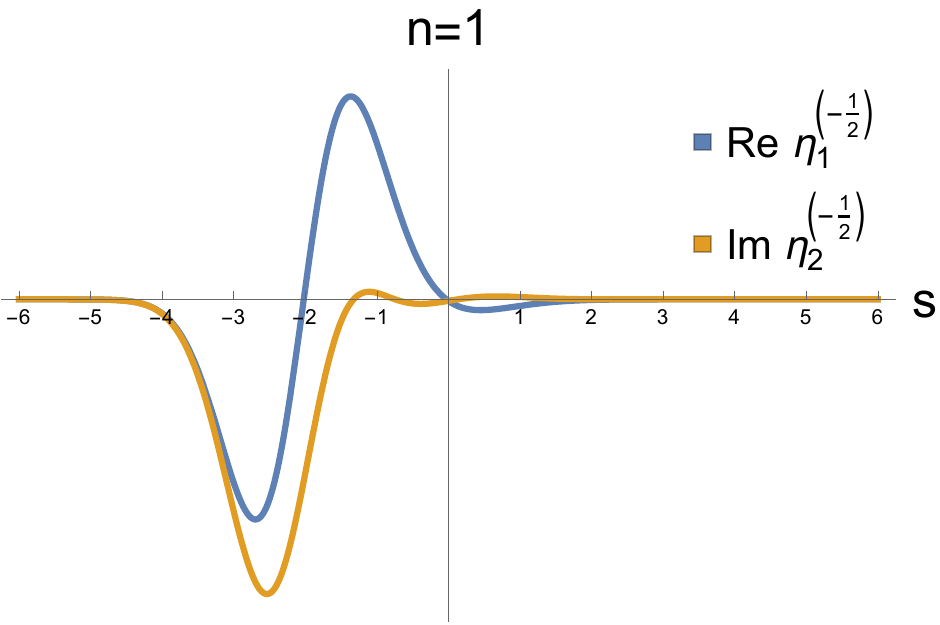}
        
    \end{subfigure}
    ~ 
    \begin{subfigure}[t]{0.3\textwidth}
        \centering
        \includegraphics[scale=0.25]{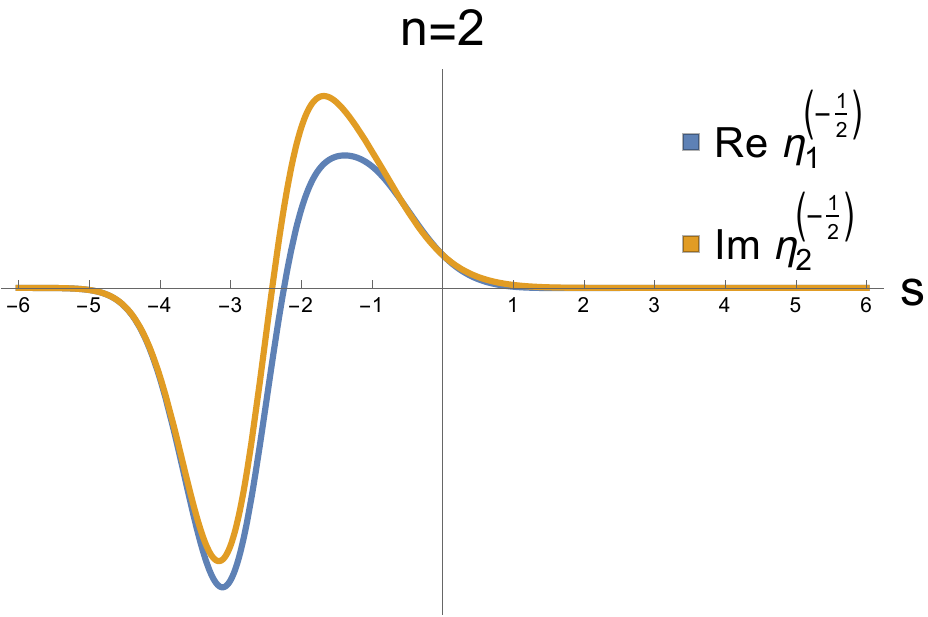}
        
    \end{subfigure}
    \begin{subfigure}[t]{0.3\textwidth}
        \centering
        \includegraphics[scale=0.25]{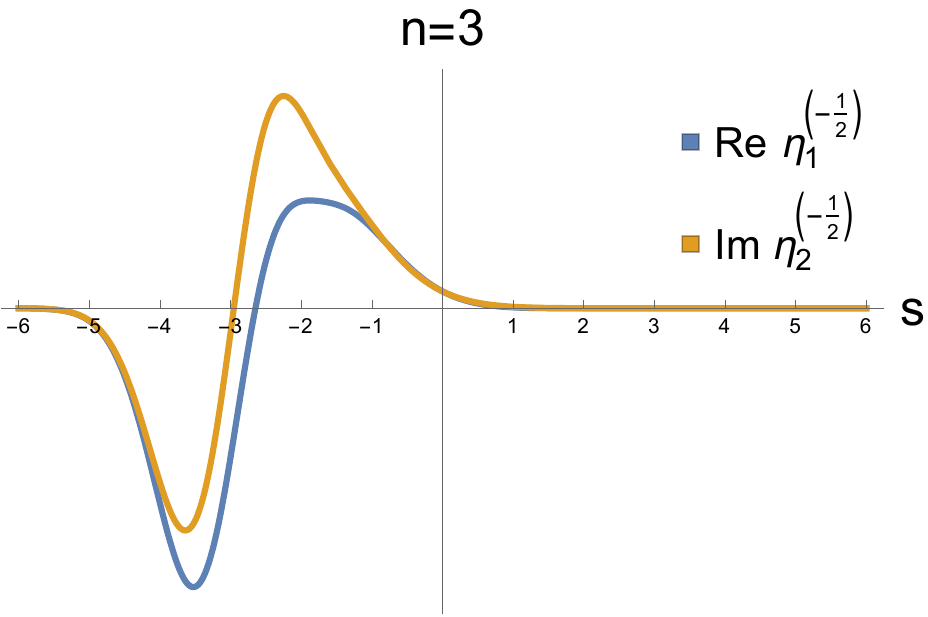}
        
    \end{subfigure}
    \caption{The subleading correction $\eta_i^{(-\frac{1}{2})}$ to $\eta_1$ and $\eta_2$ for $n=1,2,3$. While the leading parts $\eta_i^{(\frac{1}{2})}$ are proportional, the subleading parts are not as can be clearly seen. }
    \label{fig:SubEtaFig}
\end{figure*}

Our numerical analysis indicates that $\eta \sim \sqrt{g}$ for all $n$ and $S=2$, but the shape of $\eta$ is adjusted slightly depending on $n$, see Figure~\ref{fig:etaDifferentN}.
Just as for $\rho$ we find that $\eta$ can naturally be expanded at strong coupling
\begin{equation}
    \eta_{i} = \sum_{m=0,1,2\dots}^{\infty} \frac{g^{\frac{1}{2}}}{(2\pi g)^{m}}\,\eta_i^{(\frac{1}{2}-m)}\,.
\end{equation}
The main difference as compared to \eqref{eq:RhoExpansion} is that the series is now in powers of $\frac{1}{g}$. From numerics we find that $\eta^{(\frac{1}{2})}_{1} \propto \eta^{(\frac{1}{2})}_2$. Such a relation does not hold for subleading $\eta$ as can be seen in Figure~\ref{fig:SubEtaFig}.

\subsection{Scaling in $L$}
\label{sec:ScalingInL}
Apart from $g$ we also have the parameter $L$ at our disposal. It is thus natural ask the form of $\rho^{(m)}_a$ as a function of $L$. 
To investigate the dependence of $\rho$ on $L$ we will use constraints from the asymptotics of Q-functions, namely \eqref{eqn:AArelns}. Expanding $\Delta =2\sqrt{n}\sqrt{4\pi g} + \mathcal{O}(g^{0})$ we find the following constraints on the densities, using \eqref{eqn:Pfinal},
\begin{align}
    &\left(\simoint \frac{dy}{\pi \ii}\, y \,\rho_1(y) \right)\, A = -g\,\ii\frac{16\,\pi^2\,n^2}{L(L+1)} +\mathcal{O}(g^{\frac{1}{2}})\,,
    \\
    &-\left(A+\simoint \frac{dy}{\pi \ii} \rho_2(y)\right)\simoint \frac{dy}{\pi \ii} \frac{1}{y}\rho_1(y) =  -g\,\ii\frac{16\,\pi^2\,n^2}{L(L-1)} +\mathcal{O}(g^{\frac{1}{2}})\,.
\end{align}
We will assume the scaling $A \sim \simoint dy\,y^{\pm 1}\,\rho_1(y) \sim \simoint dy\,\rho_2(y)\sim \sqrt{g}$. To recast the left-hand side into an expansion in $g$ we use \eqref{eq:RhoExpansion} and expand $A=A^{(\frac{1}{2})}\sqrt{g}+\mathcal{O}(g^{0})$. We find
\begin{equation}\label{eq:MomentsFromAA}
\begin{split}
    &A^{(\frac{1}{2})}\int_{-\infty}^{\infty} \frac{dt}{(\sqrt{2\pi})^{n}\,\pi} \sum_{m=0}^{n-1} \frac{(2\ii t)^{m}}{m!} \rho^{(1+\frac{m}{2})}_1(t) = - \ii\frac{16\pi^2n^2}{L(L+1)}\,,
    \\
    &\left(A^{(\frac{1}{2})} + \int_{-\infty}^{\infty} \frac{dt}{(\sqrt{2\pi})^{n}\,\pi}\,\sum_{m=0}^{n-1}\frac{(it)^{m}}{m!}\rho_{2}^{(1+\frac{m}{2})}  \right)\,\int_{-\infty}^{\infty} \frac{dt}{(\sqrt{2\pi})^{n} \pi} \,\rho^{(1)}_{1}(t) = - \ii\frac{16\pi^2n^2}{L(L-1)}\,.
\end{split}
\end{equation}
By itself \eqref{eq:MomentsFromAA} is not sufficient to fix how $\rho$ depends on $L$. To find constraints among $\rho_a$ we can use that the PP-QQ relations \eqref{eqn:PPQQrelns}. For us it will be enough to consider these relations to order $\mathcal{O}(g^{0})$. As noted in Subsection~\ref{sec:ScalingEta} $\eta(s) \sim \sqrt{g}$ which implies that $\bP_{a}^{[m]}(\bP^{a})^{[n]} = \mathcal{O}(g)$. Furthermore, numerics shows that the constraint is even stronger, namely $\bP_{a}^{[m]}(\bP^{a})^{[n]} = \mathcal{O}(g^{0})$. Since $m,n$ are arbitrary we can consider the slightly more convenient equation
\begin{align}\label{eq:PPQQtoLO}
    \bP_{a}(s)\bP^{a}(t) = \mathcal{O}(g^{0})
\end{align}
for $s$ and $t$ continuous. From \eqref{eq:RhoExpansion} it follows that $\bP \bP \sim \mathcal{O}(g^{n+1})$ and thus \eqref{eq:PPQQtoLO} provides us with a set of constraints on the densities. For example, expanding \eqref{eq:PPQQtoLO} one find $\rho^{(\frac{n+1}{2})}_1 \propto \rho^{(\frac{n+1}{2})}_2$ which we can verify to high accuracy using our numerical results. Unfortunately, using \eqref{eq:PPQQtoLO} to higher and higher orders in $g$ is still rather cumbersome and we will refrain from attempting a general analysis. Working out the constraints for $n=1,2$ we found that
\begin{align}\label{eq:ScalingWithL}
    &\left(\int_{-\infty}^{\infty} dt\,\rho^{(1)}_{1}(t)\right)\left(\int_{-\infty}^{\infty} dt\,\rho^{(1)}_{2}(t)\right)  = 64\,\ii\,\pi^{5}\,\,\frac{1}{L^2-1}\,,\quad n=1\,,
    \\
    &\left(\int_{-\infty}^{\infty} dt \,t\,\rho^{(\frac{3}{2})}_{1}(t) \right) \left(\int_{-\infty}^{\infty} dt \,t\,\rho^{(\frac{3}{2})}_{2}(t) \right)=512\,\ii\,\pi^{6}\,\frac{L^2-18\pm\sqrt{L^4-4L^2+36}}{(L^2-9)(L^2-1)} \,,\quad n=2\,.\label{eq:ScalingWithLN2}
\end{align}
These expressions are the reason we we were able to deduce the square-root in Section~\ref{subsec:LargeGFitting}.

\begin{figure*}[h]
    \centering
    \begin{subfigure}[t]{0.5\textwidth}
        \centering
        \includegraphics[scale=0.4]{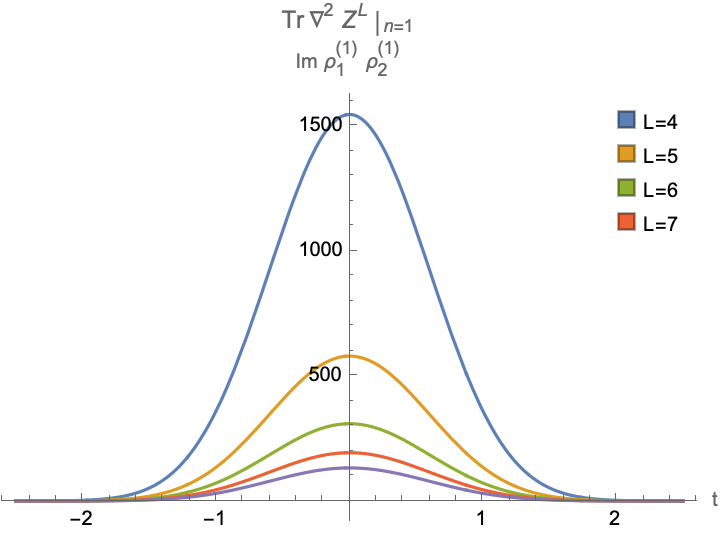}
        
    \end{subfigure}%
    ~ 
    \begin{subfigure}[t]{0.5\textwidth}
        \centering
        \includegraphics[scale=0.4]{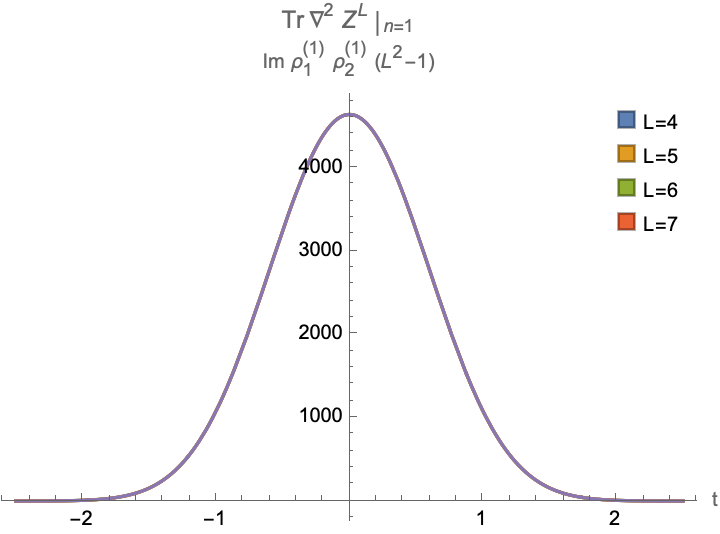}
        
    \end{subfigure}
     \caption{The combination $\rho^{(1)}_1\,\rho^{(1)}_2$ for $n=1$ scales as $\frac{1}{L^2-1}$.}
    \label{fig:DensityScalingN1}
\end{figure*}

While the constraints \eqref{eq:ScalingWithL} are only for the integrated $\rho^{(\frac{n+1}{2})}$ with some powers of $t$, our numerics clearly shows that the scaling is true on the level of densities. For $n=1$ we display this scaling in Figure~\ref{fig:DensityScalingN1}.

\begin{figure*}[th]
    \centering
    \begin{subfigure}[t]{0.5\textwidth}
        \centering
        \includegraphics[scale=0.4]{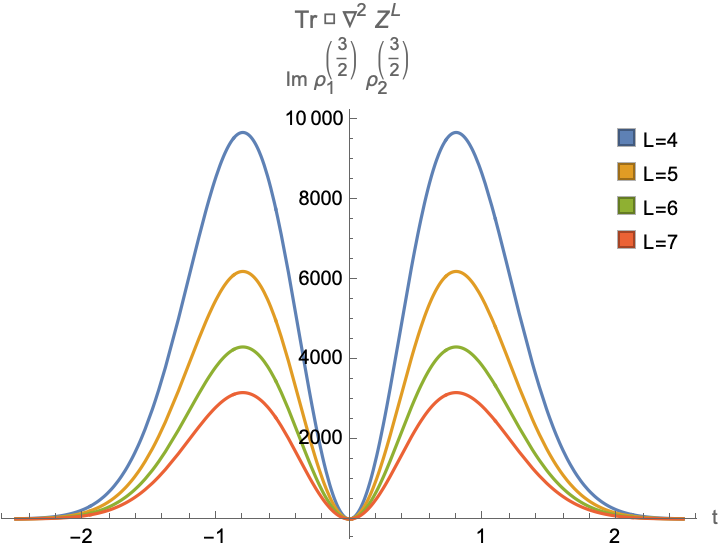}
    \end{subfigure}%
    ~ 
    \begin{subfigure}[t]{0.5\textwidth}
        \centering
        \includegraphics[scale=0.4]{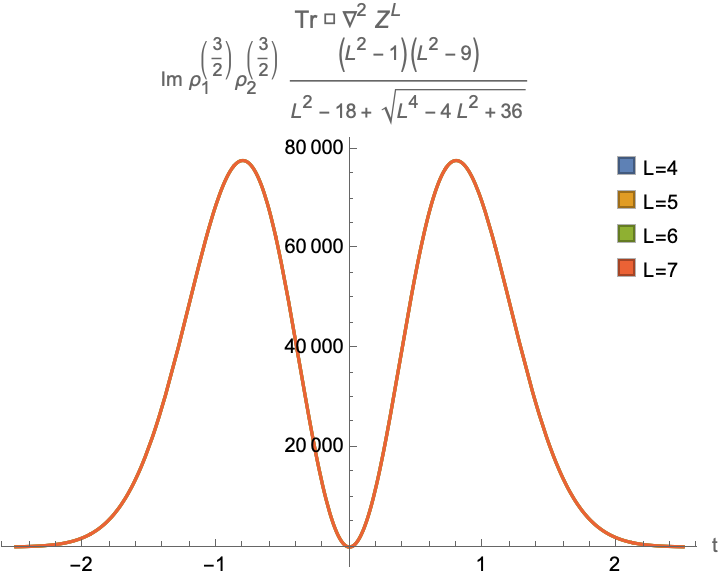}
    \end{subfigure}
    \begin{subfigure}[t]{0.5\textwidth}
        \centering
        \includegraphics[scale=0.4]{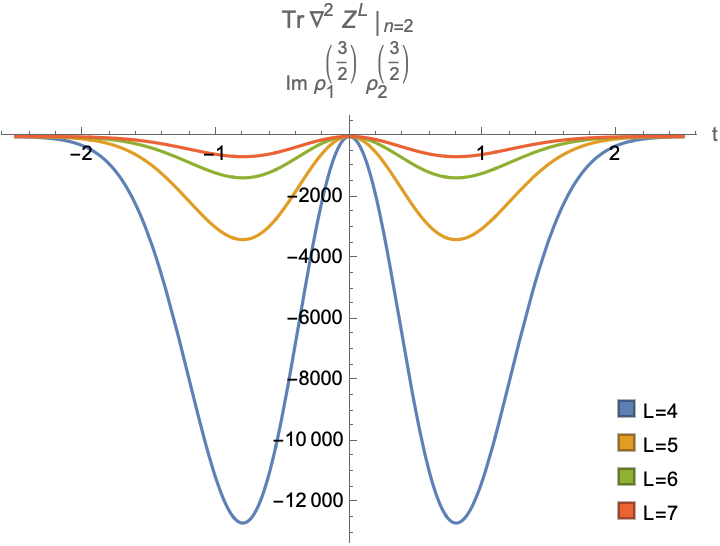}
   \end{subfigure}%
   ~
   \begin{subfigure}[t]{0.5\textwidth}
       \centering
       \includegraphics[scale=0.4]{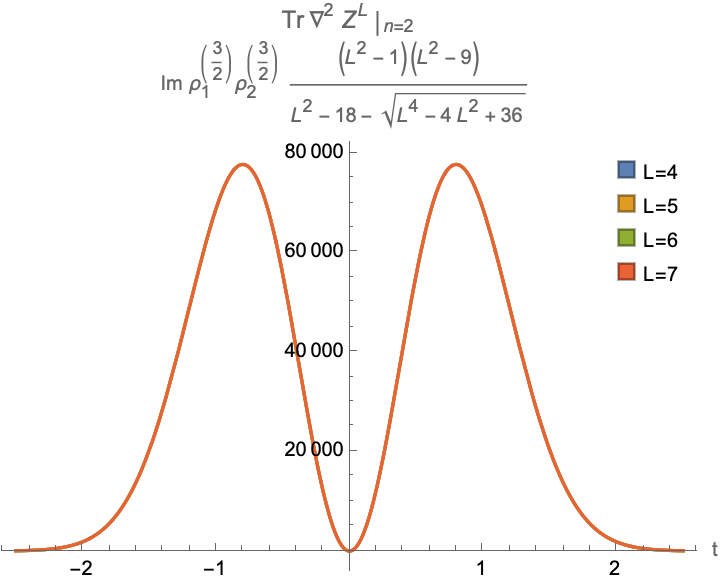}
   \end{subfigure}
    \caption{We have plotted the product of the leading densities $\rho^{(\frac{3}{2})}$ for different choices of L weighted by unity or a factor $\frac{(L^2-1)(L^2-9)}{L^2-18\pm \sqrt{L^4-4L^2+46}}$. When including the weight factor all graphs becomes identical, implying that all $L$ dependence is cancelled by the overall weight factor.}
    \label{fig:DensityScalingN2}
\end{figure*}

For $n=2$ the two sign in \eqref{eq:ScalingWithLN2} should correspond to different solutions of the QSC. As expected from the discussion in Section~\ref{sec:Results} we found from numerics that one solution is relevant for the $\algsl(2)$ $n=2$ state ($-$) and the other for ``the Laplacian" insertion like $\tr \Box D^{2} Z^{L}$ ($+$) as can be seen in Figure~\ref{fig:DensityScalingN2}.

\paragraph{Expansion in $\frac{1}{L^2}$.} 
From our numerics, see Figure~\ref{fig:DensityScalingN1}, we find that $\rho^{(1)}_{1}\rho^{(1)}_{2} \sim \frac{1}{L^2-1}$ for $n=1$. It is natural to also ask how subleading $\rho$ will depend on $L$. In the remainder of this paragraph we will investigate this expansion for $n=1$ using our numerical results. We leave the more complicated, but also more intriguing case of $n>1$ to future work. 

\begin{figure*}[h]
    \centering
    \begin{subfigure}[t]{0.5\textwidth}
        \centering
        \includegraphics[scale=0.4]{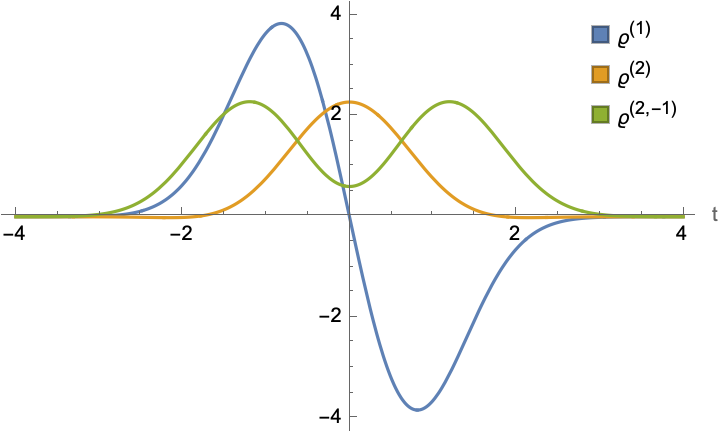}
        
    \end{subfigure}\hfill
    \begin{subfigure}[t]{0.5\textwidth}
        \centering
        \includegraphics[scale=0.4]{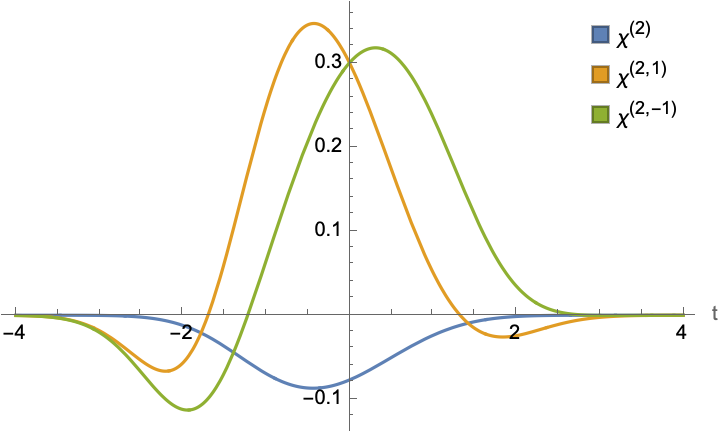}
        
    \end{subfigure}
    \caption{A collection of $\varrho$ and $\chi$ defined in \eqref{eq:ExpansionLRho} and \eqref{eq:ExpansionLEta} as obtained from our numerical algorithm.}
    \label{fig:ExpansionDensities}
\end{figure*}

As a first step we consider rescaled densities $\hrho$ defined as
\begin{equation}
    \hrho_1(x) = \bP_1(x) - \bP_3\(\frac{1}{x}\) = x^{-\frac{L}{2}+1}\,\rho_1(x)\;,
    \quad
    \hrho_2(x) = \bP_2(x) - \bP_4\(\frac{1}{x}\) = x^{-\frac{L}{2}+1}\,\rho_2(x)\;,
\end{equation}
and pick a democratic gauge $\hrho^{(1)}_{1} = \ii \hrho_{2}^{(1)}$. To leading order we find
\begin{equation}
    \hrho_{1}^{(1)} =\frac{\ii}{\sqrt{L^2-1}} \varrho^{(0)}\;, \quad  \hrho_{2}^{(1)} = \frac{1}{\sqrt{L^2-1}}\varrho^{(0)}\;,
\end{equation}
where $\varrho^{(0)}$ is a \emph{universal} real density independent of $L$. Using our numerical data we find that the subleading pieces are also very simple functions of $L$. Explicitly we found fitting our data that
\begin{subequations}\label{eq:ExpansionLRho}
\begin{eqnarray}
    &&\hrho_{1}^{(\frac{1}{2})} = \frac{L\,\varrho^{(1)}}{\sqrt{L^2-1}}\;, \;\;\;\;\;\;\;\;\;\;\;\;\;\;\;\;\;\;\;\; \hrho_{2}^{(\frac{1}{2})} = -\ii\frac{(L-\frac{2}{L})\,\varrho^{(1)}}{\sqrt{L^2-1}}\;, \label{eq:ExpansionLRho2}
    \\
    &&\hrho_{1}^{(0)} = \ii\frac{ \left(L^2\varrho^{(2)}+\varrho^{(2,1)}\right)}{\sqrt{L^2-1}}\;, \;\;\;\;\;\;\;\; \hrho_{2}^{(0)} = \frac{\left(L^2\varrho^{(2)}+\varrho^{(2,-1)}\right)}{\sqrt{L^2-1}}\;, \\
    &&\hrho_{1}^{(-\frac{1}{2})} = \frac{ \left(L^3\varrho^{(3)}+L\varrho^{(3,1)}\right)}{\sqrt{L^2-1}}\;,\quad \hrho_{2}^{(-\frac{1}{2})} = -\ii\frac{\left(L^3\varrho^{(3)}+L \varrho^{(3,-1)}+\frac{1}{L}\varrho^{(3,-2)}\right)}{\sqrt{L^2-1}}\;.
\end{eqnarray}
\end{subequations}
where all $\varrho$ are independent of $L$. We expect that this pattern will continue indefinitely, i.e $\sqrt{L^2-1}\,\hrho^{(m)}_a \sim L^{2(1-m)}\varrho^{(2(1-m))} + L^{2(1-m)-2}\varrho^{(2(1-m),\pm 1)}+\dots $ with the sum terminating at $L^{1}$ or $L^{0}$ for $a=1$ and $L^{0}$ or $L^{-1}$ for $a=2$. The factor $(L-\frac{2}{L})$ in \eqref{eq:ExpansionLRho2} can be fixed analytically by using \eqref{eq:PPQQtoLO} and we find numerically that $\varrho^{(2,1)} = \varrho^{(2,-1)}$. We have not yet fixed any of the $\varrho^{(m)}$ analytically and at the moment we have only access to them through numerics. For illustration purposes we plot a some of the $\varrho$ in Figure~\ref{fig:ExpansionDensities}. The very natural expansion in $L$ seems to hint at trying the limit $L \rightarrow \infty$ where one in principle should be able to make contact with the ABA. 

We also obtained a similar expansion for $\eta$. Defining the rescaled densities
\begin{equation}
    \heta_i(x) = x^{\frac{\Delta}{2}+1}\,\eta_i(x) \,,
\end{equation}
our numerics gave the following expansion in $L$
\begin{subequations}\label{eq:ExpansionLEta}
\begin{eqnarray}
    &&\heta^{(\frac{1}{2})}_{1} = \chi^{(0)}\;, \;\;\;\;\;\;\;\;\;\;\;\;\;\;\;\;\;\;\;\;\;\;\;\;\;\;\; \heta^{(\frac{1}{2})}_{2} =\ii \, \chi^{(0)}\;, 
    \\
    &&\heta_{1}^{(-\frac{1}{2})} = \left(L^2 \,\chi^{(2)}+\chi^{(2,1)}\right)\;, \;\;\; \heta_{2}^{(-\frac{1}{2})} =  \ii\left(L^2 \,\chi^{(2)}+\chi^{(2,-1)}\right)\;.
\end{eqnarray}
\end{subequations}
and we expect that also this structure will keep going indefinitely, that is $\heta^{(\frac{1}{2}-m)}_{a}\sim L^{2m}\chi^{(2m)}+L^{2m-2} \chi^{(2m-2,\pm 1)} + \dots$. We don't have analytic expressions for $\chi$ but we have plotted them numerically in Figure~\ref{fig:ExpansionDensities}. 

We believe that the natural expansion pattern observed in this section hints at a constructive way to approach the QSC at strong coupling. Our approach allows for a systematic expansion at strong coupling, an important first step towards a full analytic solution at strong coupling in the same spirit as the one already available at weak coupling. 

\section{Overview, Discussion, Future Directions, and Conclusion}\label{sec:discussions}
In this paper, we introduced a new efficient parametrization of the Quantum Spectral Curve (QSC) using {\it densities}, which is particularly effective for large coupling $g$. We demonstrated that, when combined with the Baxter equation, this allows to reformulate the QSC into a new closed system of equations. 

Based on this approach, we developed a numerical algorithm to solve the QSC at large $g$, testing it across a range from $g=2$ to $g=100$ for various states. This range was previously unreachable especially with such precision. 

Our precise numerical data enabled us to obtain new strong coupling analytic results for the expansion coefficients in large $g$. This includes a new term for the lowest trajectory in $\mathfrak{sl}(2)$ for each twist $L$. 

Similarly, for higher energy states with mode number $2$ we found a new type of non-analytic square-root type dependence on the twist $L$. We argue that this square-root structure originates from the mixing with operators outside the $\mathfrak{sl}(2)$ sector -- a novel feature specific to strong coupling perturbation theory. A better understanding of this mixing from the string theory side would be interesting, as we argue that the effect should be visible in the quasi-classical regime of long strings. Another possibility is that the non-polynomial expressions in $L$ that we find may have an interpretation on the string side as being a consequence of the equation for AdS energy being of higher order than quadratic (see e.g.~\cite{Roiban:2009aa})\footnote{We are grateful to A.~Tseytlin for discussing this point.}.

To reveal this new type of analytic dependence on the twist, it was crucial to study the densities analytically. We found analytically various relations between the densities at different orders. However, we have not yet found the closed analytic expressions for the densities even at the leading order, something we leave for future work. Despite this, there are clear signs of simplification in the new parametrization, which we hope can lead to a better understanding of the QSC at strong coupling and help to build a systematic way of computing the strong coupling expansion for the string spectrum in the curved background.

Our new analytic results could be useful to produce additional constraints for the strong coupling correlators obtained at the leading orders from conformal bootstrap with additional structural constraints~\cite{Alday:2023mvu,Alday:2023flc}. They could also be used to extract the analytic expressions for the OPE coefficients, disentangling the data packed into the 4-point functions like in~\cite{Gromov:2023hzc,Julius:2023hre}.

Furthermore, it would be interesting to extend our methods to the AdS$_4$/CFT$_3$ QSC \cite{Cavaglia:2014exa,Bombardelli:2017vhk} and update the strong coupling numerical results obtained in \cite{Bombardelli:2018bqz}. Another avenue to explore is the conjectured QSC for ${\rm AdS}_3\times $S$^3 \times $T$^4$ \cite{Ekhammar:2021pys,Cavaglia:2021eqr}. This QSC was recently solved in \cite{Cavaglia:2022xld}, but the tools developed were not sufficient to reach strong coupling. We hope that our new methods will be better suited for this task. Generalization of our construction may help to compare the conjectured QSC to the AdS$_3$/CFT$_2$ mirror TBA \cite{Frolov:2021bwp,Brollo:2023pkl,Brollo:2023rgp,Frolov:2023wji}. 

We anticipate that our methods, with minor modifications, should also be applicable to systems with twist. Interesting cases that deserve further investigation at strong coupling include the $\gamma$ and $\beta$-deformed QSC studied in \cite{Levkovich-Maslyuk:2020rlp,Marboe:2019wyc} and also the Hagedorn temperature for $\mathcal{N}=4$ and ABJM which can be computed using the QSC \cite{Harmark:2017yrv,Harmark:2018red,Harmark:2021qma,Ekhammar:2023cuj,Ekhammar:2023glu} and have recently recieved much interest from different point of view \cite{Urbach:2022xzw,Bigazzi:2023hxt,Harmark:2024ioq}. The twisted cases are particularly challenging for numerical study at strong coupling with the old methods, making our new approach especially attractive in this case.

Additionally, it would be interesting to generalize our methods to boundary problems, such as the cusped Wilson line, where the strongly coupled spectrum has a different behavior in the coupling. Acquiring more analytic data for the spectrum at strong coupling could boost the bootstrap program and allow for more analytic results for structure constants and $4$-point correlators 
\cite{Grabner:2017pgm,Ferrero:2021bsb,Ferrero:2023gnu,Ferrero:2023znz,Cavaglia:2021bnz,Caron-Huot:2022sdy,Cavaglia:2022qpg,Cavaglia:2022yvv,Cavaglia:2023mmu}.

Perhaps the most intriguing, but also challenging, task is to reproduce our results from a first principle analytic quantization of strings in AdS$_5 \times$S$^5$. Developing a systematic expansion of the QSC would help to provide further clues on the physics in the regime of short strings at strong coupling -- the regime which remains the most challenging in the planar limit. The densities we obtained may have a simple interpretation, for example, as wave functions of zero modes (transverse coordinates) on the short string in a slightly curved space~\cite{Passerini:2010xc}.

\paragraph{Acknowledgements} We are grateful to Julius, N. Primi and N. Sokolova collaboration on the initial stage of this project as well as to B. Basso, A. Georgoudis, Á. Hegedus, V. Kazakov, I. Kostov, J. Minahan, D. Serban, A. Tseytlin and P. Vieira for numerous discussions on related topics. 
Part of the calculations in this work were done on the ``King's Computational Research, Engineering and Technology Environment" (CREATE) cluster.
N.G. and P.R. are grateful to IPhT Saclay for  warm hospitality at an early stage of this project. 
N.G. is grateful to LPENS Paris for warm hospitality while a part of this work was done. 
P.R. is grateful to Perimeter Institute for warm hospitality during a part of this work.  
The work of S.E, N.G. and P.R was
supported by the European Research Council (ERC) under the European Union’s Horizon 2020 research and innovation program – 60 – (grant agreement No. 865075) EXACTC.

\appendix

\section{Further Details on QSC}\label{app:QSC}

Here we discuss some additional properties of the QSC, see for example \cite{Gromov:2017blm} for an in-depth introduction.

\subsection{$\mathbb{A}\mathbb{A}$ and $\mathbb{B}\mathbb{B}$ Relations}
The prefactors $\mathbb{A}$ and $\mathbb{B}$ entering the large-$u$ asymptotics of the $\bP$ and $\bQ$-functions \eqref{eqn:PQlargeu} are constrained to satisfy
\begin{equation}\label{eqn:AArelns}
\begin{split}
   & \mathbb{A}_1 \mathbb{A}_4 =-i \frac{(2+L-S-\bar{\Delta})(L+S-\bar{\Delta})(2+L-S+\bar{\Delta})(L+S+\bar{\Delta})}{16L(L+1)}, \\
   & \mathbb{A}_2 \mathbb{A}_3 =i \frac{(S+\bar{\Delta}-L)(L+S-2-\bar{\Delta})(L-S+\bar{\Delta})(L+S-2+\bar{\Delta})}{16L(L-1)}, \\
\end{split}
\end{equation}
\begin{equation}
\begin{split}\label{eqn:BBrelns}
   & \mathbb{B}_1 \mathbb{B}_4 =-i \frac{(L+S-\bar{\Delta})(L-S+\bar{\Delta})(L+S-\bar{\Delta}-2)(L-S+\bar{\Delta}+2)}{16\bar{\Delta}(S-1)(S-\bar{\Delta}-1)}, \\
   & \mathbb{B}_2 \mathbb{B}_3 =i \frac{(L+S+\bar{\Delta})(S-L+\bar{\Delta})(L+S+\bar{\Delta}-2)(S-L+\bar{\Delta}-2)}{16\bar{\Delta}(S-1)(S-\bar{\Delta}+1)}\,. \\
\end{split}
\end{equation}
We remind the reader that $\bar{\Delta}=\Delta+2$. 

\subsection{Baxter Equation and $\mathbb{P}\mathbb{Q}$-relations}

\paragraph{Baxter equation.}

The fourth-order Baxter equation relating the $\bP$ and $\bQ$-functions is given by \footnote{We use the notation $f^{[n]}:= f(u+i\tfrac{n}{2})$ for shifts of the spectral parameter.}:
\begin{equation}\label{eq:BaxterEqBQ}
\begin{split}
    & \bQ^{[+4]}D_0-\bQ^{[+2]}\left[D_1-\bP_a^{[+2]}\bP^{a[+4]} D_0\right] \\
    +\frac{1}{2} &\bQ\left[D_2 +\bar{D}_2- \bP_a\bP^{a[+2]}D_1+\bP_a\bP^{a[+4]}D_0+\bP_a\bP^{a[-2]}\bar{D}_1-\bP_a\bP^{a[-4]}\bar{D}_0\right] \\
     + & \bQ^{[-4]}\bar{D}_0-\bQ^{[-2]}\left[\bar{D}_1+\bP_a^{[-2]}\bP^{a[-4]} \bar{D}_0\right]=0
\end{split}
\end{equation}
where $D_0$ $\bar{D}_0$, $D_1$, $\bar{D}_2$ and $D_2$, $\bar{D}_2$ are given explicitly by 
\begin{equation}
\begin{split}
   & D_0 = \det\left(\begin{array}{cccc}
        \bP^{1[+2]} & \bP^{2[+2]} & \bP^{3[+2]} & \bP^{4[+2]} \\
        \bP^1\ \ \ \   & \bP^2\ \ \ \  & \bP^3\ \ \ \  & \bP^4\ \ \ \ \\
        \bP^{1[-2]} & \bP^{2[-2]} & \bP^{3[-2]} & \bP^{4[-2]} \\
        \bP^{1[-4]} & \bP^{2[-4]} & \bP^{3[-4]} & \bP^{4[-4]} 
    \end{array} \right),\quad \bar{D}_0 = \det\left(\begin{array}{cccc}
        \bP^{1[-2]} & \bP^{2[-2]} & \bP^{3[-2]} & \bP^{4[-2]} \\
        \bP^1\ \ \ \   & \bP^2\ \ \ \  & \bP^3\ \ \ \  & \bP^4\ \ \ \ \\
        \bP^{1[+2]} & \bP^{2[+2]} & \bP^{3[+2]} & \bP^{4[+2]} \\
        \bP^{1[+4]} & \bP^{2[+4]} & \bP^{3[+4]} & \bP^{4[+4]}
    \end{array} \right) \\ 
  &  D_0 = \det\left(\begin{array}{cccc}
        \bP^{1[+4]} & \bP^{2[+4]} & \bP^{3[+4]} & \bP^{4[+4]} \\
        \bP^1\ \ \ \   & \bP^2\ \ \ \  & \bP^3\ \ \ \  & \bP^4\ \ \ \ \\
        \bP^{1[-2]} & \bP^{2[-2]} & \bP^{3[-2]} & \bP^{4[-2]} \\
        \bP^{1[-4]} & \bP^{2[-4]} & \bP^{3[-4]} & \bP^{4[-4]} 
    \end{array} \right),\quad \bar{D}_0 = \det\left(\begin{array}{cccc}
        \bP^{1[-4]} & \bP^{2[-4]} & \bP^{3[-4]} & \bP^{4[-4]} \\
        \bP^1\ \ \ \   & \bP^2\ \ \ \  & \bP^3\ \ \ \  & \bP^4\ \ \ \ \\
        \bP^{1[+2]} & \bP^{2[+2]} & \bP^{3[+2]} & \bP^{4[+2]} \\
        \bP^{1[+4]} & \bP^{2[+4]} & \bP^{3[+4]} & \bP^{4[+4]}
    \end{array} \right) \\ 
  &  D_2 = \det\left(\begin{array}{cccc}
        \bP^{1[+4]} & \bP^{2[+4]} & \bP^{3[+4]} & \bP^{4[+4]} \\
        \bP^{1[+2]} & \bP^{2[+2]} & \bP^{3[+2]} & \bP^{4[+2]} \\
        \bP^{1[-2]} & \bP^{2[-2]} & \bP^{3[-2]} & \bP^{4[-2]} \\
        \bP^{1[-4]} & \bP^{2[-4]} & \bP^{3[-4]} & \bP^{4[-4]} 
    \end{array} \right),\quad \bar{D}_2 = \det\left(\begin{array}{cccc}
        \bP^{1[-4]} & \bP^{2[-4]} & \bP^{3[-4]} & \bP^{4[-4]} \\
        \bP^{1[-2]} & \bP^{2[-2]} & \bP^{3[-2]} & \bP^{4[-2]} \\
        \bP^{1[+2]} & \bP^{2[+2]} & \bP^{3[+2]} & \bP^{4[+2]} \\
        \bP^{1[+4]} & \bP^{2[+4]} & \bP^{3[+4]} & \bP^{4[+4]}
    \end{array} \right)
\end{split}
\end{equation}

In the left-right symmetric sector which we restrict to in this paper $\bP^a$ are related to $\bP_a$ by
\begin{equation}\label{eqn:defnchi}
    \bP^a = \chi^{ab}\bP_b\;,\quad \chi^{ab} = \left(\begin{array}{cccc}
        0 & 0 & 0 & -1 \\
       0 & 0 & 1 & 0 \\
       0 & -1 & 0 & 0 \\
       1 & 0 & 0 & 0
    \end{array} \right)\;,\quad \chi^{ab}\chi_{bc}=\delta^{a}_{\ c}\;.
\end{equation}

\paragraph{$\mathbb{P}\mathbb{Q}$-relations.}

The Baxter equation relating $\bP$ and $\bQ$ can be conveniently encoded in a set of $\bbPP\bbQQ$ relations. We use the following notation
\begin{align}
    &\mathbb{P}_{m}^{n} = \bP^{[m]}_{a}\chi^{ab}\bP^{[n]}_{b} \,,
    &
    &\mathbb{Q}_{m}^{n} = \bQ_{i}^{[m]}\chi^{ij} \bQ_{j}^{[n]}\,.
\end{align}
The $\bbPP\bbQQ$ relations are then given in our signs convention by~\cite{Grabner:2020nis}
\begin{align}\label{eqn:PPQQrelns}
    \bbPP_0^{2} &= \bbQQ_{0}^{2}\;,\\
    \bbPP_0^{4} &= \bbQQ_0^{4}+\bbQQ_0^2\bbQQ_2^4\;,\\
    \bbPP_0^{6} &= \bbQQ_{0}^{6} + \bbQQ_0^2 \bbQQ_2^6 + \bbQQ_4^6\bbPP_0^4\\
    &= \bbQQ_0^6 + \bbQQ_0^{2} \bbQQ_2^{6} + \bbQQ_0^4\bbQQ_4^6 + \bbQQ_0^2\bbQQ_2^4\bbQQ_4^6 \;,\\
    \bbPP_{0}^{8} &= \bbQQ_0^{8} +\bbQQ_0^2 \bbQQ_2^{8} + \bbPP_0^4\bbQQ_4^8 + \bbPP_0^6\bbQQ_6^8 \\
    &=\bbQQ_0^8 + \bbQQ_0^2\bbQQ_2^8 + \bbQQ_0^4\bbQQ_4^{8} + \bbQQ_0^2\bbQQ_2^4\bbQQ_4^8+\bbQQ_0^6 \bbQQ_6^8 + \bbQQ_0^2\bbQQ_2^6\bbQQ_6^8+\bbQQ_0^4\bbQQ_4^6\bbQQ_6^8+\bbQQ_0^2\bbQQ_2^4\bbQQ_4^6\bbQQ_6^8\,.
\end{align}
In Appendix \ref{BaxPPQQ} we show that these relations are actually equivalent to the Baxter equation~\eq{eq:BaxterEqBQ}.

\subsection{Reality Properties}
The $\bP$-functions have definite reality properties \cite{Gromov:2014caa}
\begin{equation}\label{eqn:Preality}
    \overline{\bP}_a(u) = \pm \bP_a(u),\quad u\in \mathbb{R}\,.
\end{equation}
In this paper we take $\bP_1$, $\bP_3$ to be purely imaginary and $\bP_2$, $\bP_4$ to be real. 

These definite reality properties also translate to the $\bQ$-functions, which also inherit simple transformation properties under complex conjugation: if $\bQ$ is any solution to the Baxter equation then so is $\overline\bQ$. Complex conjugation then relates UHPA and LHPA $\bQ$-functions, and so $\overline{\bQ^\downarrow}$ must be related to $\bQ^\uparrow$. By comparing asymptotics, we see that we must have
\begin{equation}\label{eqn:upconjreln}
    \overline{\bQ^\downarrow}_k = e^{i \phi_k}\bQ^\uparrow_k\;,\quad \phi_k \equiv \frac{\overline{\mathbb{B}_k}}{\mathbb{B}_k}\,.
\end{equation}
Note that as a result of \eqref{eqn:BBrelns} for real quantum numbers the phases $\phi_k$ must satisfy 
\begin{equation}
    \phi_4 = \pi -\phi_1,\quad \phi_3 = \pi-\phi_2\,.
\end{equation}

\paragraph{Reality properties of the densities.}

We now discuss the reality properties of $\rho$. For $x$ on the unit circle we have $\bar{x}=1/x$. Since $\bP_1,\bP_3$ are purely imaginary and $\bP_2,\bP_4$ are real \eqref{eqn:Preality}, it follows that
\begin{equation}
    \overline{\rho_a(x)}=(-1)^{a}\rho_a(1/x),
\end{equation}
a feature which is clearly visible in Figure \ref{fig:rhoplotsmain}. 

Recall that in our gauge the coefficients $\mathbb{B}_1$ and $\mathbb{B}_3$ are real while $\mathbb{B}_2$ and $\mathbb{B}_4$ are imaginary. Hence, as a result of complex conjugation symmetry $q_1$ and $q_3$ are mapped to themselves while $q_2$ and $q_4$ pick up a sign. More preciely, for real $x$ we must have
\begin{equation}
    \overline{q^\downarrow_k}(x) = (-1)^{k+1}q_k^\uparrow(x)
\end{equation}
which immediately implies 
\begin{equation}
    \overline{\eta_{13}}(x)=\eta_{13}(x),\quad \overline{\eta_{24}}(x)=-\eta_{24}(x)\,,
\end{equation}
meaning that $\eta_{13}$ and $\eta_{24}$ are real and imaginary, respectively. 

\subsection{Parity}

Certain gauge theory operators have a parity symmetry, reversing the order of operators under the trace. This is inherited by the Q-functions. Assuming the state is symmetric under this symmetry the $\bP$-functions should be either even or odd for even $L$'s. When $L$ is odd however things are slightly more subtle since $\bP$ have half-integer asymptotics. It is convenient thus to consider the quantities 
\begin{equation}\label{eqn:rescaledP}
    p_1 = x^{\frac{L}{2}-1}\;\bP_1\;,\quad p_3 = x^{-\frac{L}{2}+1}\;\bP_3\;,\quad p_2 = x^{\frac{L}{2}}\;\bP_2\;,\quad p_4 = x^{-\frac{L}{2}}\bP_4\;.
\end{equation}
Then parity symmetry implies that all $p_1$ and $p_3$ are even while $p_2$ and $p_4$ are odd, that is 
\begin{equation}\label{eqn:parityP}
    p_a(-u)=(-1)^{a+1}p_a(u)\,.
\end{equation}
The parameterisation \eq{eqn:rescaledP} is useful in what follows. 

The $\mu$-functions also have certain parity properties \cite{Alfimov:2014bwa}, but this parity is present in $\mu^+(u)\equiv\mu(u+\tfrac{i}{2})$, and only in the region $0<{\rm Im}\;u<1$. We have
\begin{equation}\label{eqn:muparity}
    \mu_1^{\ 1,+},\ \mu_1^{\ 3,+},\ \mu_3^{\ 1,+} - \text{even},\quad \mu_3^{\ 4,+},\ \mu_{3}^{\ 3,+} - \text{odd}\,.
\end{equation}

The parity properties of the $\bP$-functions imply that if $\bQ$ is any solution to the Baxter equation then so is $\bQ(-u)$. Clearly, $u\rightarrow -u$ must relate the UHPA and LHPA $\bQ$ functions:
\begin{equation}
    \bQ^\downarrow_k(-u) = e^{i\pi\mathtt{powQ}_k} \bQ^\uparrow_k(u)\,.
\end{equation}

In Appendix~\ref{sec:ParitySection} we discuss the implications of parity symmetry on the formulas \eqref{eqn:Pfromrho} and \eqref{eqn:Q1Q3},\eqref{eqn:Q2Q4} for reconstructing $\bP$ and $\bQ$ from $\rho$ and $\eta$.

\section{Formulas for $\bP$ and $\bQ$ in the Parity-Symmetric Sector}\label{sec:ParitySection}

We now show how to simplify the expressions \eqref{eqn:Pfromrho}, \eqref{eqn:Q1Q3} and \eqref{eqn:Q2Q4} by imposing the $u\to -u$ parity symmetry for our parametrisation in terms of densities. 

\paragraph{Effect of parity symmetry on $\rho$.}

The parity symmetry \eqref{eqn:parityP} implies
\begin{equation}
    \rho_a(x) = (-1)^{a+1}\rho_a(-x)\,,
\end{equation}
and restricting the integration in \eqref{eqn:Pfromrho} to only one half of the circle gives
\begin{equation}\label{eqn:Pfinal}
\begin{split}
   & \bP_1(x) = x^{-\frac{L}{2}+1}\displaystyle \simoint \frac{{\rm d}y}{2\pi i}\, \rho_1(y)\left(\frac{1}{x-y}-\frac{1}{x+y}\right)\;,\\
   & \bP_3(x) = x^{\frac{L}{2}-1}\displaystyle \simoint \frac{{\rm d}y}{2\pi i}\, \rho_1(y)\left(\frac{1}{\frac{1}{x}-y}-\frac{1}{\frac{1}{x}+y}\right)\;,\\
   & \bP_2(x) = x^{-\frac{L}{2}}\,A+x^{-\frac{L}{2}+1}\displaystyle \simoint \frac{ {\rm d}y}{2\pi i}\, \rho_2(y)\left(\frac{1}{x-y}+\frac{1}{x+y}\right)\;,\\
   & \bP_4(x) = x^{\frac{L}{2}}\,A+x^{\frac{L}{2}-1}\displaystyle \simoint \frac{{\rm d}y}{2\pi i} \, \rho_2(y)\left(\frac{1}{\frac{1}{x}-y}+\frac{1}{\frac{1}{x}+y}\right)\,.
\end{split}
\end{equation}
Here the symbol $\simoint$ denotes the integration path to be restricted to the right half of the unit circle and going counter-clockwise.

\paragraph{Effect of parity symmetry on $\eta$.}

Parity-symmetry implies that $\bQ_k^\downarrow(-u) \propto \bQ^\uparrow(u)$ with the proportionality factor controlled by the non-integer asymptotics in $\bQ$. We recall that we define $q_k$ by \eqref{eqn:smallq} which have asymptotics 
\begin{align}
q_k\sim u^{\left(-1,0,1-S,S-2\right)_k}\,,
\end{align}
Stripping this factor out to obtain $q_k$ as we did we then obtain the relation 
\begin{equation}
    q_k^\downarrow(-x) = (-1)^k q_k^\uparrow(x)\,.
\end{equation}
As a result, we have the following parity property for the densities $\eta_{13}$ and $\eta_{24}$:
\begin{equation}
    \eta_{13}(-x) = \eta_{13}(x),\quad \eta_{24}(-x)=-\eta_{24}(x),\,
\end{equation}
allowing us to write the integral representations as integrals over $[0,\infty)$ instead of $(-\infty, \infty)$:
\begin{equation}\label{eqn:Qfinal}
\begin{split}
   & \bQ^\downarrow_1(x) = x^{\frac{\Delta}{2}-\frac{S}{2}+2}\displaystyle \int^{\infty}_{0} \frac{{\rm d}y}{2\pi i} \,\eta_{13}(y)\left(\frac{1}{y-x}-\frac{1}{y+x}\right), \\
   & \bQ^\downarrow_3(x) = x^{-\frac{\Delta}{2}+\frac{S}{2}-2}\displaystyle \int^{\infty}_{0} \frac{{\rm d}y}{2\pi i} \,\eta_{13}(y)\left(\frac{1}{y-\frac{1}{x}}-\frac{1}{y+\frac{1}{x}}\right), \\
   & \bQ^\downarrow_2(x) = x^{\frac{\Delta}{2}+\frac{S}{2}}\displaystyle \sum_{n=0}^{ S/2-1} \frac{r_{2n}}{x^{2n}}+x^{\frac{\Delta}{2}+\frac{S}{2}}\displaystyle \int^{\infty}_{0} \frac{{\rm d}y}{2\pi i}\, \eta_{24}(y)\left(\frac{1}{y-x}+\frac{1}{y+x}\right), \\
   & \bQ^\downarrow_4(x) = x^{-\frac{\Delta}{2}-\frac{S}{2}}\displaystyle \sum_{n=0}^{ S/2-1}r_{2n} x^{2n}+x^{-\frac{\Delta}{2}-\frac{S}{2}}\displaystyle \int^{\infty}_{0} \frac{{\rm d}y}{2\pi i}\, \eta_{24}(y)\left(\frac{1}{y-\frac{1}{x}}+\frac{1}{y+\frac{1}{x}}\right),
\end{split}
\end{equation}
where as before $|x|>1$ and ${\Im}\;x>0$; formulas for $\bQ^\uparrow(x)$ are obtained by choosing ${\Im}\;x<0$. 

Note that the parameters $A$ and $r_{2n}$ are not independent constants needing to be fixed in the numerical algorithm. Instead they can be fixed analytically in terms of the moments of the densities using the Baxter equation.

\section{Equivalence of Baxter Equation and $\bbPP\bbQQ$-relations}\label{BaxPPQQ}

The Baxter equation implies the $\bbPP\bbQQ$-relations \cite{Grabner:2020nis}. Indeed, the Baxter equation serves as a definition of the $\bQ$-functions, from which all other relations follows. In this appendix we show that these two sets of equations are equivalent, by showing that the $\bbPP\bbQQ$-relations imply the Baxter equation.

The starting point is the trivial $5\times 5$ determinant
\begin{equation}
   \det\left(\begin{array}{ccccc}
      \bQ_i^{[-4]}  & \bQ_i^{[-2]} & \bQ_i & \bQ_i^{[+2]} & \bQ_i^{[+4]} \\
    \bQ_1^{[-4]}  & \bQ_1^{[-2]} & \bQ_1 & \bQ_1^{[+2]} & \bQ_1^{[+4]} \\
    \bQ_2^{[-4]}  & \bQ_2^{[-2]} & \bQ_2 & \bQ_2^{[+2]} & \bQ_2^{[+4]} \\
    \bQ_3^{[-4]}  & \bQ_3^{[-2]} & \bQ_3 & \bQ_3^{[+2]} & \bQ_3^{[+4]} \\
    \bQ_4^{[-4]}  & \bQ_4^{[-2]} & \bQ_4 & \bQ_4^{[+2]} & \bQ_4^{[+4]} \\
   \end{array} \right) = 0
\end{equation}
which vanishes for $i=1,2,3,4$. We now expand the determinant obtaining an expression of the general form
\begin{equation}
    F_{2-}\bQ_i^{[-4]} - F_{1-}\bQ_i^{[-2]} +F_0 \bQ_i - F_{1+}\bQ_i^{[2]}+F_{2+}\bQ_i^{[+4]} =0
\end{equation}
where each of the coefficients $F$ are $4\times 4$ determinants of $\bQ_1,\dots,\bQ_4$, for example we have
\begin{equation}
    F_{2-}=\det\left(\begin{array}{cccc}
     \bQ_1^{[-2]} & \bQ_2 & \bQ_1^{[+2]} & \bQ_1^{[+4]} \\
     \bQ_2^{[-2]} & \bQ_2 & \bQ_2^{[+2]} & \bQ_2^{[+4]} \\
     \bQ_3^{[-2]} & \bQ_3 & \bQ_3^{[+2]} & \bQ_3^{[+4]} \\
     \bQ_4^{[-2]} & \bQ_4 & \bQ_4^{[+2]} & \bQ_4^{[+4]} \\
    \end{array} \right)\,.
\end{equation}

We now use the $\bbPP\bbQQ$-relations to rewrite the determinants $F$ in terms of the $\bP$-functions. The first step is to rewrite the determinants $F$ into a form where the $\bbPP\bbQQ$-relations can be easily applied. Let us recall the definition
\begin{equation}
    \bbQQ_m^n\equiv\bQ_i^{[+m]} \chi^{ij}\bQ_j^{[+n]}\,.
\end{equation}
We can now easily show the following equalities
\begin{equation}
    \begin{split}
         F_{2-} & =-\bbQQ_{-2}^{4}\bbQQ_{0}^{2}+\bbQQ_{-2}^{2}\bbQQ_{0}^{4}-\bbQQ_{-2}^{0}\bbQQ_{2}^{4} \\
         F_{1-} & =-\bbQQ_{-4}^{4}\bbQQ_{0}^{2}+\bbQQ_{-4}^{2}\bbQQ_{0}^{4}-\bbQQ_{-4}^{0}\bbQQ_{2}^{4} \\
         F_{0} & =-\bbQQ_{-4}^{4}\bbQQ_{-2}^{2}+\bbQQ_{-4}^{2}\bbQQ_{-2}^{4}-\bbQQ_{-4}^{-2}\bbQQ_{2}^{4} \\
         F_{1+} & =-\bbQQ_{-4}^{4}\bbQQ_{-2}^{0}+\bbQQ_{-4}^{0}\bbQQ_{-2}^{4}-\bbQQ_{-4}^{-2}\bbQQ_{0}^{4} \\
         F_{2+} & =-\bbQQ_{-4}^{2}\bbQQ_{-2}^{0}+\bbQQ_{-4}^{0}\bbQQ_{-2}^{2}-\bbQQ_{-4}^{-2}\bbQQ_{0}^{2}
    \end{split}\,.
\end{equation}
From here we can use the $\bbPP\bbQQ$ relations to write the $F$ functions in terms of the $\bP$-functions. The result is then a finite-difference equation on $\bQ_i$ with coefficients built from $\bP$'s. It is then straightforward to check that the coefficients have exactly the form as in 
\eqref{eq:BaxterEqBQ}.

\section{Optimal Polynomials}\label{app:optpoly}

In our numerical algorithm we need to efficiently approximate functions of the form $H(u)=\mu(u)h(u)$ on some domain, where $\mu(x)$ is some measure factor and $h(u)$ is a smooth function. We do this using the theory of optimal polynomials. Namely, we try to approximate $H(u)$ by a function $P(u)=\mu(u)p(u)$ where $p(u)$ is a polynomial, and seek to minimise the maximal value of the difference $|H(u)-P(u)|$ as $u$ ranges over the given domain, to the extent which is practical for the numerical implementation.

For the case where $\mu(u)=1$ on an interval $[-1,1]$, it is well known that the nearly ``optimal" polynomial $p(u)$ of degree $N$
can be built as an interpolation polynomial at Chebychev points i.e. $p(u_i)=H(u_n)$ for $u_n$ being a set of $N+1$ roots of the degree $N+1$ Chebychev polynomial
of the first kind i.e. $u_n = \cos\left(\frac{2n-1}{2N}\pi\right)$ for $n=1,\dots,N$.
In this case the difference $H(u)-P(u)$ oscillates back and forth between $\pm \epsilon$, $\epsilon>0$, a total of $N+2$ times, with an error at most $\epsilon$, and this error decreases as $N$ increases.

The question we are trying to answer in this appendix is how to build the analog of the Chebychev points for a generic non-trivial measure $\mu$. For this we need to determine the optimal nodal or probe points $u_n$ at which to sample the function $H(u)$ and build an interpolation polynomial $p(u)$ such that $H(u)$ and $P(u)$ coincide at these points.

\paragraph{Behaviour of the difference.}
Let us consider the difference $H(u)-P(u)$ and write it as 
\begin{equation}\label{difference}
    H(u)-P(u) = Q(u)r(u)\;\;,\;\;Q(u)\equiv \mu(u)q(u), \quad q(u) = \prod_{n=1}^N (u-u_n)\,,
\end{equation}
where the roots $u_n$ of $q(u)$ are the sought-after probe points and $r(u)$ is some ``remainder" function. 

We are interested in the setting where the number of probe points is large, the r.h.s. of \eq{difference} becomes highly oscillating. These oscillations are captured in the polynomial $q(u)$, whereas $r(u)$ is a smooth, slowly changing part of the difference, which depends on the particular function $h(u)$. Our goal is then to choose the 
the optimal points which would work for a generic $h(u)$, so following the analogy with Chebishev points we require that $Q(u)$ has equal magnitude $\epsilon$ at its maxima and minima $\tilde{u}_n$, i.e. we impose
\begin{eqnarray}\label{dQQ}
    Q'(\tilde{u}_n)=0,\quad Q(\tilde{u}_n)=(-1)^n \epsilon\,,
\end{eqnarray}
where we assume the ordering $u_n < \tilde{u}_n < u_{n+1}$. In the case of $\mu=1$ we have $Q(u)=T_{N+1}(u)$ (first kind Chebychev polynomial) for which it holds that $Q'(\tilde{u}_n)=\pm 1$ in the standard normalization. Hence  the requirement \eq{dQQ} is the generalization of this defining feature of the Chebychev polynomials.

\paragraph{Scaling.}
Before continuing we need to make some comments on the behaviour of the measure. Depending on the asymptotic behaviour of the measure and its support the scaling of the roots $u_n$ could be quite different. Here, for definiteness, we assume that the measure 
has Gaussian asympotitc $e^{-\alpha u^2}$,
and also assume that the probe points scale as $\sqrt{N}$ and that the typical distance between the roots is $\sim 1/\sqrt{N}$ (near the ends of the support of distribution of the points distribution the distance is expected to become bigger). For convenience we also introduce the introduce rescaled variable $U=u/\sqrt{N}$.

First we evaluate the ratio $Q'(u)/Q(u)$ at $\tilde{u}_n$, giving 
\begin{equation}\label{eqn:electrosum}
    F(\tilde{u}_n) + \displaystyle\sum_{j=1}^N \frac{1}{\tilde{u}_n-u_j}=0,\quad F(u)\equiv\frac{\mu'(u)}{\mu(u)}\,.
\end{equation}

Next we evaluate the sum in this expression. Normally one could try to evaluate it using the standard Euler-Maclaurin formula, allowing us to express the sum as an integral. However, as a result of the condition $u_n < \tilde{u}_n < u_{n+1}$, when $N$ is large $\tilde{u}_n$ gets pinched between $u_n$ and $u_{n+1}$ and the function develops a pole. This requires a modification of the standard Euler-Maclaurin formula, in which the sum gets expressed as a principal value (PV) integral. We now quickly review this, following \cite{Gromov:2005gp,Lyness1985TheEM}.

\paragraph{Euler-Maclaurin for PV integrals.} 

For a smooth function $f(x) \in [0,1]$ the Euler-Maclaurin (EM) formula states
\begin{equation}
  \displaystyle\sum_{j=1}^N f\left(\frac{j}{N} \right)= N\displaystyle\int_0^1 f(x){\rm d}x + \text{extra}\,,
\end{equation}
where the ``extra" term includes various boundary, derivative and remainder terms which are not important for us\footnote{One can show that they are subleading in $N$.}.

Suppose now that $f(x)$ has a simple pole at $c\in(0,1)$. Consider the function $\psi(x) = \pi r\cot(\pi(x-c))$ with $r={\rm Res}_{x=c}f(x)$. The combination $f(x)-\psi(x)$ is regular on the whole interval and thus for the difference we have
\begin{equation}
  \displaystyle\sum_{j=1}^N \(f\left(\frac{j}{N} \right)-\psi\left(\frac{j}{N} \right)\)\simeq \displaystyle\int_0^1 \[f(x)-\psi(x)\]{\rm d}x = 
  \displaystyle\dashint_0^1 f(x){\rm d}x
\end{equation}
where in the last equality we used the property of $\psi$ that
$\dashint_0^1 \psi(x)=0$.

Furthermore, we have the following identity
\begin{equation}
    \frac{1}{N}\sum_{j=1}^N \psi\left(\frac{j}{N}\right)= -\pi r \cot(\pi N c)\,,
\end{equation}
we arrive at
\begin{equation}\label{fjN}
  \displaystyle\sum_{j=1}^N f\left(\frac{j}{N} \right)= N \dashint_0^1 f(x) {\rm d}x +N\pi r \cot(\pi N c) +\text{extra}\,.
\end{equation}

\paragraph{Evaluating the sum.}

We want to compute the sum 
\begin{equation}\label{eqn:EMsum}
    \displaystyle \sum_{j=1}^N \frac{1}{\tilde{u}_n-u_j},\quad u_n<\tilde{u}_n<u_{n+1}\,.
\end{equation}
In order to relate this sum to the EM formula and take the large $N$ limit it is useful to introduce a function $u(x)$ such that $u(n)=u_n$. Then the sum \eqref{eqn:EMsum} can be written as 
\begin{eqnarray}
   \displaystyle \sum_{j=1}^N \frac{1}{ \tilde{u}_n-u_j}=\sum_{j=1}^N f\left(\frac{j}{N} \right),\quad f(x)=\frac{1}{\tilde{u}_n-u(x N)}\,.
\end{eqnarray}

The function $f$ has a pole at the point $c$, where $c$ is such that $u(cN)=\tilde{u}_n$. To compute $c$, it is convenient to introduce the density $\rho$ defined by
\begin{equation}
    \frac{{\rm d}u}{{\rm d}x}=\frac{1}{\sqrt{N}\rho(U)}\simeq u_{n+1}-u_n\,.
\end{equation}
Note that due to our assumed scaling behaviour we can estimate that $\rho(U)\sim 1$.

Then we can expand $u(x)$ around $n$ 
\begin{eqnarray}
    u(n+\delta)=u_n + \delta\frac{1}{\sqrt{N}\rho}-\delta^2 \frac{\rho'}{2 N^{\frac{3}{2}} \rho^3}
    +\lO\left(\frac{1}{N^{5/2}}\right)
\end{eqnarray}
allowing us to compute 
\begin{eqnarray}
    c=\frac{n}{N}+\frac{\rho\left({U_n}\right)}{\sqrt N}(\tilde{u}_n-u_n)+\lO\left(\frac{1}{N^2} \right),\quad U_n = \frac{u_n}{\sqrt{N}}\,.
\end{eqnarray}
In the bulk of the distribution $n\sim N$, so the first term is order $1$, whereas the second term is order $1/N$ which becomes finite when we multiply it by $\pi N$ as in \eq{fjN}. On the other hand the first term becomes irrelevant due to the periodicity of $\cot$.
Additionally, the residue of $r$ of $f(x)$ at $c$ is easily computed to be 
$r=-N^{-1/2}\rho\left({\tilde{u}_n/\sqrt{N}} \right)$.
As a result we get the following relation, valid to leading order in large $N$:
\begin{eqnarray}
    F(\tilde u_n)-\dashint{\rm d}u \frac{\sqrt N\rho(U)}{\tilde{u}_n-u}+\frac{1}{\sqrt{N}}\pi\rho(U_n)   \cot(\xi_n)=0\,,
\end{eqnarray}
where $\xi_n=\pi \sqrt{N} \rho\left({U_n}\right)(\tilde{u}_n-u_n)$, which is an order $N^0$ quantity. 
Rewriting in the rescaled variables we get
\begin{eqnarray}
    \frac{1}{{\sqrt N}}F-\dashint{\rm d}U \frac{\rho(U)}{\tilde{U}_n-U}+\frac{1}{N}\pi\rho(U_n)   \cot(\xi_n)=0\,,
\end{eqnarray}
where the first two terms are in fact of the same order $N^0$ since asymptotically $F$ is a linear function of $u=\sqrt N U$.

One can think about the above equation as an equation on $\tilde u_n$. Note that the first two terms are smooth functions of $\tilde u_n$, whereas the last term changes between $+\infty$ to $-\infty$ when we very $\tilde u_n$ between $u_n$ and $u_{n+1}$, so there is always a solution for $\tilde u_n$ in this range.
When changing $n$ to $n+1$ the first two terms would not change much and thus we estimate 
\beqa\label{relationxi}
\xi_{n+1}\simeq \xi_{n}\;,
\eeqa
meaning that to the leading order $\tilde u_n-u_n\simeq \tilde u_{n+1}-u_{n+1}$.

Next we have to impose the condition \eq{dQQ} e.g. $Q(\tilde u_n)=-Q(\tilde u_{n+1})$. For that we can repeat the previous calculation for $Q'(u)/Q(u)$ for 
$u_n<u<u_{n+1}$ and integrate it over $u$ to obtain
\beq
\log Q(u) \simeq 
\log\mu(u)-\sqrt{N}\int \rho(v/\sqrt N)\log(v-u)dv
+\frac{1}{N}\log \sin (\xi_u )+C
\eeq
where $\xi_u=\pi N \rho\left(\frac{u_n}{N}\right)(u-u_n)$ and $C$ is an integration constant. We see, however, that the ``anomaly" term is $1/N$ suppressed,
and can be neglected, whereas the first two terms in the r.h.s. are order $N$. Finally, denoting $e_n= \log Q(\tilde u_n)$ we have to require that $e_{n+1}-e_n=\pm 2\pi i$.
Note that the $2\pi i$ simply comes from the imaginary part of $\log(v-u)$, which is given by $-2\pi\sqrt{N}\int_{a}^u \rho(v/\sqrt{N}) dv$ and we can use that $\int_{\tilde u_n}^{\tilde u_{n+1}}\rho
\simeq \int_{u_n}^{u_{n+1}}\rho = \frac{1}{\sqrt N}
$.
Hence, we arrive at
\begin{equation}
\log Q(\tilde u_{n+1})-
\log Q(\tilde u_{n})=\pm 2\pi i=\(\tilde u_{n+1}-\tilde u_n\)\[
\d_u \log\mu(u)+ \sqrt N \dashint \frac{\rho(V)}{u-v}dv
\]+2\pi i\;
\end{equation}
which when combined with \eq{relationxi} gives
\beqa\la{roots}\label{mudist}
\d_u \log\mu(u)+ \sqrt{N} \dashint \frac{\rho(V)}{u-v}dv=0\;.
\eeqa
This equation on $\rho$ is precisely the same equation as the density of the roots of the orthogonal polynomial with the weight $w(u)=\mu^2(u)$ (see below). Note that this result is only correct to the leading order in $N$, but corrections can be obtained by keeping more terms in the above derivation.
Another source of corrections comes from boundary terms -- near the end points of the distribution finite-size effects can become stronger and require special treatment.

\paragraph{Roots of orthogonal polynomials.}
Finally, for completeness let us quickly remind why \eq{mudist} is satisfied by the density of the roots of the orthogonal polynomials of the weight $w(u)=\mu^2(u)$. 

Following Section, the orthogonal polynomial $\pi_N$ for the weight $w(u)$ can be written as a ratio of determinants
\begin{equation}
    \pi_N(u) = D_N(u)/D_{N-1}
\end{equation}
where 
\begin{equation}
    D_{N}(u) = \displaystyle \int \prod_{k=1}^{N} {\rm d}u_k w(u_k) u_k^{k-1} \Delta(u_i,u)
\end{equation}
where $\Delta(u_i,u)$ is the Vandermonde determinant built from $u_1,\dots,u_N,u$. 
Note that we can rewrite this determinant as $\Delta(u_i,u) = \Delta(u_i)\prod_j(u_j-u)$.
Also note that because of the antisymmetrisiation we are free to replace $u_k^{k-1}$ factor with another Vandermonde determinant $\Delta(u_i)$, giving the expression
\begin{equation}\label{eqn:integralorthogpoly}
    D_N(u) = \displaystyle \int \prod_{k=1}^{N} {\rm d}u_k w(u_k)(u_k-u)  \Delta(u_i)^2\,.
\end{equation}

We can now compute this integral by saddle-point approximation. Defining $S$ by
\begin{equation}
    S= 2N \log\Delta(u_i) + \displaystyle\sum_{k=1}^N \log(w(u_k))
\end{equation}
the saddle point equations are $\partial S/\partial u_k =0$, giving 
\begin{eqnarray}\label{eqn:saddle}
   \frac{1}{2}\frac{w'(u_k)}{w(u_k)}+N \displaystyle \sum_{j\neq k} \frac{1}{u_k-u_j}=0\,.
\end{eqnarray}
In the large-$N$ limit the integral gets localised at each of the critical points $u_k =u_k^*$. Then, the rhs of \eqref{eqn:integralorthogpoly} clearly vanishes when $u=u_k^*$ implying these are the zeroes of the orthogonal polynomial $\pi_N(u)$. Comparing with \eqref{mudist}, we see that \eqref{mudist} is recovered from \eqref{eqn:saddle} upon setting $w(u) = \mu^2(u)$.

\section{Small Spin and Quasi-Classic String Theory}\label{app:foldedstring}

The strong coupling expansion of $\Delta$ for short physical operators is expected to take the general form \eqref{eq:DeltaStrongCoupling}. For general states not much is known about the structure of this expansion but some information can be deduced by considering either small spin or quasi-classic string theory. In this appendix we briefly review these results. In particular, our goal will not be to treat \eqref{eq:DeltaStrongCoupling} directly but first to constrain the coefficients $A_i,B_i,C_i,\dots$ defined in \eqref{eq:SquareDeltaStrongMain}, which reads
\begin{equation}\label{eq:SquareDeltaStrongAppendix}
    \bar\Delta^2=L^2 + S(\sqrt{\lambda}\,A_1 + A_2 + \dots) + S^2\(B_1+\frac{B_2}{\sqrt{\lambda}} + \dots \) + S^3\(\frac{C_1}{\sqrt{\lambda}} + \frac{C_2}{\lambda}+\dots\)+\mathcal{O}(S^4)\;,
\end{equation}
and thereafter deduce consequences for $\Delta$. We emphasise that \eqref{eq:SquareDeltaStrongAppendix} must be treated with caution for $n>1$ as for these mode numbers we currently do not know how to perform an analytic continuation on the level of the QSC which reproduces \eqref{eq:SquareDeltaStrongMain} for non-integer $S$.

\subsection{Quasi-Classics}
At strong coupling and large quantum numbers operators of type $\tr D^{S}\,Z^{L}$ are described by classical folded strings. A powerful integrability method to describe these solutions is the classical spectral curve; a set of $4+4$ complex functions known as quasi-momenta. Four of these quasi-momenta have a single branch cut. It is located at $(a,b)$, $1<a<b$, for two of the quasi-momenta and at $(\frac{1}{b},\frac{1}{a})$ for the remaining two, see for example \cite{Gromov:2011de} for details. The spectral curve relates the finite objects $\{\mathcal{D},\mathcal{J},\mathcal{S}\}=\{\frac{\Delta}{n\sqrt{\lambda}},\frac{L}{n\sqrt{\lambda}},\frac{S}{n\sqrt{\lambda}}\}$ and the branch-points as
\begin{subequations}\label{eq:ClassicsAppendix}
\begin{align}
    &2\pi\,\mathcal{S} = \frac{a b+1}{a b}\left(b\,E(1-\frac{a^2}{b^2})-a\,K(1-\frac{a^2}{b^2}) \right)\,, \\
    &2\pi\,\mathcal{J} = \frac{\sqrt{(a^2-1)(b^2-1)}}{b}\,K(1-\frac{a^2}{b^2})\,, \\
    &2\pi\,\mathcal{D}_{\text{classical}} = \frac{a b-1}{a b}\left(b\,E(1-\frac{a^2}{b^2})+a\,K(1-\frac{a^2}{b^2}) \right) \,.
\end{align}
\end{subequations}

Taking the limit $\mathcal{S}\simeq 0$ we can expand the classical energy order by order in $\mathcal{S}$. However, it turns out to be more economical to consider $\mathcal{D}^2$ and from \eqref{eq:ClassicsAppendix} we find
\begin{equation}\label{eq:DeltaClassic}
\begin{split}
    \mathcal{D}^2_{\text{classical}} &= \mathcal{J}^2 + \mathcal{S} \, 2\sqrt{\mathcal{J}^2+1}+\mathcal{S}^2 \, \frac{2\,\mathcal{J}+3}{2(\mathcal{J}^2+1)} - \mathcal{S}^3 \frac{3+\mathcal{J}^2}{8\left(\mathcal{J}^2+1\right)^{\frac{5}{2}}} \\
    &\quad \quad + \mathcal{S}^{4}\, \frac{31+18\,\mathcal{J}^2+3\,\mathcal{J}^4}{64\,\left(\mathcal{J}^2+1 \right)^{4}} - \mathcal{S}^{5} \frac{3\,(137+109\,\mathcal{J}^2+31\,\mathcal{J}^4+3\,\mathcal{J}^6)}{512\left(\mathcal{J}^2+1\right)^{\frac{11}{2}}}\\
    &\quad \quad +\mathcal{S}^{6} \frac{1572+1567\,\mathcal{J}^2+597\,\mathcal{J}^4+93\,\mathcal{J}^6+3\,\mathcal{J}^{8}}{1024(\mathcal{J}^2+1)^{7}}+\mathcal{O}(\mathcal{S}^{6})\,.
\end{split}
\end{equation}
The $1$-loop quantisation of the spectral curve is under good control \cite{Gromov:2007aq,Gromov:2007ky,Gromov:2009zza,Gromov:2008ec,Schafer-Nameki:2010qho} and was explored in detail for folded strings in \cite{Gromov:2011bz}. The resulting expression is
\begin{equation}\label{eq:Delta1Loop}
\begin{split}
    &\Delta_{\text{1-loop}} = -\mathcal{S}\,\frac{1}{2(\mathcal{J}^3+\mathcal{J})}+\sum_{a=2}\mathcal{S}^{a}\left(R_{a}(\mathcal{J})+ \sum_{\substack{m>0\\m\neq n}} P_{a}(n,m,\mathcal{J})\right)\,.
\end{split}
\end{equation}    
where $R_{a},P_{a}$ can be found in \cite{Gromov:2011bz}. Expanding the expression for small $\mathcal{J}$ all sums can be readily performed.

To constrain \eqref{eq:SquareDeltaStrongAppendix} using \eqref{eq:DeltaClassic} and \eqref{eq:Delta1Loop} we first expand for small $S$ and match orders. Thereafter we expand for large $\lambda$ and match the leading $\frac{1}{L}$ terms which fixes $A_i,B_i,C_i$ to the leading order in $L$. This procedure gives for $n=1,2$ the following results:
\paragraph{n=1} 
\begin{subequations}
\begin{eqnarray}
    &&A_1 = 2\;, \;\;\;\;\;\;\; A_2=-1\;,
    \\
    &&B_1 = \frac{3}{2}\;, \;\;\;\;\;\;
    B_2 = -3\,\zeta_3+\frac{3}{8}\,,\\
    &&C_1 = -\frac{3}{8}\;,\;\;\;\; C_2 = \frac{3}{16}\left(20\,\zeta_3+20\,\zeta_5-3 \right)\;,\\
    &&D_1 = \frac{31}{64}\;,\;\;\;\; D_2 = \frac{1}{512}\left(-4720\,\zeta_3-4160\,\zeta_5-2520\,\zeta_7+337 \right)\,,\\
    &&E_1 = -\frac{411}{512}\;.
\end{eqnarray}
\end{subequations}
\paragraph{n=2}
\begin{subequations}
\begin{eqnarray}
    &&A_1 = 4\;,\;\;\;\;\;\;\;
    A_2=-1\;,
    \\
    &&B_1 = \frac{3}{2}\;,\;\;\;\;\;\; B_2 =-\frac{13}{16}-12\, \zeta_3\,,\\
    &&C_1 = -\frac{3}{16}\;,\;\;
    C_2 = \frac{71}{64}+\frac{15\,\zeta_3}{2}+30 \,\zeta_5\;,\\
    &&D_1 = \frac{31}{256}\;.
\end{eqnarray}
\end{subequations}

\subsection{Small-Spin}
For $n=1$ the coefficients $A_i$ and $B_i$ can be fixed by expanding $\Delta$ for $S\ll1$ according to
\begin{equation}
    \Delta - L -S = S \gamma^{(1)}_L + S^2 \, \gamma^{(2)}_L + \mathcal{O}(S^3)
\end{equation}
where $\gamma^{(1)}$ is the slope function and $\gamma^{(2)}$ the so-called curvature function. The curvature function is expected to take the form
\begin{equation}
    \gamma^{(1)}_L = \frac{4\pi \,g\, n \, I_{L+1}(4\pi \,g\, n)}{L\, I_{L}(4\pi \, g \,n)}\;,
\end{equation}
for all $n$, however at the moment only $n=1$ has been reliable computed from the QSC. The curvature functions is more cumbersome, the full expression can be found \cite{Gromov:2014bva}. As explained in the main text to extract $B$ from the curvature function we currently have to rely on high precision numerics. Using this method the following coefficients have been extracted
\beqa
\begin{array}{ccl}
 B_1 & = & \frac{3}{2} \\
 B_2 & = & \frac{3}{8}-3 \,\zeta _3 \\
 B_3 & = & \left(\frac{5}{16}-\frac{9 \zeta _3}{2}\right)-\frac{L^2}{2}  \\
 B_4 & = & -\frac{3}{16} \left(62 \,\zeta _3+40\, \zeta _5+1\right)+\frac{3}{16} \left(16\, \zeta _3+20 \,\zeta _5-9\right) L^2  \\
 B_5 & = & -\frac{1}{64} \left(2362\, \zeta _3+1580 \,\zeta _5+203\right)+\frac{1}{8} \left(116 \,\zeta _3+100\, \zeta _5-39\right) L^2+\frac{L^4}{2}
\end{array} \,.
\eeqa

\bibliographystyle{JHEP}
\bibliography{references}

\providecommand{\href}[2]{#2}\begingroup\raggedright\begin{thebibliography}{10}

\bibitem{Minahan:2002ve}
J.~A. Minahan and K.~Zarembo, \emph{{The Bethe ansatz for N=4
  superYang-Mills}},
  \href{https://doi.org/10.1088/1126-6708/2003/03/013}{\emph{JHEP} {\bfseries
  03} (2003) 013} [\href{https://arxiv.org/abs/hep-th/0212208}{{\ttfamily
  hep-th/0212208}}].

\bibitem{Lipatov:2009nt}
L.~N. Lipatov, \emph{{Integrability of scattering amplitudes in N=4 SUSY}},
  \href{https://doi.org/10.1088/1751-8113/42/30/304020}{\emph{J. Phys. A}
  {\bfseries 42} (2009) 304020}
  [\href{https://arxiv.org/abs/0902.1444}{{\ttfamily 0902.1444}}].

\bibitem{Beisert:2010jr}
N.~Beisert et~al., \emph{{Review of AdS/CFT Integrability: An Overview}},
  \href{https://doi.org/10.1007/s11005-011-0529-2}{\emph{Lett. Math. Phys.}
  {\bfseries 99} (2012) 3} [\href{https://arxiv.org/abs/1012.3982}{{\ttfamily
  1012.3982}}].

\bibitem{Gromov:2009bc}
N.~Gromov, V.~Kazakov, A.~Kozak and P.~Vieira, \emph{{Exact Spectrum of
  Anomalous Dimensions of Planar N = 4 Supersymmetric Yang-Mills Theory: TBA
  and excited states}},
  \href{https://doi.org/10.1007/s11005-010-0374-8}{\emph{Lett. Math. Phys.}
  {\bfseries 91} (2010) 265} [\href{https://arxiv.org/abs/0902.4458}{{\ttfamily
  0902.4458}}].

\bibitem{Drukker:2012de}
N.~Drukker, \emph{{Integrable Wilson loops}},
  \href{https://doi.org/10.1007/JHEP10(2013)135}{\emph{JHEP} {\bfseries 10}
  (2013) 135} [\href{https://arxiv.org/abs/1203.1617}{{\ttfamily 1203.1617}}].

\bibitem{Basso:2015zoa}
B.~Basso, S.~Komatsu and P.~Vieira, \emph{{Structure Constants and Integrable
  Bootstrap in Planar N=4 SYM Theory}},
  \href{https://arxiv.org/abs/1505.06745}{{\ttfamily 1505.06745}}.

\bibitem{Komatsu:2017buu}
S.~Komatsu, \emph{{Three-point functions in $\mathcal N=$ 4 supersymmetric
  Yang\textendash{}Mills theory}},
  \href{https://arxiv.org/abs/1710.03853}{{\ttfamily 1710.03853}}.

\bibitem{deLeeuw:2017cop}
M.~de~Leeuw, A.~C. Ipsen, C.~Kristjansen and M.~Wilhelm, \emph{{Introduction to
  integrability and one-point functions in $\mathcal N=$ 4 supersymmetric
  Yang\textendash{}Mills theory and its defect cousin}},
  \href{https://arxiv.org/abs/1708.02525}{{\ttfamily 1708.02525}}.

\bibitem{Gromov:2013pga}
N.~Gromov, V.~Kazakov, S.~Leurent and D.~Volin, \emph{{Quantum Spectral Curve
  for Planar $\mathcal{N} = 4$ Super-Yang-Mills Theory}},
  \href{https://doi.org/10.1103/PhysRevLett.112.011602}{\emph{Phys. Rev. Lett.}
  {\bfseries 112} (2014) 011602}
  [\href{https://arxiv.org/abs/1305.1939}{{\ttfamily 1305.1939}}].

\bibitem{Gromov:2014caa}
N.~Gromov, V.~Kazakov, S.~Leurent and D.~Volin, \emph{{Quantum spectral curve
  for arbitrary state/operator in AdS$_{5}$/CFT$_{4}$}},
  \href{https://doi.org/10.1007/JHEP09(2015)187}{\emph{JHEP} {\bfseries 09}
  (2015) 187} [\href{https://arxiv.org/abs/1405.4857}{{\ttfamily 1405.4857}}].

\bibitem{Marboe:2014gma}
C.~Marboe and D.~Volin, \emph{{Quantum spectral curve as a tool for a
  perturbative quantum field theory}},
  \href{https://doi.org/10.1016/j.nuclphysb.2015.08.021}{\emph{Nucl. Phys.}
  {\bfseries B899} (2015) 810}
  [\href{https://arxiv.org/abs/1411.4758}{{\ttfamily 1411.4758}}].

\bibitem{Gromov:2015vua}
N.~Gromov, F.~Levkovich-Maslyuk and G.~Sizov, \emph{{Pomeron Eigenvalue at
  Three Loops in $\mathcal N=$ 4 Supersymmetric Yang-Mills Theory}},
  \href{https://doi.org/10.1103/PhysRevLett.115.251601}{\emph{Phys. Rev. Lett.}
  {\bfseries 115} (2015) 251601}
  [\href{https://arxiv.org/abs/1507.04010}{{\ttfamily 1507.04010}}].

\bibitem{Gromov:2014bva}
N.~Gromov, F.~Levkovich-Maslyuk, G.~Sizov and S.~Valatka, \emph{{Quantum
  spectral curve at work: from small spin to strong coupling in $ \mathcal{N} $
  = 4 SYM}}, \href{https://doi.org/10.1007/JHEP07(2014)156}{\emph{JHEP}
  {\bfseries 07} (2014) 156} [\href{https://arxiv.org/abs/1402.0871}{{\ttfamily
  1402.0871}}].

\bibitem{Gromov:2015dfa}
N.~Gromov and F.~Levkovich-Maslyuk, \emph{{Quantum Spectral Curve for a cusped
  Wilson line in $ \mathcal{N}=4 $ SYM}},
  \href{https://doi.org/10.1007/JHEP04(2016)134}{\emph{JHEP} {\bfseries 04}
  (2016) 134} [\href{https://arxiv.org/abs/1510.02098}{{\ttfamily
  1510.02098}}].

\bibitem{Gromov:2015wca}
N.~Gromov, F.~Levkovich-Maslyuk and G.~Sizov, \emph{{Quantum Spectral Curve and
  the Numerical Solution of the Spectral Problem in AdS5/CFT4}},
  \href{https://doi.org/10.1007/JHEP06(2016)036}{\emph{JHEP} {\bfseries 06}
  (2016) 036} [\href{https://arxiv.org/abs/1504.06640}{{\ttfamily
  1504.06640}}].

\bibitem{Gromov:2023hzc}
N.~Gromov, A.~Hegedus, J.~Julius and N.~Sokolova, \emph{{Fast QSC Solver: tool
  for systematic study of N=4 Super-Yang-Mills spectrum}},
  \href{https://arxiv.org/abs/2306.12379}{{\ttfamily 2306.12379}}.

\bibitem{Giombi:2018qox}
S.~Giombi and S.~Komatsu, \emph{{Exact Correlators on the Wilson Loop in
  $\mathcal{N}=4$ SYM: Localization, Defect CFT, and Integrability}},
  \href{https://doi.org/10.1007/JHEP11(2018)123,
  10.1007/JHEP05(2018)109}{\emph{JHEP} {\bfseries 05} (2018) 109}
  [\href{https://arxiv.org/abs/1802.05201}{{\ttfamily 1802.05201}}].

\bibitem{Giombi:2018hsx}
S.~Giombi and S.~Komatsu, \emph{{More Exact Results in the Wilson Loop Defect
  CFT: Bulk-Defect OPE, Nonplanar Corrections and Quantum Spectral Curve}},
  \href{https://doi.org/10.1088/1751-8121/ab046c}{\emph{J. Phys.} {\bfseries
  A52} (2019) 125401} [\href{https://arxiv.org/abs/1811.02369}{{\ttfamily
  1811.02369}}].

\bibitem{Cavaglia:2018lxi}
A.~Cavaglia, N.~Gromov and F.~Levkovich-Maslyuk, \emph{{Quantum spectral curve
  and structure constants in $ \mathcal{N}=4 $ SYM: cusps in the ladder
  limit}}, \href{https://doi.org/10.1007/JHEP10(2018)060}{\emph{JHEP}
  {\bfseries 10} (2018) 060}
  [\href{https://arxiv.org/abs/1802.04237}{{\ttfamily 1802.04237}}].

\bibitem{McGovern:2019sdd}
J.~McGovern, \emph{{Scalar Insertions in Cusped Wilson Loops in the Ladders
  Limit of Planar $N$=4 SYM}},
  \href{https://arxiv.org/abs/1912.00499}{{\ttfamily 1912.00499}}.

\bibitem{Cavaglia:2021mft}
A.~Cavagli\`a, N.~Gromov and F.~Levkovich-Maslyuk, \emph{{Separation of
  variables in AdS/CFT: functional approach for the fishnet CFT}},
  \href{https://doi.org/10.1007/JHEP06(2021)131}{\emph{JHEP} {\bfseries 06}
  (2021) 131} [\href{https://arxiv.org/abs/2103.15800}{{\ttfamily
  2103.15800}}].

\bibitem{Basso:2022nny}
B.~Basso, A.~Georgoudis and A.~K. Sueiro, \emph{{Structure Constants of Short
  Operators in Planar N=4 Supersymmetric Yang-Mills Theory}},
  \href{https://doi.org/10.1103/PhysRevLett.130.131603}{\emph{Phys. Rev. Lett.}
  {\bfseries 130} (2023) 131603}
  [\href{https://arxiv.org/abs/2207.01315}{{\ttfamily 2207.01315}}].

\bibitem{Bercini:2022jxo}
C.~Bercini, A.~Homrich and P.~Vieira, \emph{{Structure Constants in
  $\mathcal{N} = 4$ SYM and Separation of Variables}},
  \href{https://arxiv.org/abs/2210.04923}{{\ttfamily 2210.04923}}.

\bibitem{Giombi:2022anm}
S.~Giombi, S.~Komatsu and B.~Offertaler, \emph{{Large charges on the Wilson
  loop in $ \mathcal{N} $ = 4 SYM. Part II. Quantum fluctuations, OPE, and
  spectral curve}}, \href{https://doi.org/10.1007/JHEP08(2022)011}{\emph{JHEP}
  {\bfseries 08} (2022) 011}
  [\href{https://arxiv.org/abs/2202.07627}{{\ttfamily 2202.07627}}].

\bibitem{Roiban:2009aa}
R.~Roiban and A.~A. Tseytlin, \emph{{Quantum strings in AdS(5) x S**5:
  Strong-coupling corrections to dimension of Konishi operator}},
  \href{https://doi.org/10.1088/1126-6708/2009/11/013}{\emph{JHEP} {\bfseries
  11} (2009) 013} [\href{https://arxiv.org/abs/0906.4294}{{\ttfamily
  0906.4294}}].

\bibitem{Roiban:2011fe}
R.~Roiban and A.~A. Tseytlin, \emph{{Semiclassical string computation of
  strong-coupling corrections to dimensions of operators in Konishi
  multiplet}},
  \href{https://doi.org/10.1016/j.nuclphysb.2011.02.016}{\emph{Nucl. Phys. B}
  {\bfseries 848} (2011) 251}
  [\href{https://arxiv.org/abs/1102.1209}{{\ttfamily 1102.1209}}].

\bibitem{Vallilo:2011fj}
B.~C. Vallilo and L.~Mazzucato, \emph{{The Konishi multiplet at strong
  coupling}}, \href{https://doi.org/10.1007/JHEP12(2011)029}{\emph{JHEP}
  {\bfseries 12} (2011) 029} [\href{https://arxiv.org/abs/1102.1219}{{\ttfamily
  1102.1219}}].

\bibitem{Gromov:2011de}
N.~Gromov, D.~Serban, I.~Shenderovich and D.~Volin, \emph{{Quantum folded
  string and integrability: From finite size effects to Konishi dimension}},
  \href{https://doi.org/10.1007/JHEP08(2011)046}{\emph{JHEP} {\bfseries 08}
  (2011) 046} [\href{https://arxiv.org/abs/1102.1040}{{\ttfamily 1102.1040}}].

\bibitem{Frolov:2013lva}
S.~Frolov, M.~Heinze, G.~Jorjadze and J.~Plefka, \emph{{Static gauge and energy
  spectrum of single-mode strings in AdS$_{5} ×$ S$^{5}$}},
  \href{https://doi.org/10.1088/1751-8113/47/8/085401}{\emph{J. Phys. A}
  {\bfseries 47} (2014) 085401}
  [\href{https://arxiv.org/abs/1310.5052}{{\ttfamily 1310.5052}}].

\bibitem{Alday:2022uxp}
L.~F. Alday, T.~Hansen and J.~A. Silva, \emph{{AdS Virasoro-Shapiro from
  dispersive sum rules}},
  \href{https://doi.org/10.1007/JHEP10(2022)036}{\emph{JHEP} {\bfseries 10}
  (2022) 036} [\href{https://arxiv.org/abs/2204.07542}{{\ttfamily
  2204.07542}}].

\bibitem{Alday:2023mvu}
L.~F. Alday and T.~Hansen, \emph{{The AdS Virasoro-Shapiro amplitude}},
  \href{https://doi.org/10.1007/JHEP10(2023)023}{\emph{JHEP} {\bfseries 10}
  (2023) 023} [\href{https://arxiv.org/abs/2306.12786}{{\ttfamily
  2306.12786}}].

\bibitem{Alday:2023flc}
L.~F. Alday, T.~Hansen and J.~A. Silva, \emph{{On the spectrum and structure
  constants of short operators in N=4 SYM at strong coupling}},
  \href{https://doi.org/10.1007/JHEP08(2023)214}{\emph{JHEP} {\bfseries 08}
  (2023) 214} [\href{https://arxiv.org/abs/2303.08834}{{\ttfamily
  2303.08834}}].

\bibitem{Gromov:2009zb}
N.~Gromov, V.~Kazakov and P.~Vieira, \emph{{Exact Spectrum of Planar ${\cal
  N}=4$ Supersymmetric Yang-Mills Theory: Konishi Dimension at Any Coupling}},
  \href{https://doi.org/10.1103/PhysRevLett.104.211601}{\emph{Phys. Rev. Lett.}
  {\bfseries 104} (2010) 211601}
  [\href{https://arxiv.org/abs/0906.4240}{{\ttfamily 0906.4240}}].

\bibitem{Gromov:2011bz}
N.~Gromov and S.~Valatka, \emph{{Deeper Look into Short Strings}},
  \href{https://doi.org/10.1007/JHEP03(2012)058}{\emph{JHEP} {\bfseries 03}
  (2012) 058} [\href{https://arxiv.org/abs/1109.6305}{{\ttfamily 1109.6305}}].

\bibitem{Hegeds_2016}
A.~Hegedus and J.~Konczer, \emph{Strong coupling results in the ads5/cft4
  correspondence from the numerical solution of the quantum spectral curve},
  \href{https://doi.org/10.1007/jhep08(2016)061}{\emph{Journal of High Energy
  Physics} {\bfseries 2016} (2016) }.

\bibitem{Berenstein:2002jq}
D.~E. Berenstein, J.~M. Maldacena and H.~S. Nastase, \emph{{Strings in flat
  space and pp waves from N=4 superYang-Mills}},
  \href{https://doi.org/10.1088/1126-6708/2002/04/013}{\emph{JHEP} {\bfseries
  04} (2002) 013} [\href{https://arxiv.org/abs/hep-th/0202021}{{\ttfamily
  hep-th/0202021}}].

\bibitem{Kazakov:2004qf}
V.~Kazakov, A.~Marshakov, J.~Minahan and K.~Zarembo, \emph{{Classical/quantum
  integrability in AdS/CFT}},
  \href{https://doi.org/10.1088/1126-6708/2004/05/024}{\emph{JHEP} {\bfseries
  05} (2004) 024} [\href{https://arxiv.org/abs/hep-th/0402207}{{\ttfamily
  hep-th/0402207}}].

\bibitem{Beisert:2005bm}
N.~Beisert, V.~A. Kazakov, K.~Sakai and K.~Zarembo, \emph{{The Algebraic curve
  of classical superstrings on AdS(5) x S**5}},
  \href{https://doi.org/10.1007/s00220-006-1529-4}{\emph{Commun. Math. Phys.}
  {\bfseries 263} (2006) 659}
  [\href{https://arxiv.org/abs/hep-th/0502226}{{\ttfamily hep-th/0502226}}].

\bibitem{Gromov:2017blm}
N.~Gromov, \emph{{Introduction to the Spectrum of $N=4$ SYM and the Quantum
  Spectral Curve}},  \href{https://arxiv.org/abs/1708.03648}{{\ttfamily
  1708.03648}}.

\bibitem{Kazakov:2018ugh}
V.~Kazakov, \emph{{Quantum Spectral Curve of $\gamma$-twisted ${\cal N}=4$ SYM
  theory and fishnet CFT}},  \href{https://arxiv.org/abs/1802.02160}{{\ttfamily
  1802.02160}}.

\bibitem{Levkovich-Maslyuk:2019awk}
F.~Levkovich-Maslyuk, \emph{{A review of the AdS/CFT Quantum Spectral Curve}},
  \href{https://arxiv.org/abs/1911.13065}{{\ttfamily 1911.13065}}.

\bibitem{Gubser:2002tv}
S.~S. Gubser, I.~R. Klebanov and A.~M. Polyakov, \emph{{A Semiclassical limit
  of the gauge / string correspondence}},
  \href{https://doi.org/10.1016/S0550-3213(02)00373-5}{\emph{Nucl. Phys. B}
  {\bfseries 636} (2002) 99}
  [\href{https://arxiv.org/abs/hep-th/0204051}{{\ttfamily hep-th/0204051}}].

\bibitem{Frolov:2010wt}
S.~Frolov, \emph{{Konishi operator at intermediate coupling}},
  \href{https://doi.org/10.1088/1751-8113/44/6/065401}{\emph{J. Phys. A}
  {\bfseries 44} (2011) 065401}
  [\href{https://arxiv.org/abs/1006.5032}{{\ttfamily 1006.5032}}].

\bibitem{Frolov:2012zv}
S.~Frolov, \emph{{Scaling dimensions from the mirror TBA}},
  \href{https://doi.org/10.1088/1751-8113/45/30/305402}{\emph{J. Phys. A}
  {\bfseries 45} (2012) 305402}
  [\href{https://arxiv.org/abs/1201.2317}{{\ttfamily 1201.2317}}].

\bibitem{Julius:2023hre}
J.~Julius and N.~Sokolova, \emph{{Conformal field theory-data analysis for
  $\mathcal{N}$ = 4 Super-Yang-Mills at strong coupling}},
  \href{https://doi.org/10.1007/JHEP03(2024)090}{\emph{JHEP} {\bfseries 03}
  (2024) 090} [\href{https://arxiv.org/abs/2310.06041}{{\ttfamily
  2310.06041}}].

\bibitem{Basso:2011rs}
B.~Basso, \emph{{An exact slope for AdS/CFT}},
  \href{https://arxiv.org/abs/1109.3154}{{\ttfamily 1109.3154}}.

\bibitem{Beccaria:2012xm}
M.~Beccaria, S.~Giombi, G.~Macorini, R.~Roiban and A.~A. Tseytlin,
  \emph{{'Short' spinning strings and structure of quantum $AdS_5 \times S^5$
  spectrum}}, \href{https://doi.org/10.1103/PhysRevD.86.066006}{\emph{Phys.
  Rev. D} {\bfseries 86} (2012) 066006}
  [\href{https://arxiv.org/abs/1203.5710}{{\ttfamily 1203.5710}}].

\bibitem{Brower:2014wha}
R.~C. Brower, M.~S. Costa, M.~Djuri\'c, T.~Raben and C.-I. Tan, \emph{{Strong
  Coupling Expansion for the Conformal Pomeron/Odderon Trajectories}},
  \href{https://doi.org/10.1007/JHEP02(2015)104}{\emph{JHEP} {\bfseries 02}
  (2015) 104} [\href{https://arxiv.org/abs/1409.2730}{{\ttfamily 1409.2730}}].

\bibitem{Klabbers:2023zdz}
R.~Klabbers, M.~Preti and I.~M. Sz\'ecs\'enyi, \emph{{Regge Spectroscopy of
  Higher-Twist States in N=4 Supersymmetric Yang-Mills Theory}},
  \href{https://doi.org/10.1103/PhysRevLett.132.191601}{\emph{Phys. Rev. Lett.}
  {\bfseries 132} (2024) 191601}
  [\href{https://arxiv.org/abs/2307.15107}{{\ttfamily 2307.15107}}].

\bibitem{Cavaglia:2014exa}
A.~Cavagli\`a, D.~Fioravanti, N.~Gromov and R.~Tateo, \emph{{Quantum Spectral
  Curve of the $\mathcal N=$ 6 Supersymmetric Chern-Simons Theory}},
  \href{https://doi.org/10.1103/PhysRevLett.113.021601}{\emph{Phys. Rev. Lett.}
  {\bfseries 113} (2014) 021601}
  [\href{https://arxiv.org/abs/1403.1859}{{\ttfamily 1403.1859}}].

\bibitem{Bombardelli:2017vhk}
D.~Bombardelli, A.~Cavagli\`a, D.~Fioravanti, N.~Gromov and R.~Tateo,
  \emph{{The full Quantum Spectral Curve for $AdS_4/CFT_3$}},
  \href{https://doi.org/10.1007/JHEP09(2017)140}{\emph{JHEP} {\bfseries 09}
  (2017) 140} [\href{https://arxiv.org/abs/1701.00473}{{\ttfamily
  1701.00473}}].

\bibitem{Bombardelli:2018bqz}
D.~Bombardelli, A.~Cavagli\`a, R.~Conti and R.~Tateo, \emph{{Exploring the
  spectrum of planar AdS$_{4}$/CFT$_{3}$ at finite coupling}},
  \href{https://doi.org/10.1007/JHEP04(2018)117}{\emph{JHEP} {\bfseries 04}
  (2018) 117} [\href{https://arxiv.org/abs/1803.04748}{{\ttfamily
  1803.04748}}].

\bibitem{Ekhammar:2021pys}
S.~Ekhammar and D.~Volin, \emph{{Monodromy bootstrap for SU(2|2) quantum
  spectral curves: from Hubbard model to AdS$_{3}$/CFT$_{2}$}},
  \href{https://doi.org/10.1007/JHEP03(2022)192}{\emph{JHEP} {\bfseries 03}
  (2022) 192} [\href{https://arxiv.org/abs/2109.06164}{{\ttfamily
  2109.06164}}].

\bibitem{Cavaglia:2021eqr}
A.~Cavagli\`a, N.~Gromov, B.~Stefa\'nski, Jr., Jr. and A.~Torrielli,
  \emph{{Quantum Spectral Curve for AdS$_{3}$/CFT$_{2}$: a proposal}},
  \href{https://doi.org/10.1007/JHEP12(2021)048}{\emph{JHEP} {\bfseries 12}
  (2021) 048} [\href{https://arxiv.org/abs/2109.05500}{{\ttfamily
  2109.05500}}].

\bibitem{Cavaglia:2022xld}
A.~Cavagli\`a, S.~Ekhammar, N.~Gromov and P.~Ryan, \emph{{Exploring the Quantum
  Spectral Curve for AdS$_{3}$/CFT$_{2}$}},
  \href{https://doi.org/10.1007/JHEP12(2023)089}{\emph{JHEP} {\bfseries 12}
  (2023) 089} [\href{https://arxiv.org/abs/2211.07810}{{\ttfamily
  2211.07810}}].

\bibitem{Frolov:2021bwp}
S.~Frolov and A.~Sfondrini, \emph{{Mirror thermodynamic Bethe ansatz for
  AdS3/CFT2}}, \href{https://doi.org/10.1007/JHEP03(2022)138}{\emph{JHEP}
  {\bfseries 03} (2022) 138}
  [\href{https://arxiv.org/abs/2112.08898}{{\ttfamily 2112.08898}}].

\bibitem{Brollo:2023pkl}
A.~Brollo, D.~le~Plat, A.~Sfondrini and R.~Suzuki, \emph{{Tensionless Limit of
  Pure\textendash{}Ramond-Ramond Strings and AdS3/CFT2}},
  \href{https://doi.org/10.1103/PhysRevLett.131.161604}{\emph{Phys. Rev. Lett.}
  {\bfseries 131} (2023) 161604}
  [\href{https://arxiv.org/abs/2303.02120}{{\ttfamily 2303.02120}}].

\bibitem{Brollo:2023rgp}
A.~Brollo, D.~le~Plat, A.~Sfondrini and R.~Suzuki, \emph{{More on the
  tensionless limit of pure-Ramond-Ramond AdS3/CFT2}},
  \href{https://doi.org/10.1007/JHEP12(2023)160}{\emph{JHEP} {\bfseries 12}
  (2023) 160} [\href{https://arxiv.org/abs/2308.11576}{{\ttfamily
  2308.11576}}].

\bibitem{Frolov:2023wji}
S.~Frolov, A.~Pribytok and A.~Sfondrini, \emph{{Ground state energy of twisted
  AdS$_{3}$ \texttimes{} S$^{3}$ \texttimes{} T$^{4}$ superstring and the
  TBA}}, \href{https://doi.org/10.1007/JHEP09(2023)027}{\emph{JHEP} {\bfseries
  09} (2023) 027} [\href{https://arxiv.org/abs/2305.17128}{{\ttfamily
  2305.17128}}].

\bibitem{Levkovich-Maslyuk:2020rlp}
F.~Levkovich-Maslyuk and M.~Preti, \emph{{Exploring the ground state spectrum
  of \ensuremath{\gamma}-deformed N = 4 SYM}},
  \href{https://doi.org/10.1007/JHEP06(2022)146}{\emph{JHEP} {\bfseries 06}
  (2022) 146} [\href{https://arxiv.org/abs/2003.05811}{{\ttfamily
  2003.05811}}].

\bibitem{Marboe:2019wyc}
C.~Marboe and E.~Wid\'en, \emph{{The fate of the Konishi multiplet in the
  $\beta$-deformed Quantum Spectral Curve}},
  \href{https://doi.org/10.1007/JHEP01(2020)026}{\emph{JHEP} {\bfseries 01}
  (2020) 026} [\href{https://arxiv.org/abs/1902.01248}{{\ttfamily
  1902.01248}}].

\bibitem{Harmark:2017yrv}
T.~Harmark and M.~Wilhelm, \emph{{Hagedorn Temperature of AdS$_5$/CFT$_4$ via
  Integrability}},
  \href{https://doi.org/10.1103/PhysRevLett.120.071605}{\emph{Phys. Rev. Lett.}
  {\bfseries 120} (2018) 071605}
  [\href{https://arxiv.org/abs/1706.03074}{{\ttfamily 1706.03074}}].

\bibitem{Harmark:2018red}
T.~Harmark and M.~Wilhelm, \emph{{The Hagedorn temperature of AdS$_5$/CFT$_4$
  at finite coupling via the Quantum Spectral Curve}},
  \href{https://doi.org/10.1016/j.physletb.2018.09.033}{\emph{Phys. Lett. B}
  {\bfseries 786} (2018) 53}
  [\href{https://arxiv.org/abs/1803.04416}{{\ttfamily 1803.04416}}].

\bibitem{Harmark:2021qma}
T.~Harmark and M.~Wilhelm, \emph{{Solving the Hagedorn temperature of
  AdS$_{5}$/CFT$_{4}$ via the Quantum Spectral Curve: chemical potentials and
  deformations}}, \href{https://doi.org/10.1007/JHEP07(2022)136}{\emph{JHEP}
  {\bfseries 07} (2022) 136}
  [\href{https://arxiv.org/abs/2109.09761}{{\ttfamily 2109.09761}}].

\bibitem{Ekhammar:2023cuj}
S.~Ekhammar, J.~A. Minahan and C.~Thull, \emph{{The ABJM Hagedorn Temperature
  from Integrability}},
  \href{https://doi.org/10.1007/JHEP10(2023)066}{\emph{JHEP} {\bfseries 10}
  (2023) 066} [\href{https://arxiv.org/abs/2307.02350}{{\ttfamily
  2307.02350}}].

\bibitem{Ekhammar:2023glu}
S.~Ekhammar, J.~A. Minahan and C.~Thull, \emph{{The asymptotic form of the
  Hagedorn temperature in planar super Yang-Mills}},
  \href{https://doi.org/10.1088/1751-8121/acf9d0}{\emph{J. Phys. A} {\bfseries
  56} (2023) 435401} [\href{https://arxiv.org/abs/2306.09883}{{\ttfamily
  2306.09883}}].

\bibitem{Urbach:2022xzw}
E.~Y. Urbach, \emph{{String stars in anti de Sitter space}},
  \href{https://doi.org/10.1007/JHEP04(2022)072}{\emph{JHEP} {\bfseries 04}
  (2022) 072} [\href{https://arxiv.org/abs/2202.06966}{{\ttfamily
  2202.06966}}].

\bibitem{Bigazzi:2023hxt}
F.~Bigazzi, T.~Canneti and A.~L. Cotrone, \emph{{Higher order corrections to
  the Hagedorn temperature at strong coupling}},
  \href{https://doi.org/10.1007/JHEP10(2023)056}{\emph{JHEP} {\bfseries 10}
  (2023) 056} [\href{https://arxiv.org/abs/2306.17126}{{\ttfamily
  2306.17126}}].

\bibitem{Harmark:2024ioq}
T.~Harmark, \emph{{Hagedorn temperature from the thermal scalar in AdS and
  pp-wave backgrounds}},  \href{https://arxiv.org/abs/2402.06001}{{\ttfamily
  2402.06001}}.

\bibitem{Grabner:2017pgm}
D.~Grabner, N.~Gromov, V.~Kazakov and G.~Korchemsky, \emph{{Strongly
  $\gamma$-Deformed $\mathcal{N}=4$ Supersymmetric Yang-Mills Theory as an
  Integrable Conformal Field Theory}},
  \href{https://doi.org/10.1103/PhysRevLett.120.111601}{\emph{Phys. Rev. Lett.}
  {\bfseries 120} (2018) 111601}
  [\href{https://arxiv.org/abs/1711.04786}{{\ttfamily 1711.04786}}].

\bibitem{Ferrero:2021bsb}
P.~Ferrero and C.~Meneghelli, \emph{{Bootstrapping the half-BPS line defect CFT
  in $\mathcal{N}=4$ SYM at strong coupling}},
  \href{https://arxiv.org/abs/2103.10440}{{\ttfamily 2103.10440}}.

\bibitem{Ferrero:2023gnu}
P.~Ferrero and C.~Meneghelli, \emph{{Unmixing the Wilson line defect CFT. Part
  II: analytic bootstrap}},  \href{https://arxiv.org/abs/2312.12551}{{\ttfamily
  2312.12551}}.

\bibitem{Ferrero:2023znz}
P.~Ferrero and C.~Meneghelli, \emph{{Unmixing the Wilson line defect CFT. Part
  I. Spectrum and kinematics}},
  \href{https://doi.org/10.1007/JHEP05(2024)090}{\emph{JHEP} {\bfseries 05}
  (2024) 090} [\href{https://arxiv.org/abs/2312.12550}{{\ttfamily
  2312.12550}}].

\bibitem{Cavaglia:2021bnz}
A.~Cavagli\`a, N.~Gromov, J.~Julius and M.~Preti, \emph{{Integrability and
  conformal bootstrap: One dimensional defect conformal field theory}},
  \href{https://doi.org/10.1103/PhysRevD.105.L021902}{\emph{Phys. Rev. D}
  {\bfseries 105} (2022) L021902}
  [\href{https://arxiv.org/abs/2107.08510}{{\ttfamily 2107.08510}}].

\bibitem{Caron-Huot:2022sdy}
S.~Caron-Huot, F.~Coronado, A.-K. Trinh and Z.~Zahraee, \emph{{Bootstrapping $
  \mathcal{N} $ = 4 sYM correlators using integrability}},
  \href{https://doi.org/10.1007/JHEP02(2023)083}{\emph{JHEP} {\bfseries 02}
  (2023) 083} [\href{https://arxiv.org/abs/2207.01615}{{\ttfamily
  2207.01615}}].

\bibitem{Cavaglia:2022qpg}
A.~Cavagli\`a, N.~Gromov, J.~Julius and M.~Preti, \emph{{Bootstrability in
  defect CFT: integrated correlators and sharper bounds}},
  \href{https://doi.org/10.1007/JHEP05(2022)164}{\emph{JHEP} {\bfseries 05}
  (2022) 164} [\href{https://arxiv.org/abs/2203.09556}{{\ttfamily
  2203.09556}}].

\bibitem{Cavaglia:2022yvv}
A.~Cavagli\`a, N.~Gromov, J.~Julius and M.~Preti, \emph{{Integrated correlators
  from integrability: Maldacena-Wilson line in $ \mathcal{N} $ = 4 SYM}},
  \href{https://doi.org/10.1007/JHEP04(2023)026}{\emph{JHEP} {\bfseries 04}
  (2023) 026} [\href{https://arxiv.org/abs/2211.03203}{{\ttfamily
  2211.03203}}].

\bibitem{Cavaglia:2023mmu}
A.~Cavagli\`a, N.~Gromov and M.~Preti, \emph{{Computing Four-Point Functions
  with Integrability, Bootstrap and Parity Symmetry}},
  \href{https://arxiv.org/abs/2312.11604}{{\ttfamily 2312.11604}}.

\bibitem{Passerini:2010xc}
F.~Passerini, J.~Plefka, G.~W. Semenoff and D.~Young, \emph{{On the Spectrum of
  the $AdS_{5}$ x $S^{5}$ String at large $\lambda$}},
  \href{https://doi.org/10.1007/JHEP03(2011)046}{\emph{JHEP} {\bfseries 03}
  (2011) 046} [\href{https://arxiv.org/abs/1012.4471}{{\ttfamily 1012.4471}}].

\bibitem{Grabner:2020nis}
D.~Grabner, N.~Gromov and J.~Julius, \emph{{Excited States of One-Dimensional
  Defect CFTs from the Quantum Spectral Curve}},
  \href{https://doi.org/10.1007/JHEP07(2020)042}{\emph{JHEP} {\bfseries 07}
  (2020) 042} [\href{https://arxiv.org/abs/2001.11039}{{\ttfamily
  2001.11039}}].

\bibitem{Alfimov:2014bwa}
M.~Alfimov, N.~Gromov and V.~Kazakov, \emph{{QCD Pomeron from AdS/CFT Quantum
  Spectral Curve}}, \href{https://doi.org/10.1007/JHEP07(2015)164}{\emph{JHEP}
  {\bfseries 07} (2015) 164} [\href{https://arxiv.org/abs/1408.2530}{{\ttfamily
  1408.2530}}].

\bibitem{Gromov:2005gp}
N.~Gromov and V.~Kazakov, \emph{{Double scaling and finite size corrections in
  sl(2) spin chain}},
  \href{https://doi.org/10.1016/j.nuclphysb.2005.12.006}{\emph{Nucl. Phys. B}
  {\bfseries 736} (2006) 199}
  [\href{https://arxiv.org/abs/hep-th/0510194}{{\ttfamily hep-th/0510194}}].

\bibitem{Lyness1985TheEM}
J.~N. Lyness, \emph{The euler maclaurin expansion for the cauchy principal
  value integral}, {\emph{Numerische Mathematik} {\bfseries 46} (1985) 611}.

\bibitem{Gromov:2007aq}
N.~Gromov and P.~Vieira, \emph{{The AdS(5) x S**5 superstring quantum spectrum
  from the algebraic curve}},
  \href{https://doi.org/10.1016/j.nuclphysb.2007.07.032}{\emph{Nucl. Phys. B}
  {\bfseries 789} (2008) 175}
  [\href{https://arxiv.org/abs/hep-th/0703191}{{\ttfamily hep-th/0703191}}].

\bibitem{Gromov:2007ky}
N.~Gromov and P.~Vieira, \emph{{Complete 1-loop test of AdS/CFT}},
  \href{https://doi.org/10.1088/1126-6708/2008/04/046}{\emph{JHEP} {\bfseries
  04} (2008) 046} [\href{https://arxiv.org/abs/0709.3487}{{\ttfamily
  0709.3487}}].

\bibitem{Gromov:2009zza}
N.~Gromov, \emph{{Integrability in AdS/CFT correspondence: Quasi-classical
  analysis}}, \href{https://doi.org/10.1088/1751-8113/42/25/254004}{\emph{J.
  Phys. A} {\bfseries 42} (2009) 254004}.

\bibitem{Gromov:2008ec}
N.~Gromov, S.~Schafer-Nameki and P.~Vieira, \emph{{Efficient precision
  quantization in AdS/CFT}},
  \href{https://doi.org/10.1088/1126-6708/2008/12/013}{\emph{JHEP} {\bfseries
  12} (2008) 013} [\href{https://arxiv.org/abs/0807.4752}{{\ttfamily
  0807.4752}}].

\bibitem{Schafer-Nameki:2010qho}
S.~Schafer-Nameki, \emph{{Review of AdS/CFT Integrability, Chapter II.4: The
  Spectral Curve}},
  \href{https://doi.org/10.1007/s11005-011-0525-6}{\emph{Lett. Math. Phys.}
  {\bfseries 99} (2012) 169} [\href{https://arxiv.org/abs/1012.3989}{{\ttfamily
  1012.3989}}].

\end{thebibliography}\endgroup

\end{document}